\documentclass[twocolumn,superscriptaddress,showpacs]{revtex4}
\usepackage[british,UKenglish,USenglish,english,american]{babel}
\usepackage{amsmath}
\usepackage{amssymb}
\usepackage{latexsym}
\usepackage{enumerate}
\usepackage{amssymb}
\usepackage{graphicx}
\usepackage{slashed} 

\title{{\bf Single and double polarization observables in timelike Compton scattering off proton} }
\begin{document}
\title{{Timelike Compton scattering off the proton and Generalized Parton Distributions.} }

\author{M. Bo\"er$^1$, M. Guidal$^1$ and M. Vanderhaeghen$^2$ \\
\textit{$^1$Institut de Physique Nucl\'eaire d'Orsay, CNRS-IN2P3, Universit\'e Paris-Sud, F-91406 Orsay, France}\\
\textit{$^2$Institut f\"ur Kernphysik and PRISMA Cluster of Excellence,  Johannes Gutenberg Universit\"at, D-55099 Mainz, Germany.}
}

\begin{abstract}
We study the exclusive photoproduction of a lepton pair off the proton
with the aim of studying the proton quark structure via the Generalized 
Parton Distributions (GPD) formalism. After deriving the amplitudes 
of the processes contributing to the $\gamma P\to P' e^+e^-$, the Timelike Compton Scattering and the Bethe-Heitler
process, we calculate all 
unpolarized, single- and double- beam-target spin observables in the 
valence region in terms of GPDs.
\end{abstract}

\maketitle
 
%\newpage
%########################################################################
\section{Introduction}
\label{sec:intro}

The scattering of light on matter, which can generically be called Compton scattering,
is a powerful tool to investigate its inner structure. Nowadays, 
understanding the structure of hadrons in terms of quark and gluon (partons)
degrees of freedom, i.e. the basic constituents of matter known to this day,
is the subject of an intense research effort. Only these past fifteen years or so,
thanks to the emergence of high intensity, high energy (multi-GeV) and
high duty-cycle lepton accelerators, Compton scattering at the partonic level starts
 to be investigated experimentally in an efficient way. 

\begin{figure}[htbp]
\begin{center}
\includegraphics[width=9.5cm,height=6cm]{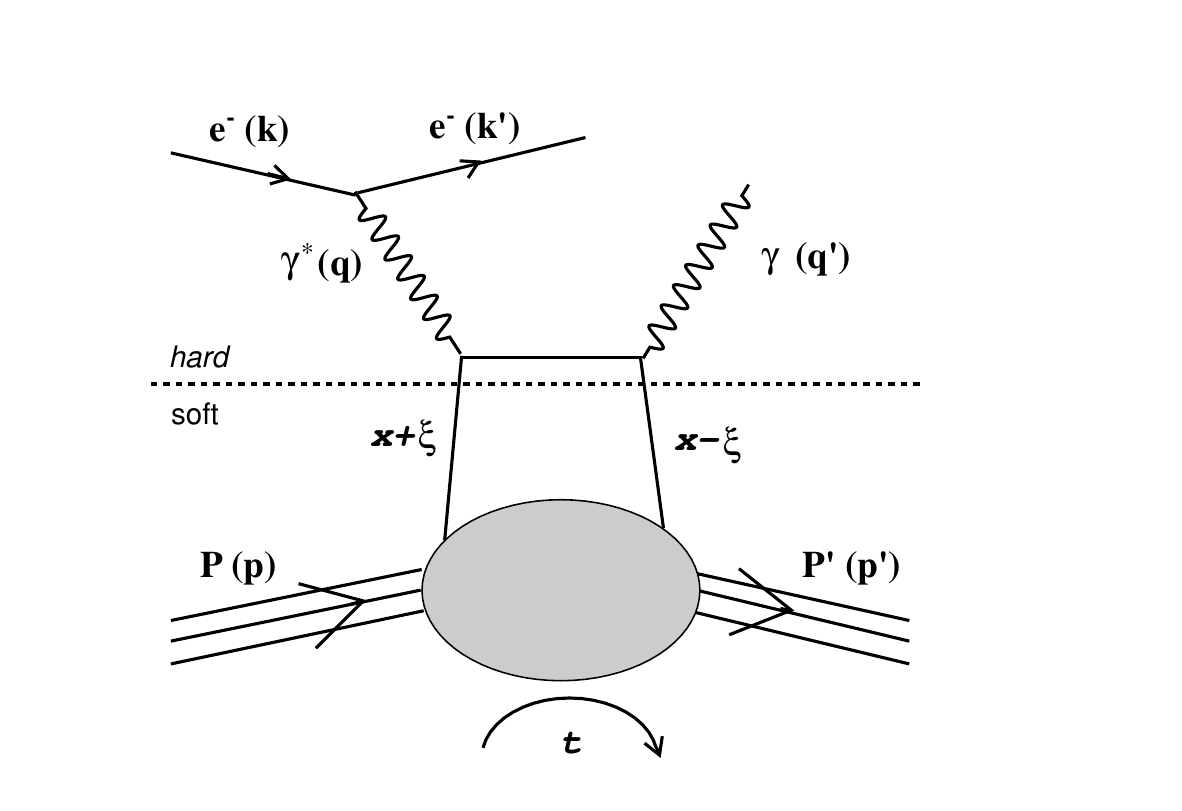}
%DVCS_handbag_bold.pdf}
\includegraphics[width=9.5cm,height=6cm]
{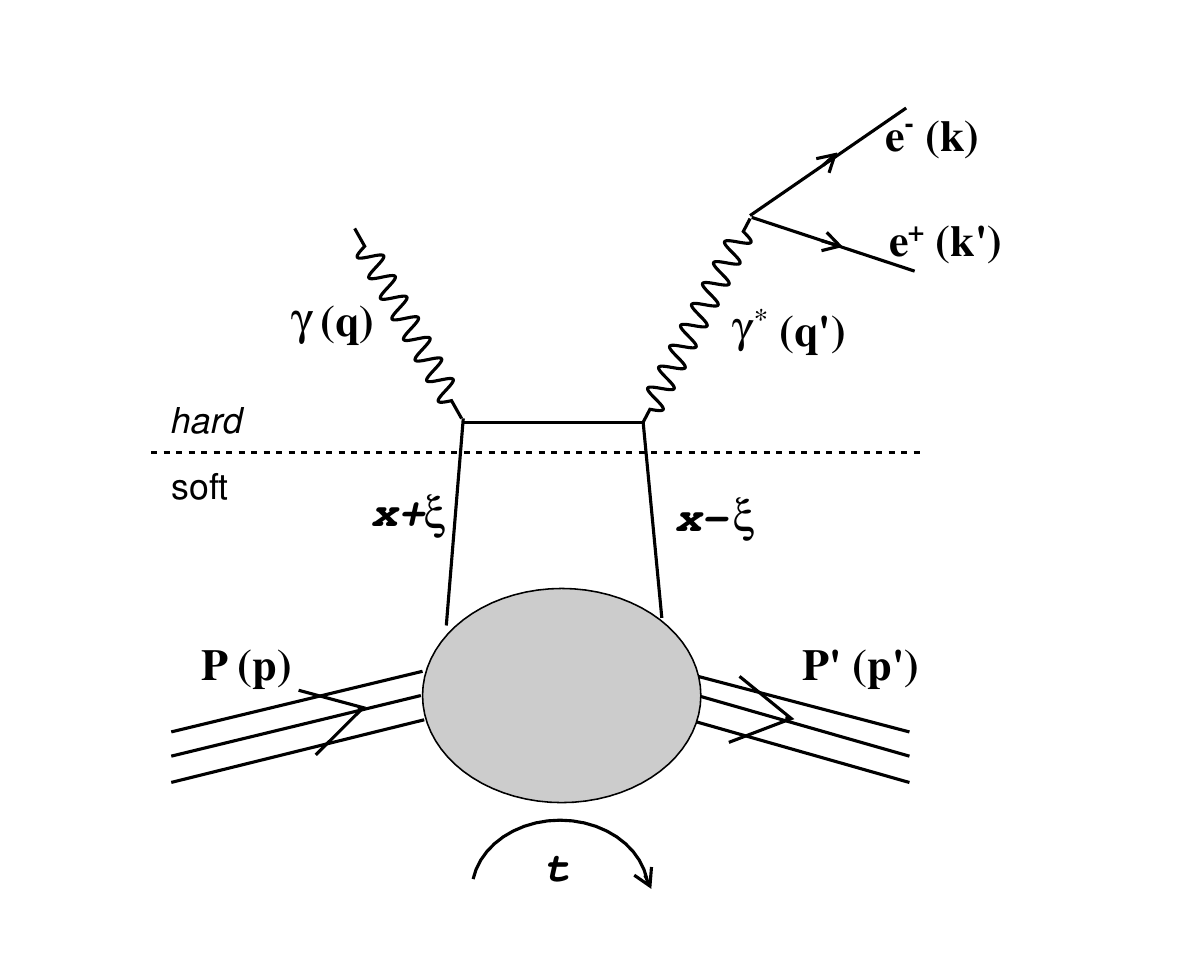}  % TCS_handbag_bold_bis.pdf}
\caption{Top: QCD leading-twist DVCS diagram; Bottom: QCD leading-twist 
TCS diagram (the crossed diagrams of both processes are not shown).
The dashed line illustrates the factorization between }
\label{fig:TCSdiag}
\end{center}
\end{figure}

A particular case of Compton scattering at the partonic level is the Deeply Virtual 
Compton Scattering (DVCS) process on the proton $P$, i.e. $\gamma^* P\to \gamma P'$
where the initial virtual photon $\gamma^*$ is radiated from an incoming lepton
beam (see Fig.~\ref{fig:TCSdiag}-top). It is of particular interest as the amplitude 
of the process allows
to access some essential operators of Quantum Chromo-Dynamics (QCD, the fundamental
theory governing the interactions of quarks and gluons).
Indeed, at sufficiently large virtuality of the
initial photon ($Q^2=(k-k')^2$), the DVCS amplitude can be factorized into
an elementary ``hard" (perturbative) scattering process $\gamma^* q\to \gamma q$ 
(where $q$ is a
quark of the proton), exactly calculable from Quantum Electro-Dynamics as well as perturbative QCD, and a ``soft" (non-perturbative) 
QCD bilocal matrix elements. 
The Fourier transforms into momentum space of these QCD matrix elements 
are the so-called Generalized Parton Distributions (GPDs). In DVCS on the nucleon, at 
QCD leading-twist order there are four quark helicity conserving
 GPDs ($H$, $E$, $\tilde H$ and $\tilde E$) which 
can be accessed and which correspond to the four independent
helicity-spin transitions between the initial quark-proton system and the final one.
The GPDs contain 
a wealth of information on the partonic structure of the proton: 
the longitudinal momentum and transverse space distributions of the quarks and gluons, 
the correlation between these momentum and space
distributions, sensitivity to the quark-antiquark content in the proton, the
quark orbital momentum contribution to the proton spin, etc... 
We refer the reader to the reviews~\cite{goeke,revdiehl,revrady,rpp} on GPDs
for the details of the formalism and of their properties.

DVCS is currently widely investigated, experimentally as well as
theoretically. Ref.~\cite{rpp} compiles all   existing data in the 
valence region and presents some first informations on the partonic
structure of the proton that can be extracted, within some
approximations, from the first DVCS data through the GPD formalism. One example 
of a pioneering result is the first quantitative evaluation of the increase of the 
transverse size of the proton as smaller longitudinal momentum 
fractions of partons are probed. 

However, the GPD information is difficult to extract:
DVCS cross sections are small (of the order of pb), there is the competing 
Bethe-Heitler (BH) process which leads to the same final state $e P \gamma$
but where the final photon is emitted from the incoming or scattered lepton and 
which is therefore not related to GPDs but nevertheless interferes. Furthermore, there is the 
need of
measuring a series of spin (beam and/or target) observables to constrain
the different GPDs, etc... It would therefore be useful to investigate if supplementary and/or 
complementary constraints on GPDs could be obtained from processes other than DVCS.
In this spirit, we investigate in this article the related process of 
Timelike Compton Scattering (TCS) which corresponds to the exclusive photoproduction
of a lepton pair on the proton: 
$\gamma P\to P'\gamma^*\hookrightarrow e^+ e^-$
%$\gamma N\to e^+e^- N'$
and which is displayed in Fig.~\ref{fig:TCSdiag}-bottom. Like for DVCS, 
at sufficiently large virtuality of the  final virtual  photon ($Q'^2=(k+k')^2$), 
it is predicted that the process factorizes 
and is sensitive to GPDs, the same ones   accessed in DVCS. 

The TCS process was originally investigated in terms of GPDs about ten years ago
in Ref.~\cite{Berger:2001xd}. In this pioneering work, analytical formulas  
in terms of GPDs were derived for the unpolarized and the circularly polarized beam
cross sections of the process
% $\gamma p\to \gamma^* p\hookrightarrow e^+ e^-$,
 $\gamma P\to P'e^+e^-$,
i.e. on a proton target. Very recently, a second article continued the investigation
by studying the linearly polarized beam cross section~\cite{Goritschnig}.

However, in order to obtain simple analytical expressions, a few approximations 
were used in the calculation of the TCS amplitude (for instance mass correction terms
of the order of ${m^2} / {Q'^2}$ where $m$ is the mass of the proton were neglected).
In the present work, we waive some of these approximations and present calculations of
 different observables. In addition, 
besides unpolarized cross sections, we study all single and 
double beam-target spin observables. We focus in this article
on a proton target.

This article is organized as follows: in the next  section, %(Sec. ~\ref{sec:Formalism}), 
we present the
general theoretical formalism of the TCS process, in particular the expression 
of the 
QCD leading-twist amplitude in terms of GPDs, and of the accompanying BH process. 
We discuss some experimental considerations in the third section.
In the fourth section, we present our numerical results for the unpolarized cross 
section of the 
$\gamma P\to P'e^+e^-$
%$\gamma N\to N'\gamma^*\hookrightarrow e^+ e^-$ 
process and 
we compare them to the previous work of Ref.~\cite{Berger:2001xd}. In the
fifth section, we present our results for single spin observables (beam
and target) and we compare our work for the beam polarization observables
to the results of Ref.~\cite{Berger:2001xd}. In the sixth section, we 
present our results
for the double-spin beam-target observables.
For each case, we will show the dependence of the observables 
on different GPDs and its sensitivity on different kinematics. 
In the seventh section we show the impact of next-to-leading-twist corrections 
on the cross sections and on the asymmetries. 
We will conclude in the eighth section.

%%%%%%%%%%%%%%%%%%%%%%%%%%%%%%%%%%%%%%%%%%%%%%%
\section{Formalism}
\label{sec:Formalism}

We are studying the process:
\begin{equation}
\gamma(q) \: P(p) \to \:\: P'(p')\: e^- (k)\: e^+(k')\: 
\end{equation}
in a GPD framework, i.e.  when the final photons virtuality
$Q'^2=(k+k')^2$ is sufficiently large and the proton momentum
transfer $t=(p-p')^2$ is sufficiently small so that the factorization 
illustrated in Fig.~\ref{fig:TCSdiag}-bottom can be applied. 
From DVCS, Deep Inelastic Scattering (DIS) and Drell-Yan analysis 
and experiments, it is believed that $Q'^2>$ 2 GeV$^2$ and $-t<$ 1 GeV$^2$
(or $\frac{-t}{Q'^2}<$ 30\%)
should define such a reasonable phase space. Regarding $Q'^2$, one should
also avoid regions where one can have the production of vector mesons,
decaying into $e^+e^-$ pairs (for instance, the broad $\rho^\prime(1700)$). 
Finally, one should consider
the squared center-of-mass energy of the incoming photon and target proton
$s=(q+p)^2 \gtrsim$ 4 GeV$^2$, in order to minimize possible contributions
from the Dalitz decay of proton resonances. 

\subsection{Kinematics}
\label{subsec:kin}

We will use the notation of Ji \cite{Ji97} for GPDs, i.e. GPDs depend
on the three variables $x$, $\xi$ and $t$ where the quark longitudinal momentum 
fractions $x$ and $\xi$ are defined w.r.t. the average proton momentum 
$P$ and proton momentum transfer $\Delta$, respectively. We therefore define: 
\begin{eqnarray}
&&P = {1 \over 2} \left(p + p' \right), \\
&&\Delta=(p'-p)=(q-q'),
\end{eqnarray}
and we also introduce the average photon momentum 
\begin{eqnarray}
&&\bar{q} = {1 \over 2} \left(q + q' \right) .
\end{eqnarray}
GPDs are matrix elements of QCD operators which are defined at equal light-cone time.
It is therefore convenient to use a frame where the 
$\bar{q}$ and $P$ momenta
are collinear along the $z$-axis and in opposite directions.
We define the lightlike vectors along the positive and negative $z$ directions 
as:  
\begin{eqnarray}
&&\tilde p^\mu = P^+/\sqrt{2} (1,0,0,1), \\  
&&n^\mu = 1/P^+ \cdot 1/\sqrt{2} (1,0,0,-1),
\end{eqnarray}
and we define the light-cone components $a^\pm$ by $a^{\pm} 
\equiv (a^0 \pm a^3)/\sqrt{2}$.
We have $\tilde{p}^2=n^2=0$ and $\tilde{p}\cdot n=1$. In this light-cone frame,
introducing the $+$ components of the $\Delta$ and $\bar{q}$ four-vectors, $\tilde{\xi}$ and 
$\tilde{\xi'}$ respectively, the various four-vectors involved can be decomposed as:
\begin{eqnarray}
&&P^{\mu} = \tilde{p}^{\mu}+\frac{\bar{m}^2}{2}\:n^{\mu}, \\
&&\bar{q}^{\mu} = -\tilde{\xi'}\:\tilde{p}^{\mu} - \frac{\bar{q}^2}{2\tilde{\xi'}}\:n^{\mu},\\ 
&&\Delta^{\mu} = -2\tilde{\xi}\tilde{p}^{\mu} + \tilde{\xi}\bar{m}^2n^{\mu} + \Delta_{\perp}^{\mu},  
%%&&q'^{\mu} = -2\xi'\:\tilde{p}^{\mu} - \frac{Q'^2}{4\xi'}\:n^{\mu} \\
%%&&q^{\mu} = -2(\xi'+\xi)\:\tilde{p}^{\mu} + (\xi\bar{m}^2-\frac{Q'^2}{4\xi'})\:n^{\mu}\\
\end{eqnarray}
with $\bar{m}^2 = m^2-\frac{\Delta^2}{4}$ and $m$ is the proton mass. 
 We relate the final photon virtuality to the average photon momentum: 
\begin{eqnarray}
\bar{q}^2=Q'^2 /2 -\Delta^2/4.
\end{eqnarray}

We have to relate the light-cone momentum fractions $\tilde{\xi}$
and $\tilde{\xi'}$ to the kinematical variables which are experimentally
accessible. To do so, we introduce the variables $\xi$ and $\xi'$:
\begin{eqnarray}
&&\xi' = - \frac{\bar{q}^2}{2P.\bar{q}} = \frac{-Q'^2+\Delta^2/2}{2(s-m^2)+\Delta^2-Q'^2},\\
&&\xi = -\frac{\Delta.\bar{q}}{2P.\bar{q}} = \frac{Q'^2}{2(s-m^2)+\Delta^2-Q'^2}.
\end{eqnarray}

The light-cone momentum fractions $\tilde{\xi}$
and $\tilde{\xi'}$ are related to the kinematical variables $\xi$ and $\xi'$ by:
\begin{eqnarray} 
&&\tilde{\xi}  = \xi \cdot \frac{
1+\tilde{\xi}'^2\: {\bar{m}^2}/{\bar{q}^2}
}{
1-\tilde{\xi}'^2\: {\bar{m}^2}/{\bar{q}^2}
}, \\
&&\tilde{\xi}' = \xi' \cdot \frac{2}{
1+\sqrt{ 1-4\xi'^2\: {\bar{m}^2}/{\bar{q}^2} }
}. 
\end{eqnarray}

In the asymptotic limit, where mass and $\Delta$ terms are 
neglected relatively to $Q'^2$, we have: 
\begin{equation}
\tilde{\xi} = \xi = -\tilde{\xi}' = -\xi' = \frac{Q'^2}{2s-Q'^2}.
%\xi = -\xi^{'}=\frac{Q'^2}{2s - Q'^2}
\end{equation}

%%%%%%%%%
\subsection{Timelike Compton amplitude}
\label{subsec:Comptonampl}

\begin{figure}[htbp]
\begin{center}
%\includegraphics[width=5.2cm,height=4.5cm]
%{Figures/fey_TCS_Amp1.pdf} \hspace*{-1.2cm}
\includegraphics[width=7cm,height=6cm]
{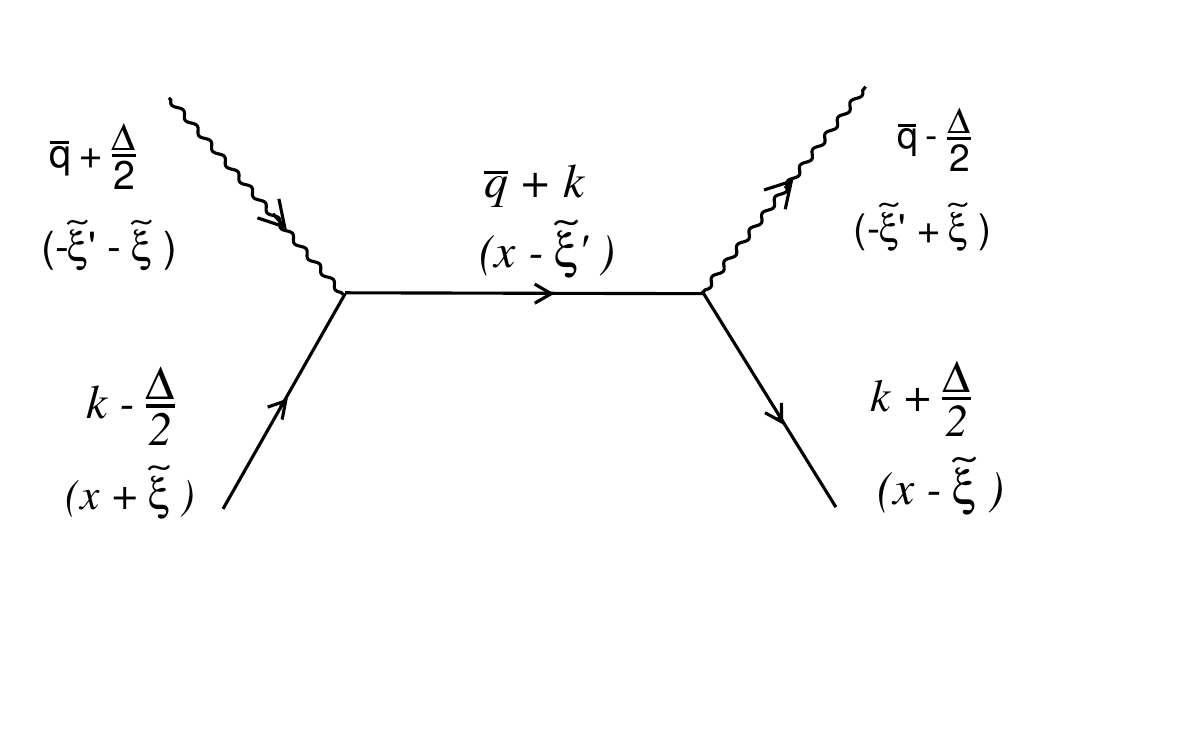}%gr1.png} 
\\
\vspace*{-1cm}
\includegraphics[width=7cm,height=6cm]
{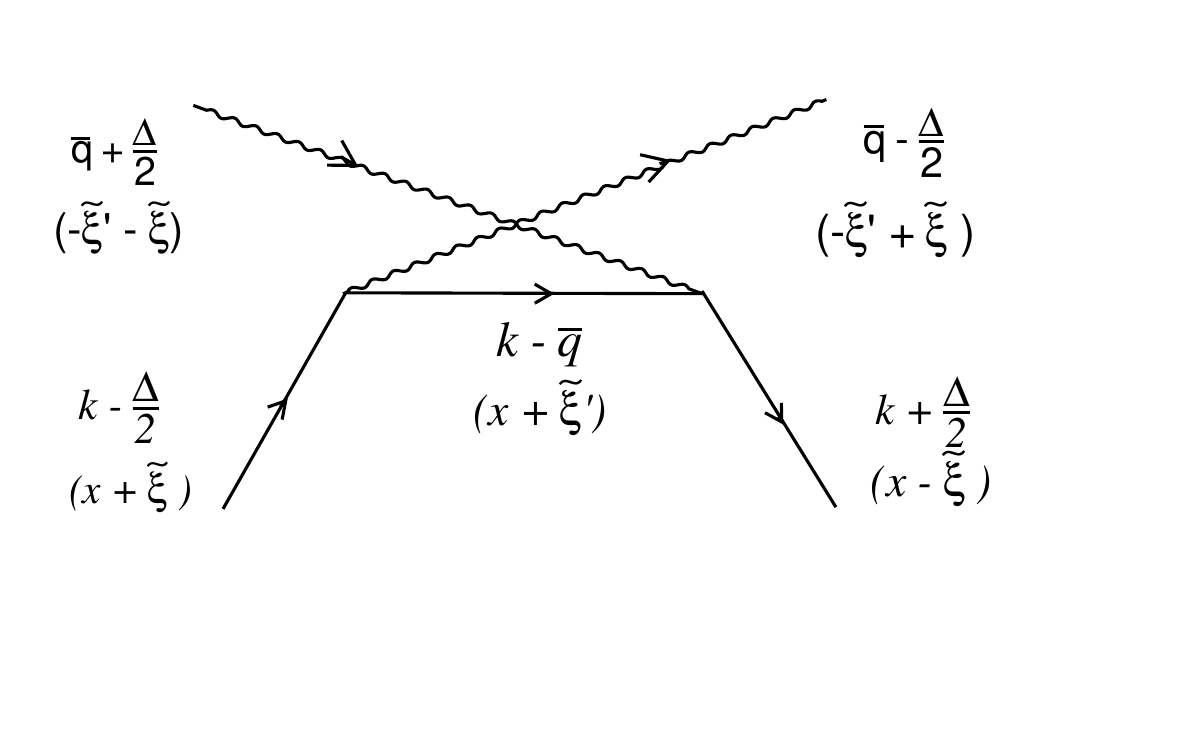} %gr2.png}
\vspace*{-1cm}
%{Figures/fey_TCS_AmpCross.pdf} 
\caption{Leading order diagrams for the process $\gamma q\to\gamma q$.
We indicate between parenthesis the $+$ components of the four-vectors.
}
\label{fig:TCS_diag}
\end{center}
\end{figure}

The two diagrams of Fig.~\ref{fig:TCS_diag} have to be calculated.
The leading order amplitude reads:
\begin{eqnarray} 
\label{eq:TH_TCS}
&&T_H^{TCS} \propto
\frac{1}{(k+\bar{q})^2 +i\epsilon } \gamma^{\nu}\: (\slashed{k}+\slashed{\bar{q}})\:\gamma^{\mu}\nonumber \\
&&+ \frac{1}{(k-\bar{q})^2 +i\epsilon } \gamma^{\mu}\: (\slashed{k}-\slashed{\bar{q}})\:\gamma^{\nu} \nonumber \\
&&\propto \frac{1}{2} \left[
\slashed{n}\: ( \tilde{p}^{\mu}n^{\nu} + \tilde{p}^{\nu}n^{\mu} -g^{\mu\nu}) 
 \:.\: (  \frac{1}{x-\tilde{\xi'}+i\epsilon} +\frac{1}{x+\tilde{\xi'}-i\epsilon}  )\right. \nonumber \\
&&\left. - \slashed{n}\gamma_5\:i\epsilon^{\mu\nu\kappa\lambda}\:n_{\lambda}\:\tilde{p}_{\kappa} 
\:\: (  
 \frac{1}{x-\tilde{\xi'}+i\epsilon} -\frac{1}{x+\tilde{\xi'}-i\epsilon} )\right].
\end{eqnarray}
%\begin{eqnarray}
%\label{eq:TH_TCS}
%&&T_H^{TCS} \propto \frac{1}{(k+q)^2 +i\epsilon } \gamma^{\nu}\: (\slashed{k}+\slashed{q})\:\gamma^{\mu} \nonumber\\
%&&+ \frac{1}{(k-q)^2 +i\epsilon } \gamma^{\nu}\: (\slashed{k}-\slashed{q})\:\gamma^{\mu}  \nonumber\\
%&&\propto \frac{1}{2} \left[
%\slashed{n}\: ( \tilde{p}^{\mu}n^{\nu} + \tilde{p}^{\nu}n^{\mu} -g^{\mu\nu}) 
% \:.\: (  \frac{1}{x+\xi'+i\epsilon} +\frac{1}{x-\xi'-i\epsilon}  )\right. \nonumber \\
%&&\left. - \slashed{n}\gamma_5\:i\epsilon^{\mu\nu\kappa\lambda}\:n_{\lambda}\:\tilde{p}_{\kappa} 
%\:\: (  
% \frac{1}{x+\xi'+i\epsilon} -\frac{1}{x-\xi'-i\epsilon} )\right]
%\end{eqnarray}

The full TCS amplitude, corresponding to the diagram of Fig.~\ref{fig:TCSdiag}-bottom 
(plus the associated crossed diagram) reads then 
\begin{equation}
\label{eq:T_TCS}
T^{TCS} = -\frac{e^3}{q'^2} \:\bar{u}(k)\:
\gamma^{\nu} \: v(k')\:
\epsilon^{\mu}(q)\:
H_{\mu\nu}^{TCS},
\end{equation}
with, in the Bjorken limit where $\xi = -\tilde{\xi}'$,  
\begin{eqnarray}
\label{eq:H_TCS}
&&H_{\mu\nu}^{TCS} \\ \nonumber
&&= \frac{1}{2}\:(-g_{\mu\nu})_{\perp}
\int\limits_{-1}^{1}dx\: \left(
\frac{1}{x-\xi-i\epsilon} + \frac{1}{x+\xi + i\epsilon}
\right) \\ \nonumber
&&.\left(
H(x,\xi,t) \bar{u}(p')\slashed{n}u(p)+ E(x,\xi,t)\bar{u}(p') i \sigma^{\alpha\beta}n_{\alpha} \frac{\Delta_{\beta}}{2m}\:u(p)
\right) \\ \nonumber
&&-\frac{i}{2}(\epsilon_{\nu\mu})_{\perp} 
\int\limits_{-1}^{1}dx\: \left(
\frac{1}{x-\xi-i\epsilon} - \frac{1}{x+\xi + i\epsilon}
\right) \\ \nonumber
&&.\left(
\tilde{H}(x,\xi,t)\bar{u}(p')\slashed{n}\gamma_5 \:u(p) 
 + \tilde{E}(x,\xi,t)\bar{u}(p')\gamma_5 \frac{\Delta.n}{2m}\:u(p)
\right),
\end{eqnarray}
where we used the metric: 
\begin{eqnarray}
%\begin{split}
&&(-g_{\mu\nu})_{\perp}= -g_{\mu\nu} + \tilde{p}_{\mu}n_{\nu}   + \tilde{p}_{\nu} n_{\mu} \:\:,\\ \nonumber
&&(\epsilon_{\nu\mu})_{\perp}=\epsilon_{\nu\mu\alpha\beta}\: n^{\alpha}\: \tilde{p}^{\beta}.
%\end{split}
\end{eqnarray}

The GPDs entering Eq.~\ref{eq:H_TCS} are proton GPDs, i.e. they
read, in terms of quark flavors:
\begin{equation}
H_{TCS}(x,\xi,t) \,=\,{4 \over 9} H^{u/p} \,+\, {1 \over 9} H^{d/p}\,+\, {1 \over 9} H^{s/p}.
\label{eq:dvcsfla}
\end{equation}
In this work, we will take the GPD parametrizations from the
VGG model~\cite{Vanderhaeghen:1998uc, Vanderhaeghen:1999xj,goeke,Guidal:2004nd},
which are summarized in Ref.~\cite{rpp} and based on the Radyushkin double-distribution
ansatz for the ($x$,$\xi$)-dependence~\cite{Radyushkin:1998es,Radyushkin:1998bz,
Mueller:1998fv} and on a Reggeized ansatz for the $t$-distribution~\cite{goeke,Guidal:2004nd}. At a couple of instances, in order to 
estimate the model dependence of our calculations, we will use the factorized 
ansatz for the $t$-dependence of the $H$ GPD~\cite{Vanderhaeghen:1998uc}.
Also, we will occasionally study the sensitivity of observables to the so-called D-term~\cite{Polyakov:1999gs},
which is included in the VGG model and whose parametrization can be found as well in Refs.~\cite{Vanderhaeghen:1998uc, Vanderhaeghen:1999xj,goeke,Guidal:2004nd}.

\subsection{Gauge invariance}
\label{subsec:gi}

The TCS amplitude is not exactly gauge invariant, only in the Bjorken limit. 
To restore gauge invariance, twist-3 corrections of order $\Delta_{\perp}/Q$
are needed. As a first step in adressing this issue and estimating its effect, 
we propose to add a correction term to the twist-2 vector part tensor
as follows:
\begin{eqnarray}
&&H^{\mu\nu} = H^{\mu\nu}_{LO} 
- \frac{P^{\mu}}{2P\cdot\bar{q}  } \cdot (\Delta_{\perp})_{\kappa}  \cdot H^{\kappa\nu}_{LO} \\ \nonumber
&&
+ \frac{P^{\nu}}{2P\cdot\bar{q}  } \cdot (\Delta_{\perp})_{\lambda}  \cdot H^{\mu\lambda}_{LO} 
\\ \nonumber
&&- \frac{P^{\mu} P^{\nu}}{ 4 (P\cdot\bar{q})^2  } \cdot (\Delta_{\perp})_{\kappa}\cdot (\Delta_{\perp})_{\lambda} \cdot H^{\kappa\lambda}_{LO} \:, 
\end{eqnarray}
where $H_{LO}^{\mu\nu}$ stands for the tensor of Eq.~\ref{eq:H_TCS}.
This is a generalization of the prescription proposed in Refs.~\cite{Vanderhaeghen:1999xj,Guichon:1998xv} for DVCS
(we also refer the reader to Refs.~\cite{Anikin:2000em,Belitsky:2000vx,Radyushkin:2000ap} 
for further discussions on the issue of gauge invariance in the DVCS amplitude).
One can readily check that $H^{\mu \nu}$ respects gauge invariance both w.r.t. initial and final photons, i.e.
$q_\mu H^{\mu \nu} = 0$, and $q'_\nu H^{\mu \nu} = 0$.
The impact of adding this correction  to the observables is shown 
at the end of this paper, in the section~\ref{sec:Corrections}.

%%%%
\subsection{The Bethe-Heitler amplitude}
\label{subsec:BHampl}

\begin{figure}[htbp]
\begin{center}
\includegraphics[width=8.5cm,height=6cm]
{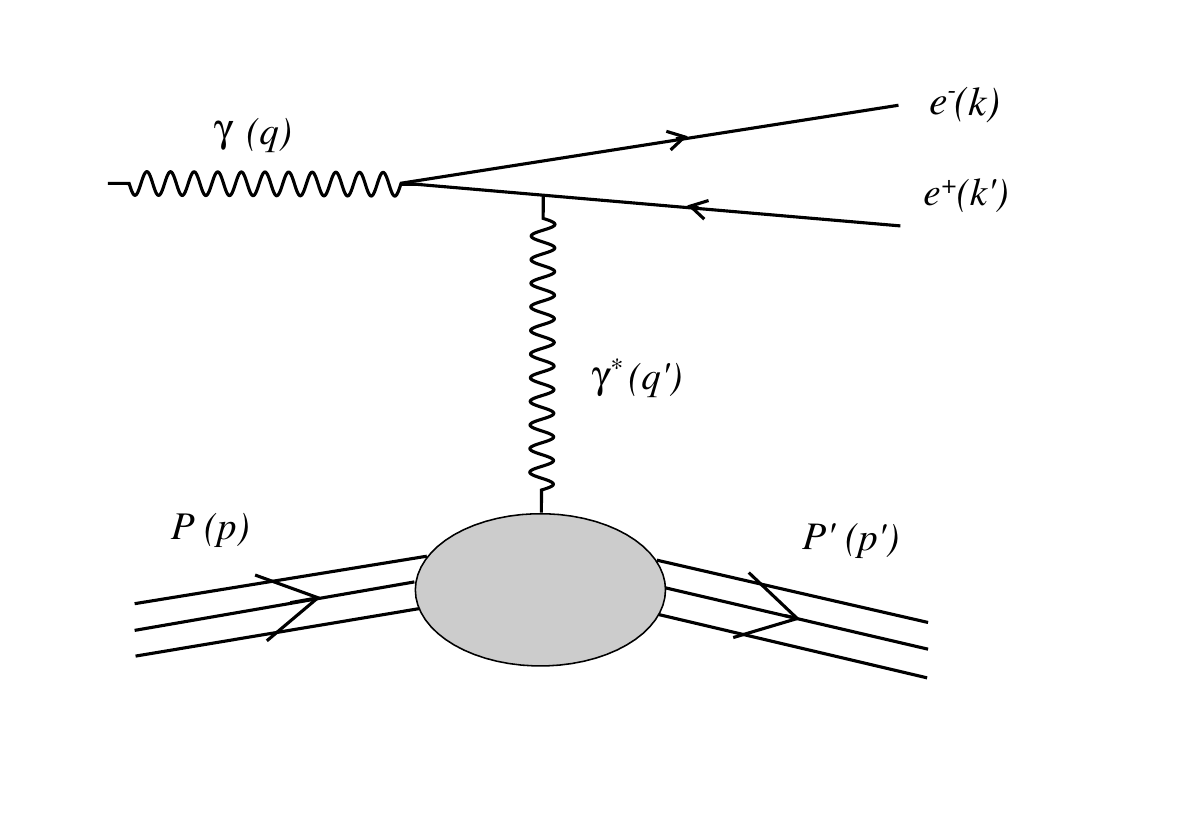}
\includegraphics[width=8.5cm,height=6cm]
{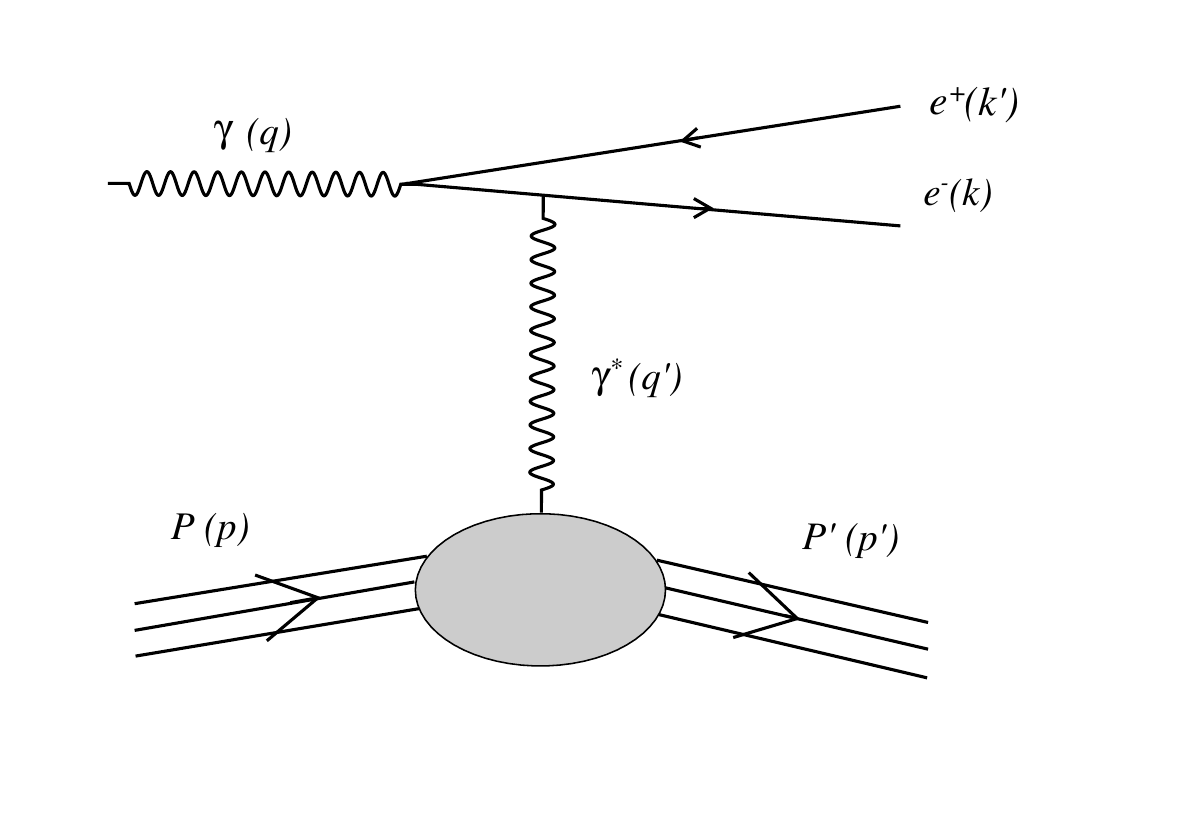}
\caption{The Bethe-Heitler diagrams.}
\label{fig:BHdiag}
\end{center}
\end{figure}

The TCS process is accompanied by the BH process, involving the  two diagrams
which are presented in Fig.~\ref{fig:BHdiag}. Their amplitude reads:
\begin{eqnarray}
&&T^{BH} = -\frac{e^3}{\Delta^2} \:\bar{u}(p')\: \Gamma^{\nu}\: u(p)\: \epsilon^{\mu}(q) \:
 \:\\
&& \bar{u}(k)\left(
\gamma_{\mu}   
\frac{\slashed{k}-\slashed{q}}{(k-q)^2}
\gamma_{\nu}
+
\gamma_{\nu}   
\frac{\slashed{q}-\slashed{k'}}{(q-k')^2}
\gamma_{\mu}
\right)
\: v(k'), 
\end{eqnarray}
with the virtual photon-proton electromagnetic vertex matrix  
\begin{equation}
\Gamma^{\nu} = \gamma^{\nu} \:F_1(t)  + \frac{i \sigma^{\nu\rho} \Delta_{\rho} }{2\:m}\:F_2(t). \:
\end{equation}
The BH amplitude depends on the proton Dirac and Pauli form 
factors $F_1(t)$ and $F_2(t)$ which, at small $t$ can be considered to be 
known with good accuracy. In this work, we take the parametrizations
issued from Refs.~\cite{Gayou:2001qd,Brash:2001qq}.

%%%%%
\subsection{Cross section}

\begin{figure}[htbp]
\begin{center}
\vspace*{-1cm}\hspace*{-1cm}
\includegraphics[width=10cm,height=8.5cm,page=1]
{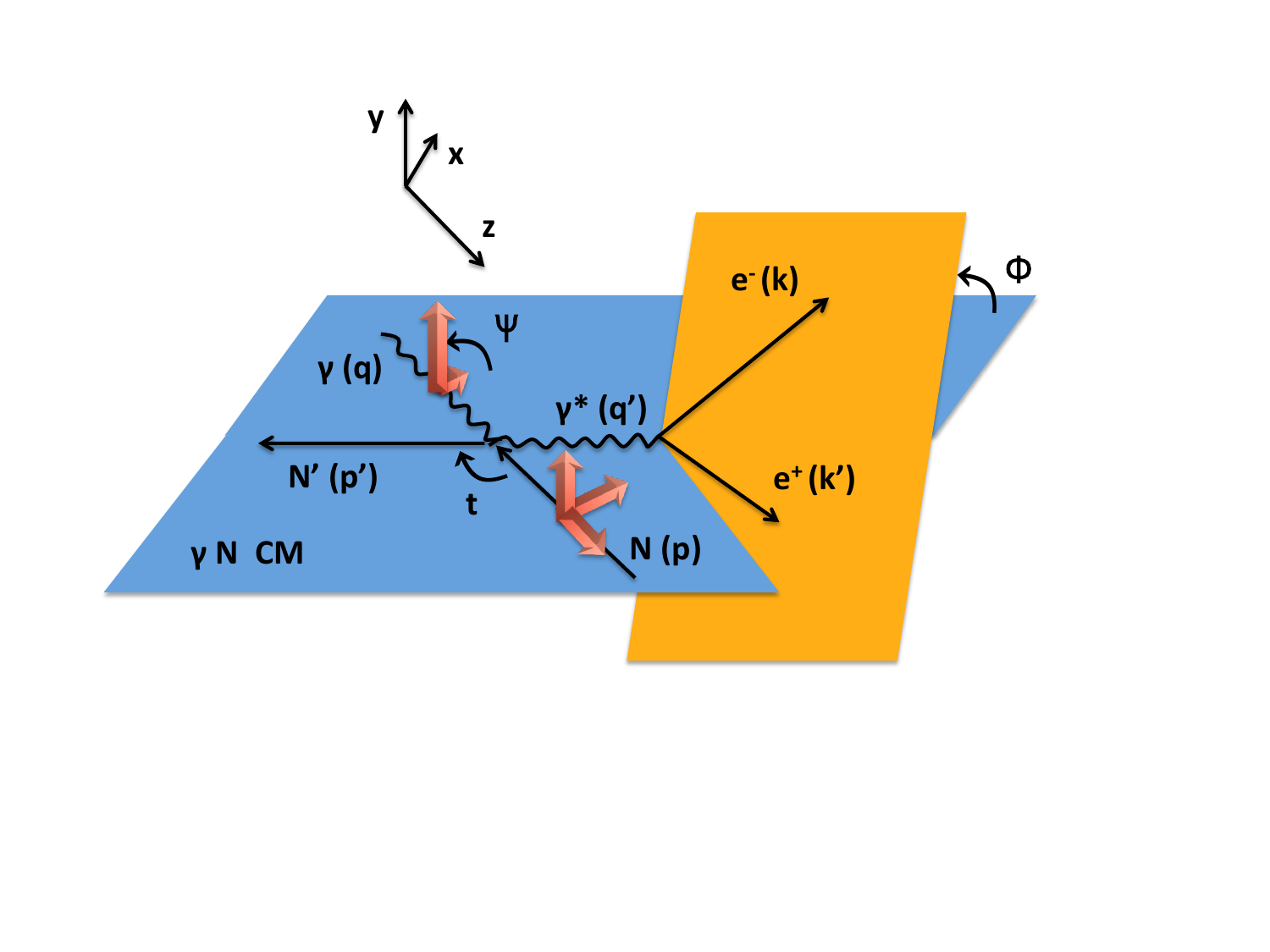}
\vspace*{-1.5cm}\hspace*{-1cm}
\includegraphics[width=10cm,height=8.5cm,page=2]
{imageangles_bis.pdf}
\vspace*{-1cm} 
\caption{Top panel: scheme of the reaction plane $\gamma P\to \gamma^* P'$ (for TCS) in 
the $\gamma P$ CM frame and illustration of some of the kinematic variables.
Red arrows show the polarization vectors of the photon beam and proton target.\\ 
Bottom panel: decay angles in the $e^-e^+$ CM frame. $\phi$ and $\theta$ are respectively 
the azimuthal and polar angles between the $e^-$ direction and the virtual photon direction in the $\gamma^*-P'$ CM frame.
}
\label{fig:angles}
\end{center}
\end{figure}

At fixed beam energy $E_\gamma$ or longitudinal 
momentum transfer $\xi$, there are four independent kinematical variables for the process
%$\gamma N(p) \to N' \gamma^* \hookrightarrow e^- e^+$. 
$\gamma(q) P(p) \to P'(p')  e^-(k) e^+(k')$. 
A natural choice is to
take: $Q'^2$ and $t=\Delta^2$ that we already defined,  
and the two angles $\theta$ and $\phi$ of the electron in the $\gamma^*$ 
center-of-mass
(with the $z'$-axis along the direction of the $\gamma^*$ in the $\gamma^*-P'$ %check
center of mass).
We illustrate in Fig.~\ref{fig:angles} the different variables involved.
In addition, we display the polarization
angle $\Psi$ between 
%the linear polarization
the polarization vector of the incoming photon 
 and the scattering plane  in the $\gamma^*-p$ C.M..
%and the target nucleon polarization vectors, which will be of later use. 

The 4-differential unpolarized cross section is then expressed as:
\begin{eqnarray}
%\begin{split} 
&&\frac{d^4\sigma}{dQ'^2dtd\Omega }({\gamma P \to P' e^+e^-}) \\ \nonumber
&&= \frac{1}{(2\pi)^4}\frac{1}{64}\frac{1}{(2m E_\gamma)^2}\mid T^{BH} + T^{TCS}\mid^2,
%\end{split}
\end{eqnarray}
where $\mid T^{BH} + T^{TCS}\mid^2$ is averaged over the target proton helicities 
and beam polarizations and summed over the final proton helicities.

%The angular dependance in $\Psi_s$ reads %ref
%\begin{equation} \label{eq:psiexpress}
%\begin{split}
%\frac{d\sigma}{d\Omega} = \sigma_{unpol} \:\: [\:\:
%1 - P_L\: a_{LU} \:cos2\Psi 
%+ P_x \: ( -P_L \:a_{Lx} \:sin2\Psi - P_0\:a_{0x}) \\
%- P_y\: (-a_{Uy} + P_L\:  a_{Ly} \:cos2\Psi)
%- P_z\: (-P_L \: a_{Lz} \: sin2\Psi + P_0\: a_{0z}    )
%\:\:]
%\end{split}
%\end{equation}
%with the coefficients $a_{ii'}$ and the corresponding polarization degrees, $P_L$ for the linear beam polarization,
%$P_0$ for the circular beam polarization and $P_x$, $P_y$, $P_z$ for the target polarization degree along $x,y,z$ axis.
%This dependancies are modifyed when we are looking at different values of $\phi$ and of $\theta$ (see
%section \ref{subsubsec:psi}). %% + explication

%%%%%

%\pagebreak
\newpage
\section{Experimental considerations}
\label{sec:expkine}

The only data for the process $\gamma P \to P' e^- e^+$ which are
in the phase space of concern for our study, have been collected 
and analyzed a few years ago by the CLAS collaboration using the
$\approx$ 6 GeV electron beam of Jefferson Lab. The data were actually
obtained in ``quasi-photoproduction" mode. This means that the scattered electron 
from the beam is not detected in CLAS and is considered to
be in almost the same direction as the beamline. This results 
in very low $Q^2$ electroproduction, i.e. ``quasi-photoproduction".
This pilot   analysis 
can be found in Ref.~\cite{Rafo}. The $\approx$ 6 GeV beam energy,
combined with the $\approx$ 10$^{34}$cm$^{-2}$s$^{-1}$ luminosity,
allowed to reach maximum $Q'^2$ values of 3 GeV$^2$, which corresponds 
to an invariant mass
of the $e^+e^-$ system $M_{e^+e^-}$ of $\approx$ 1.8 GeV. This is close to the mass
of several vector mesons decaying into $e^+e^-$, in particular the 
broad $\rho '$(1700).
In order to have a TCS interpretation as clean as possible, it is of course advisable
to avoid such resonances which contribute to the
process $\gamma P \to P' e^- e^+$. The data analysis of Ref.~\cite{Rafo}
is therefore difficult to interpret in terms of GPDs but
it nevertheless demonstrates that the process $\gamma P \to P' e^- e^+$ 
can be  measured at JLab.

The forthcoming JLab energy upgrade to 11 GeV allows to explore a $Q'^2$ region 
between 4 and 9 GeV$^2$ which corresponds to a $M_{e^+e^-}$ region between 2 and 
3 GeV,
i.e. a vector meson resonance-free region between the $\rho '$ and the $J/\Psi$.
In the case of CLAS12, the upgraded CLAS detector associated to the
JLab energy increase, there is also a luminosity gain of a factor 10.
These upgrades have led to the first dedicated TCS accepted
 proposal at JLAB ~\cite{JLABprop}. It will use a similar
``quasi-photoproduction" technique as used in the pioneer 6 GeV analysis.
In addition, there will be the improvement of the detection of the low 
$Q^2$ scattered electron via a dedicated tagging equipement, supplementing
CLAS12. This allows, besides the measurement of unpolarized cross sections, 
to obtain linearly polarized photons observables, by measuring 
the azimuthal dependence of the scattered electron. Finally,
with a polarized electron beam, which is available at JLab,
one can access circularly polarized
photons observables. It is therefore expected 
that in the next few years, numerous $\gamma P \to P' e^- e^+$ data
which can lend   to GPD interpretation will be available. 

%We show in Fig.~\ref{fig:kine} the kinematical domain which can be access by the JLAB-12 GeV.
%We display in blue the ($\xi$, $Q'^2$) phase space accessible with an 11 GeV
%beam with the two cuts: $Q'^2\in[4,9]$ GeV$^2$ and $-t\in[0,2]$ GeV$^2$. The motivation
%is to, respectively, stay in a vector mesons resonance free region and minimize 
%higher-twists corrections to the TCS formalism which grow with $\frac{t}{Q'^2}$.
%We overlap in red in this same figure the ($\xi$, $Q^2$) phase space accessible with an 11 GeV
%beam for DVCS. We have applied the cuts: $-t\in[0,2]$ GeV$^2$, for the same reason
%as for TCS, and $s>4$ GeV$^2$ (where $s=(\gamma^*+p)^2$) 
%in order to stay above the baryon resonance region.

\begin{figure}[htbp]
\begin{center}
\hspace*{-0.7cm}
\includegraphics[width=8.5cm,height=7cm]
{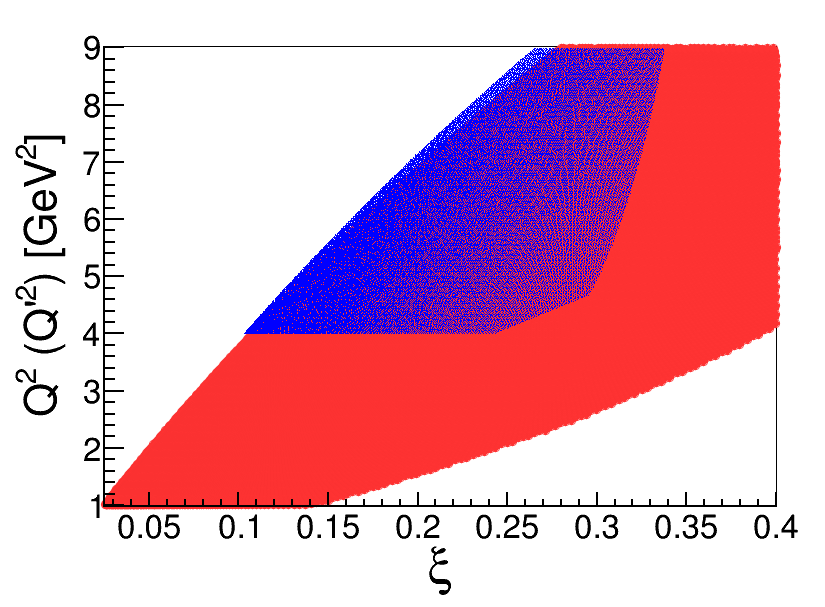}
\caption{Kinematical domain accessible as a function of $\xi$
and $Q^2$ for DVCS (red plain surface) and in $\xi$ and $Q'^2$
 for TCS (blue dotted surface) with an 11 GeV electron beam.
For DVCS, the cuts $-t\in[0,1]$ GeV$^2$ and $s>4$ GeV$^2$
have been applied and for TCS, the cuts $E_{\gamma}\in[5,11]$ GeV,
$-t\in[0,1]$ GeV$^2$ and $Q'^2\in[4,9]$ GeV$^2$ have been applied.
}
\label{fig:kine}
\end{center}
\end{figure}

We show in Fig.~\ref{fig:kine} the kinematical domain which can be accessed 
with the upgraded JLab.
We display in blue the ($\xi$, $Q'^2$) phase space accessible for TCS
with an 11 GeV electron 
beam, assuming that the real photon is provided by bremsstrahlung of the electron and
that its  energy is in $E_{\gamma}\in[5,11]$ GeV. We have applied
two cuts: $Q'^2\in[4,9]$ GeV$^2$ and $-t\in[0,1]$ GeV$^2$. The motivations
are respectively to stay in the region 
free of  vector mesons resonances  
and minimize 
higher twist  corrections to the TCS formalism, which grow with $\frac{t}{Q'^2}$.
We overlap in red in this same figure the ($\xi$, $Q^2$) phase space accessible with an 11 GeV
beam for DVCS. We have applied the cuts: $-t\in[0,1]$ GeV$^2$, for the same reason
as for TCS, and $s>4$ GeV$^2$ in order to stay above the baryon resonance region.

One notes the large intersection between the DVCS and the TCS phase spaces.
Measurements of observables sensitive to GPDs in the 
common ($\xi, Q^2$) region by both processes should bring strong constraints on the extraction of GPDs and tests of factorization and universality.

%%%ICI AJOUTER QQ CHOSE
%But one has to realize that the $t$ region that is accessed for TCS and for DVCS 
%in a certain ($\xi,Q^2$) domain is not the same. 
%Indeed, in DVCS, large $Q^2$ accesses large $-t$ while, for TCS, 
%large $\xi$ accesses large $-t$.

%Indeed, in DVCS, large $Q^2$ accesses large $-t$ while, for TCS, 
%large $\xi$ accesses large $-t$. 

 %% on va etendre le domaine en t a un xi donne qu'on a avec le DVCS grace au TCS car c'est complementaire => en angla	s. 
% on peut discuter sur la figure: sur DVCS, la cut en t coupe le "haut", a droite c'est la cut en W et a gauche c'est la cut en nu
% pour TCS: coup?? ?? droite (grand xi) par la cut en t et ?? gauche par la cut en energie, le haut est coup?? par la cut en s (grand Q'2). 

We now present our results for the calculations of the unpolarized
cross sections, single spin asymmetries and double-spin asymmetries,
respectively in sections ~\ref{sec:comparison}, ~\ref{sec:BSA} and ~\ref{sec:BTSA}.
In these sections, all calculations will be done in the Bjorken limit
and at leading order (we refer the reader to Refs.~\cite{Pire:2011st,Muller:2012yq,Moutarde:2013qs} for works on 
Next-To-Leading order corrections in $\alpha_s$ to the TCS amplitude). 
In section ~\ref{sec:Corrections}, we will study
the effect of keeping the exact kinematics presented in section~\ref{subsec:kin}
and of the gauge-invariance restoration prescription described
in section~\ref{subsec:gi}.

%########################################################################
\section{Unpolarized cross section}
\label{sec:comparison}

\begin{figure}[htbp]
\begin{center}
%\hspace*{-1.1cm}
\includegraphics[width=5cm,height=4.5cm]
{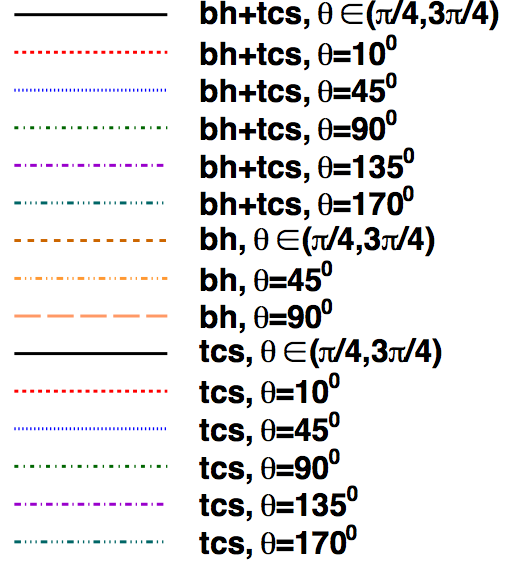}
\hspace*{-1.1cm}
\includegraphics[width=10.5cm,height=11cm]
%%%{Figures/evolphitheta.pdf}
%{figs/cross_phi_asymp_alltheta_sumbhtcs_Q7eta2t4_logy_0.pdf}
{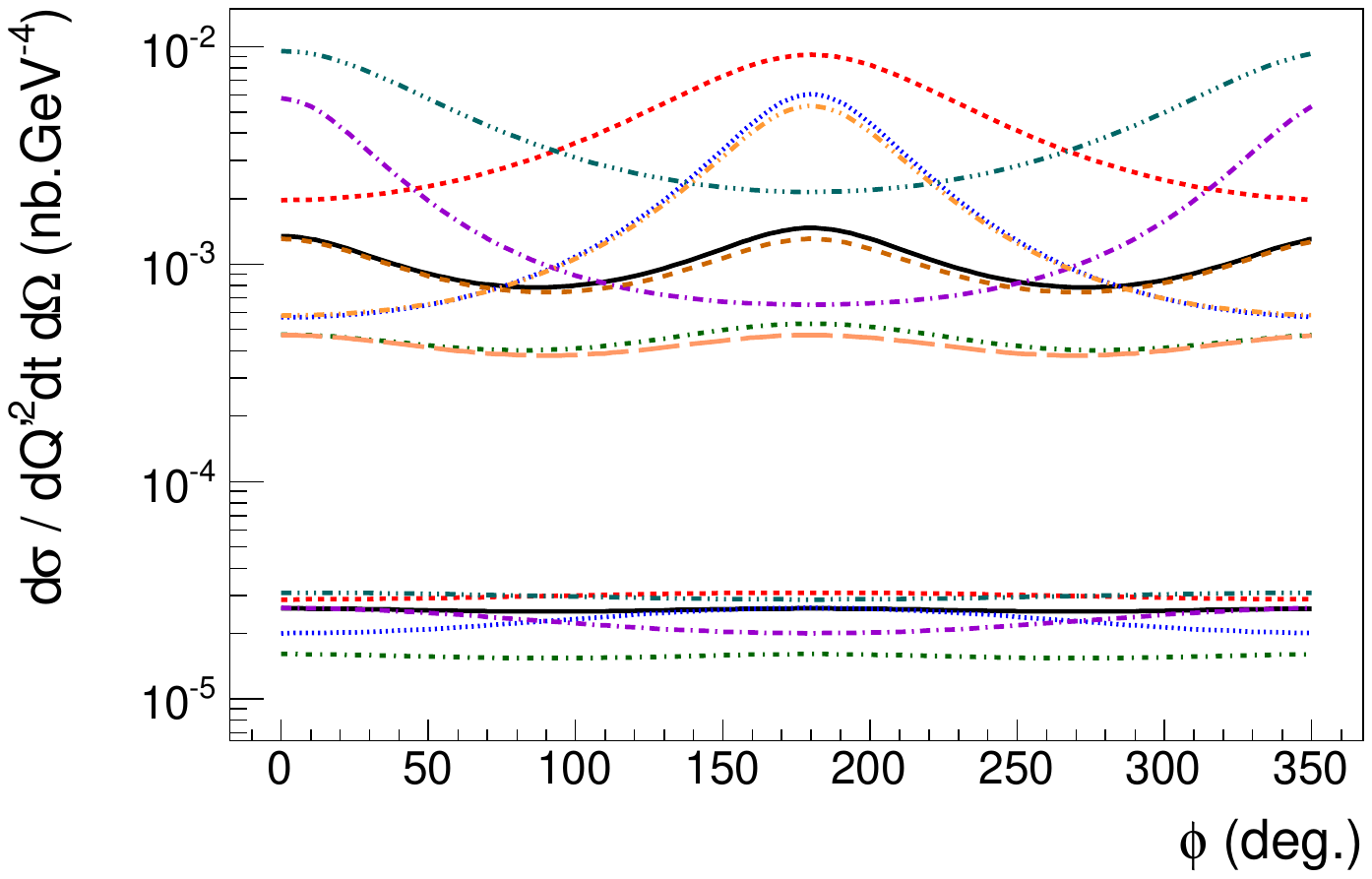}
\vspace*{-1.5cm}
\caption{
%Solid curves: unpolarized cross section $\frac{d\sigma}{dQ'^2\:dt\:d\phi\:d(cos\theta)}$ for
%the process $\gamma p\to \gamma^* p\hookrightarrow e^+ e^-$ (i.e. BH + TCS) as a function of $\phi$,
%at $E_gamma$=12 GeV, $-t=0.3$ GeV$^2$ and for different $\theta$ values ($20^\circ, 
%45^\circ,\:60^\circ,\:90^\circ,\:120^\circ,\:135^\circ$ et $\:160^\circ$). Dashed curves:
%TCS contribution alone.
Unpolarized cross section $\frac{d\sigma}{dQ'^2\:dt\:d\phi\:d(cos\theta)}$ for
the process $\gamma P\to P' e^+ e^-$ (for BH + TCS and TCS alone) as a function 
of $\phi$,
at $\xi$=0.2, $-t=0.4$ GeV$^2$, $Q'^2=7$ GeV$^2$ for different fixed $\theta$ 
values: $10^\circ, 45^\circ,\:90^\circ,\:135^\circ,\:170^\circ$
and for $\theta$ integrated over $[\pi/4,3\pi/4]$. For two values of $\theta$
($45^\circ$ and $90^\circ$) and for $\theta$ integrated over $[\pi/4,3\pi/4]$,
 we also show the calculation of the BH alone.
}
\label{fig:xsec_theta_phi}
\end{center}
\end{figure}

We discuss in this section the unpolarized cross section of the 
$\gamma P\to P' e^+ e^-$ process, which therefore includes the BH and TCS
processes. Fig.~\ref{fig:xsec_theta_phi} shows our 
calculation of the $\phi$-dependence of the 4-fold differential
cross section $\frac{d\sigma_{BH}}{dQ'^2\:dt\:d\phi\:d(cos\theta)}$ 
at $\xi=0.2$, $-t=0.4$ GeV$^2$, $Q'^2=7$ GeV$^2$
and for different $\theta$ values. 
The $\phi$-shape is strongly dependent on the $\theta$ value. As $\theta$ 
tends to $0^\circ$, the $\phi$ distribution peaks towards $\phi$=180$^\circ$ 
and as $\theta$ tends to 180$^\circ$, the $\phi$-distribution
peaks towards $\phi$=0$^\circ$ (or 360$^\circ$). There is a smooth
transition between these two singularities for the intermediate
$\theta$ values. For instance, at $\theta$=90$^\circ$, the calculation shows
only two small ``bumps" at $\phi$=0$^\circ$ and $\phi$=180$^\circ$.

These particular shapes are due to the BH process and its singularities.
Indeed, in the BH diagrams of Fig.~\ref{fig:BHdiag}, when the electron 
(positron) is emitted in the direction of the initial photon, i.e. 
$\theta$=0$^\circ$ ($\theta$=180$^\circ$), the propagator of the positron 
(electron) becomes singular and creates a peak in the $\phi$ distribution
at $\phi$=180$^\circ$ ($\phi$=0$^\circ$). Intuitively, $\theta$=0$^\circ$ 
($\theta$=180$^\circ$) forces all particles to be in the same plane, 
i.e. $\phi$=180$^\circ$ ($\phi$=0$^\circ$). The kinematics $\theta$=0$^\circ$, i.e.
the electron is in the direction of the photon beam, 
corresponds to $\phi$=180$^\circ$ because the virtual photon is
emitted by the positron, not the electron (see Fig.~\ref{fig:xsec_theta_phi}).

We display also in Fig.~\ref{fig:xsec_theta_phi} the contribution of TCS alone.
In this calculation, we have used only the GPD $H$. The inclusion
of the other GPDs barely changes the curves. In contrast to the
BH, the TCS is almost flat in $\phi$ for all $\theta$ values.
It is clear that the process $\gamma P\to P' e^+ e^-$ is largely dominated 
by the BH. There is never less than an order of magnitude between BH and TCS.

In Fig.~\ref{fig:xsec_theta_phi}, we show the curves BH+TCS for
a series of $\theta$ angles as well as the BH alone for
$\theta$=45$^\circ$ and $\theta$=90$^\circ$. 
Only at $\theta$=90$^\circ$, where one is far 
from the two BH singularities,   we have a visible difference between the
two curves and therefore a sensitivity to TCS. It is of the order 
of 30\% at $\phi$=180$^\circ$. As one gets closer
to one of the two BH singularities ($\theta$=45$^\circ$ for instance), the 
two curves BH and BH+TCS are essentially indistinguishable and
there is no sensitivity to TCS.

Finally, we show in Fig.~\ref{fig:xsec_theta_phi}, our calculations
of BH+TCS (and of BH alone) for $\theta$ integrated over the
range $[\pi/4,3\pi/4]$. In order to maximize count rates, from
an experimental point of view, it is interesting to integrate over $\theta$.
We still have a sensitivity to TCS, however it is of the order of 
5\%, i.e. lesser than at fixed $\theta$=90$^\circ$: 
the integration over $\theta$ dilutes the sensitivity to TCS.

\begin{figure}[htbp]
\begin{center}
\includegraphics[width=8.5cm,height=7.cm]
{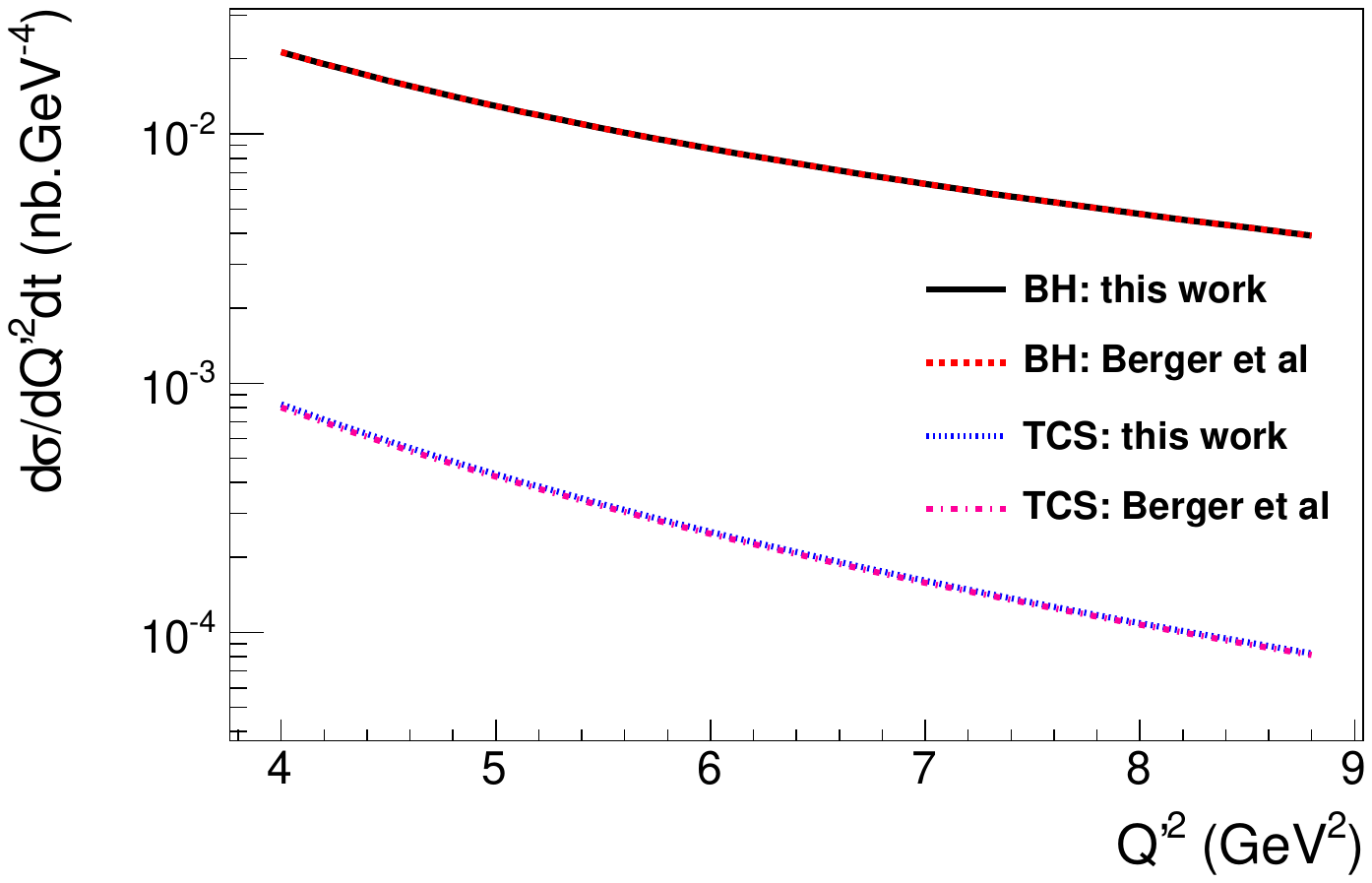}
\includegraphics[width=8.5cm,height=7cm]
{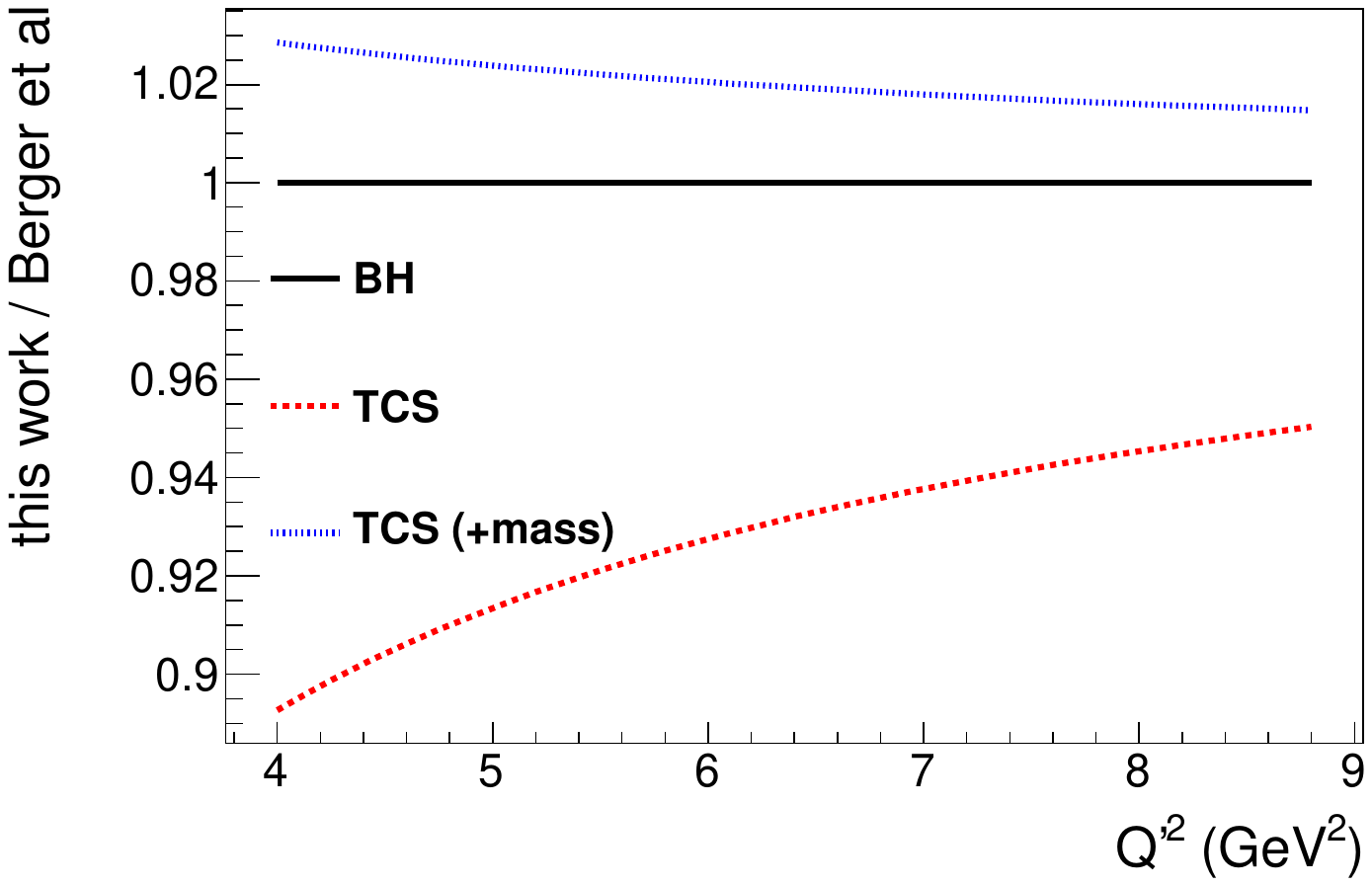}
\caption{
Top panel: the two-fold differential cross section $d\sigma/dQ'^2dt$ for BH (highest curves)
and TCS (lowest curves) as a function of $Q'^2$ for $\xi=0.2$ and $-t=0.4$ GeV$^2$. The calculations have been integrated 
over $\phi\in[0,2\pi]$ and $\theta\in[\frac{\pi}{4}, \frac{3\pi}{4}]$.
The solid curves are our calculations while the dashed curves are from Ref.~\cite{Berger:2001xd}.
Bottom panel: ratio of our calculations to those of Ref.~\cite{Berger:2001xd}. For the
TCS process, the blue dotted curve
contains the nucleon mass term in the phase space factor in the calculation of Ref.~\cite{Berger:2001xd}
while the red dashed curve does not.}
\label{fig:compar_BH}
\end{center}
\end{figure}

%We display in the top panel of Fig.~\ref{fig:compar_BH} the $Q'^2$-dependence 
%(at $-t=0.4$ GeV$^2$), and in the bottom panel the $t$-dependence (at $Q'^2=7$ GeV$^2$), 
%of $d\sigma/dQ'^2dt$ for the BH and TCS processes. 
We display in the top panel of Fig.~\ref{fig:compar_BH} the $Q'^2$-dependence, 
at $\xi=0.2$ and $-t=0.4$ GeV$^2$, of $d\sigma/dQ'^2dt$ for the BH and TCS processes. 
The 2-fold cross section has been integrated over the decay angles: $\phi\in[0,2\pi]$ 
and $\theta\in[\frac{\pi}{4}, \frac{3\pi}{4}]$. We also display in this figure the results 
of the analytical formulaes of Ref.~\cite{Berger:2001xd}. For TCS, we have of course used the same 
GPD parameterization for both calculations (only $H$ in this case). 
%In order to better appreciate the comparison, we display in Fig.~\ref{fig:gfwxc_BH} 
%the ratios of our calculations to those of Ref.~\cite{Berger:2001xd}. 

In order to better appreciate the comparison, we display in the bottom panel of Fig.~\ref{fig:compar_BH} 
the ratios of our calculations to those of Ref.~\cite{Berger:2001xd}. 
The agreement is excellent for the BH process, as it should, since there was no approximation
done in the derivation of the analytical formula for this process in Ref.~\cite{Berger:2001xd}. 
For TCS, we show two curves for the ratio. Indeed, in Ref.~\cite{Berger:2001xd}, in the analytical formula 
for the TCS process, the nucleon mass was neglected in the phase space factor, i.e. the cross section 
is proportional to ${1}\over{s^2}$ rather than ${1}\over{s-m^2}^2$ (this is not the case for the BH process 
where the phase space factor is exact in Ref.~\cite{Berger:2001xd}). 
%the differences orinigiating from the phase space factor from those coming from the TCS amplitude,
%In order properly compare the TCS amplitudes of  to ours,
%and also in order to distinguish the approximations done in the TCS amplitude and the more trivial
%one in the phase space factor in Ref.~\cite{Berger:2001xd}'s work, 
%we calculate the TCS cross sections of Ref.~\cite{Berger:2001xd} 
In the bottom panel of Fig.~\ref{fig:compar_BH}, we plot the TCS cross section of
Ref.~\cite{Berger:2001xd} with (blue dotted curve) and without (red dashed curve) the nucleon 
mass term in the phase space factor. In this way, one can distinguish the differences 
between the two calculations originating from the trivial phase space factor from those, more subtle,
coming from the TCS amplitude.
%It is seen that at the lowest $Q^'2$ and largest $t$ values considered here, the difference in the
%cross section calculations due to the prescription taken for the phase space factor can reach more than 10\%.
It is seen in this figure that at the lowest $Q'^2$ values and at the presently considered 
kinematics, the difference in the cross section calculations depending on the prescription taken 
for the phase space factor can reach more than 10\%.
In both cases, it is also seen that the difference between the Berger et al.'s calculations
and ours diminishes as $Q'^2$ increases, as expected since terms of the order 
of ${m^2}\over{Q'^2}$ were dropped in the TCS analytical formula of Ref.~\cite{Berger:2001xd}.

%\begin{figure}[htbp]
%\begin{center}
%\includegraphics[width=8.5cm,height=7.cm]
%{Figures/fig7_compvsQ_bis_logy_2.pdf}
%\includegraphics[width=8.5cm,height=7cm]
%{Figures/ratioQ2_4.pdf}
%\caption{
%The 2-fold differential cross section $d\sigma/dQ'^2dt$ as a function
%of $Q'^2$ for $-t=0.4$ GeV$^2$ (top panel) and as a function of $t$
%at $Q'^2=7$ GeV$^2$ (bottom panel). The calculations have been carried out
%at $\xi=0.2$ and integrated 
%over $\phi\in[0,2\pi]$ and $\theta\in[\frac{\pi}{4}, \frac{3\pi}{4}]$.
%In both panels, the highest cross section
%curves are for BH and the lowest ones for TCS. The solid curves
%are our calculations while the dashed curves are from Ref.~\cite{Berger:2001xd}.}
%\label{fig:compar_BH}
%\end{center}
%\end{figure}

%\begin{figure}[htbp]
%\begin{center}
%\includegraphics[width=8.5cm,height=7.cm]
%{figs/comparBDPratio_Q2_0.pdf}
%{Figures/ratioQ2_4.pdf}
%\includegraphics[width=8.5cm,height=7cm]
%{Figures/ratiot_4.pdf}
%{figs/comparBDPratio_t_2.pdf}
%\caption{Ratio of our calculations to those of Ref.~\cite{Berger:2001xd}
%for the 2-fold differential cross section 
%$d\sigma/dtdQ'^2$ as a function of $Q'^2$ (top) and as a function of  $t$ (bottom)
%at the same kinematics of Fig.~\ref{fig:compar_BH}.}
%\label{fig:gfwxc_BH}
%\end{center}
%\end{figure}

%\pagebreak

\begin{figure}
\begin{center}
\includegraphics[width=8.5cm,height=7.5cm]{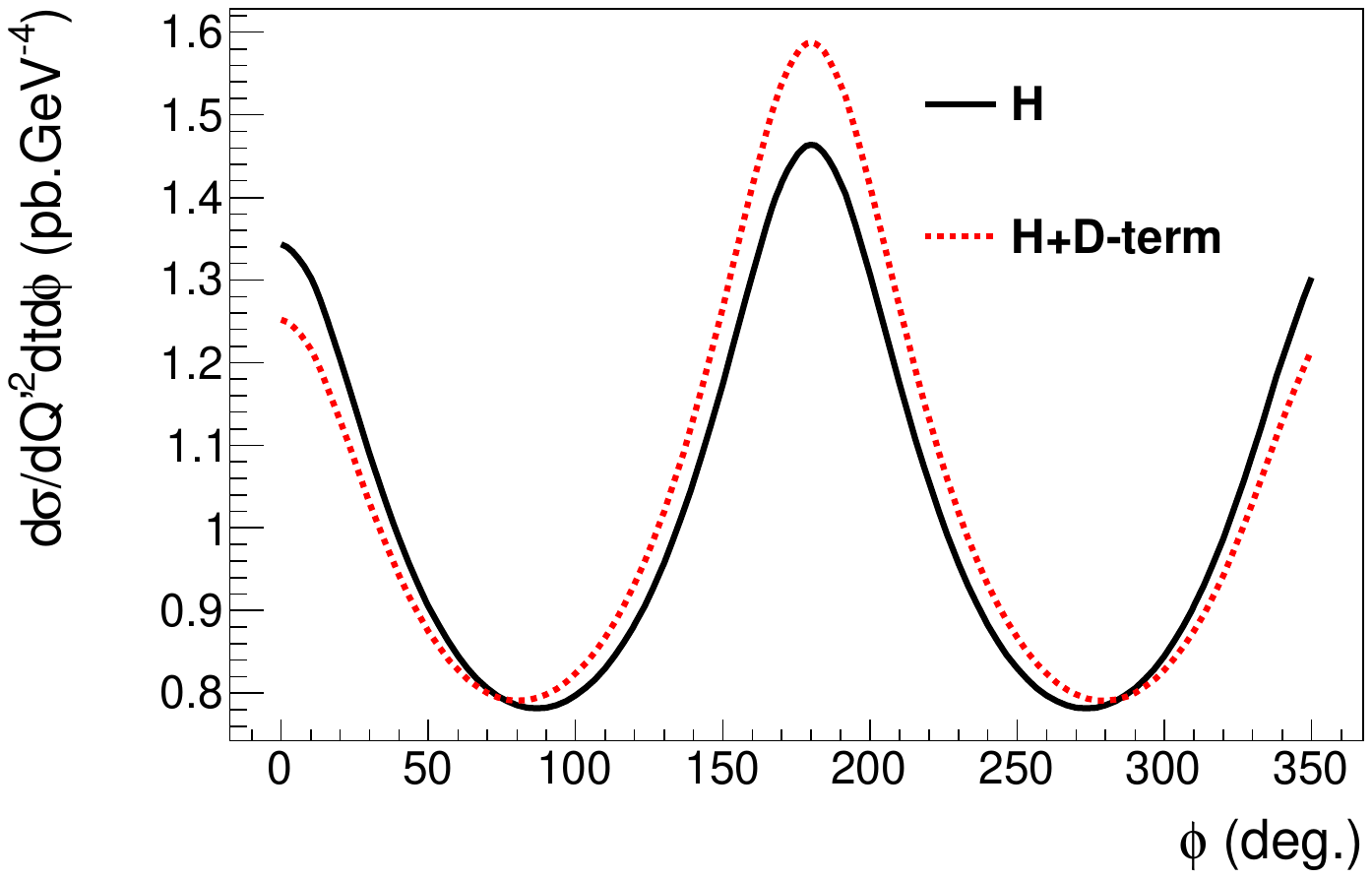}
\caption{The three-fold differential cross section $d\sigma/dQ'^2dtd\phi$ as a function of $\phi$ 
with (red dashed curve) and without (black solid curve) the D-term,
calculated at $\xi=0.2$, $-t=0.4$ GeV$^2$ and with $\theta$ integrated
over $[\frac{\pi}{4}, \frac{3\pi}{4}]$.}
\label{fig:Dtermadd}
\end{center}
\end{figure}

To end this section concerning the unpolarized cross section, we show in Fig.~\ref{fig:Dtermadd} the influence
of the D-term on the three-fold differential cross section $d\sigma/dQ'^2dtd\phi$, 
calculated for $\xi=0.2$, $-t=0.4$ GeV$^2$ and with $\theta$ integrated
over $[\frac{\pi}{4}, \frac{3\pi}{4}]$. It modifies the amplitude of the cross section at $\phi=0^\circ$
and $\phi=180^\circ$ by about 10\%.

\newpage
%%%%
\section{Single spin asymmetries}
\label{sec:BSA}

We now turn to the single spin asymmetries: beam or target.
Photons beams can be polarized  linearly  or circularly.
Target polarization vectors can be oriented along the $x-$, $y-$
or $z-$axis in the $\gamma P\to \gamma^* P'$ plane (see Fig.~\ref{fig:angles}).
For polarization observables, we will calculate spin asymmetries and,
following notations used for DVCS, we will tag them
$A_{ij}$, i.e. with two indices $i$ and $j$. The first index $i$ refers to 
the polarization type of the beam: $U$ for an unpolarized beam, $\odot$ 
for a circularly polarized beam and $\ell$ for a linearly polarized beam.
The second index $j$ refers to the polarization of the target and
can take the values $U$, $x$, $y$ or $z$, with obvious meanings. 
In this section, dedicated to single-polarization observables,
we will therefore consider successively the five independent
asymmetries $A_{\ell U}$, $A_{\odot U}$, $A_{Ux}$, $A_{Uy}$ and $A_{Uz}$.

\subsection{Linearly polarized photons}
\label{subsubsec:BSA_psi}

\begin{figure}[htbp]
\begin{center}
%\includegraphics[width=8.5cm,height=7.5cm]
%{Figures/figp_BSAlin_Psi_3.pdf}
%\includegraphics[width=5.cm,height=4.5cm]
%{Figures/figp_BSAlin_Psi_6.pdf}
%\hspace*{-1.1cm}
%\includegraphics[width=8.5cm,height=7.5cm]
%{Figures/figp_BSAlin_Psi_4.pdf}
%\vspace*{-1.5cm}
\includegraphics[width=8.5cm,height=7.5cm]
{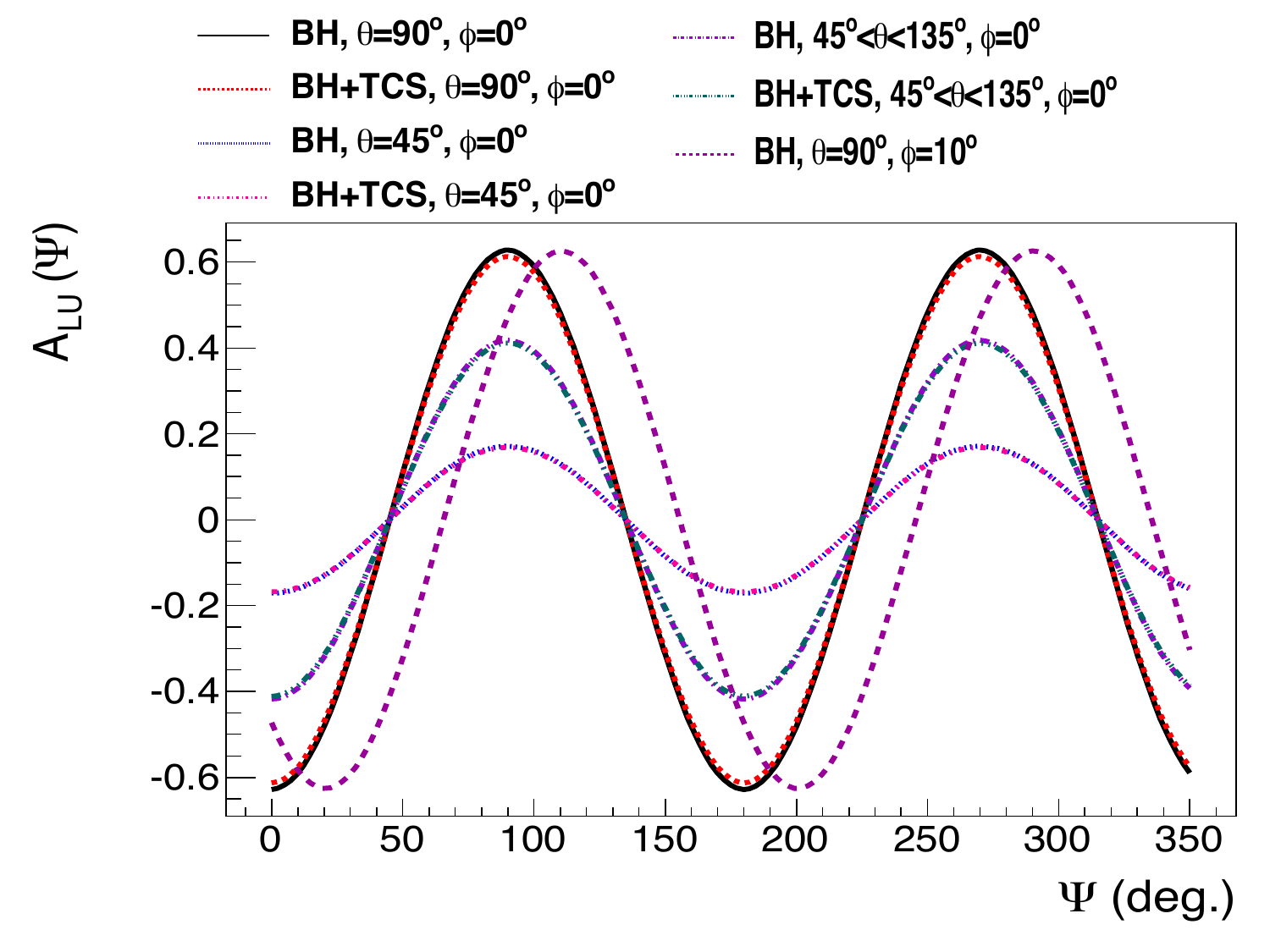}
\caption{$A_{\ell U}$ asymmetry as a function of $\Psi$
for BH and for BH+TCS (only GPD $H$ contribution) 
at $\xi$=0.2, $-t=0.4$  % $-t=0.2$ 
GeV$^2$, $Q'^2=7$ GeV$^2$ 
and for different sets of $\theta$ and $\phi$ angles.
}
\label{fig:BSA_psi}
\end{center}
\end{figure}

We introduce the angle $\Psi$ between the polarization vector of the
photon and the plane spanned by the photon beam and
the ($e^+e^-$) system, which contains 
the $x$-axis (see Fig.~\ref{fig:angles}). Then, we define:
\begin{equation}
A_{\ell U}(\Psi) = \frac{\sigma_{x}(\Psi) - \sigma_{y}(\Psi)}{\sigma_{x}(\Psi) 
+ \sigma_{y}(\Psi)},
\end{equation}
where $\sigma_{x}$ ($\sigma_{y}$) stands 
 for the 4-fold differential cross sections 
$\frac{d\sigma}{dQ'^2\:dt\:d\phi\:d(cos\theta)d\Psi}$ 
  with
a photon beam polarized in the $x$-($y$-)direction.
We display in Fig.~\ref{fig:BSA_psi} the $\Psi$-dependence of $A_{\ell U}$
for $\xi$=0.2, $-t=0.2$ GeV$^2$, $Q'^2=7$ GeV$^2$
for $\theta=45^\circ$, $\theta=90^\circ$ and $\theta$ integrated
over $[\pi/4,3\pi/4]$ and $\phi=0^\circ$ and $\phi=10^\circ$.
The approximate shape of the asymmetry is a $\cos(2\Psi)$ which is reminiscent
of the modulation which is predicted for the so-called $\Sigma$ asymmetry 
in single meson photoproduction. This $\cos(2\Psi)$ modulation
appears explicitely in the analytical expressions of Ref.~\cite{Goritschnig}.
 
We note that the BH alone produces an asymmetry. It is actually the
dominant contribution. The TCS produces only variations of the amplitude
at the percent level around the BH, making this observable not very favorable
to study TCS and GPDs. The amplitude and phase of the 
asymmetry depend strongly on the decay angles: it is the strongest 
as $\theta$ approaches 90$^\circ$ and the phase increases as $\phi$ increases.
This phase shift due to $\phi$ is also apparent in 
Ref.~\cite{Goritschnig}.

\begin{figure}[htbp]
\begin{center} 
\includegraphics[width=5.cm,height=4.5cm]
{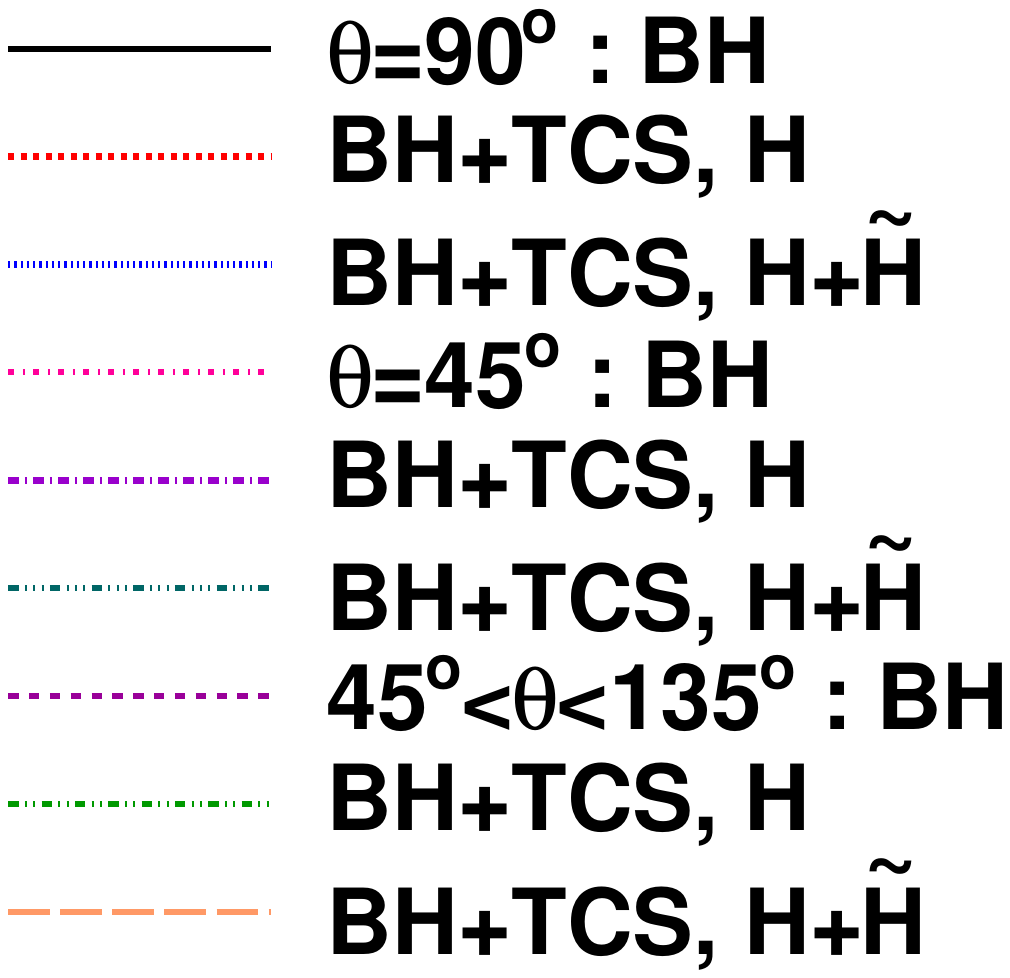}
\hspace*{-1.1cm}
\includegraphics[width=8.5cm,height=7cm]
{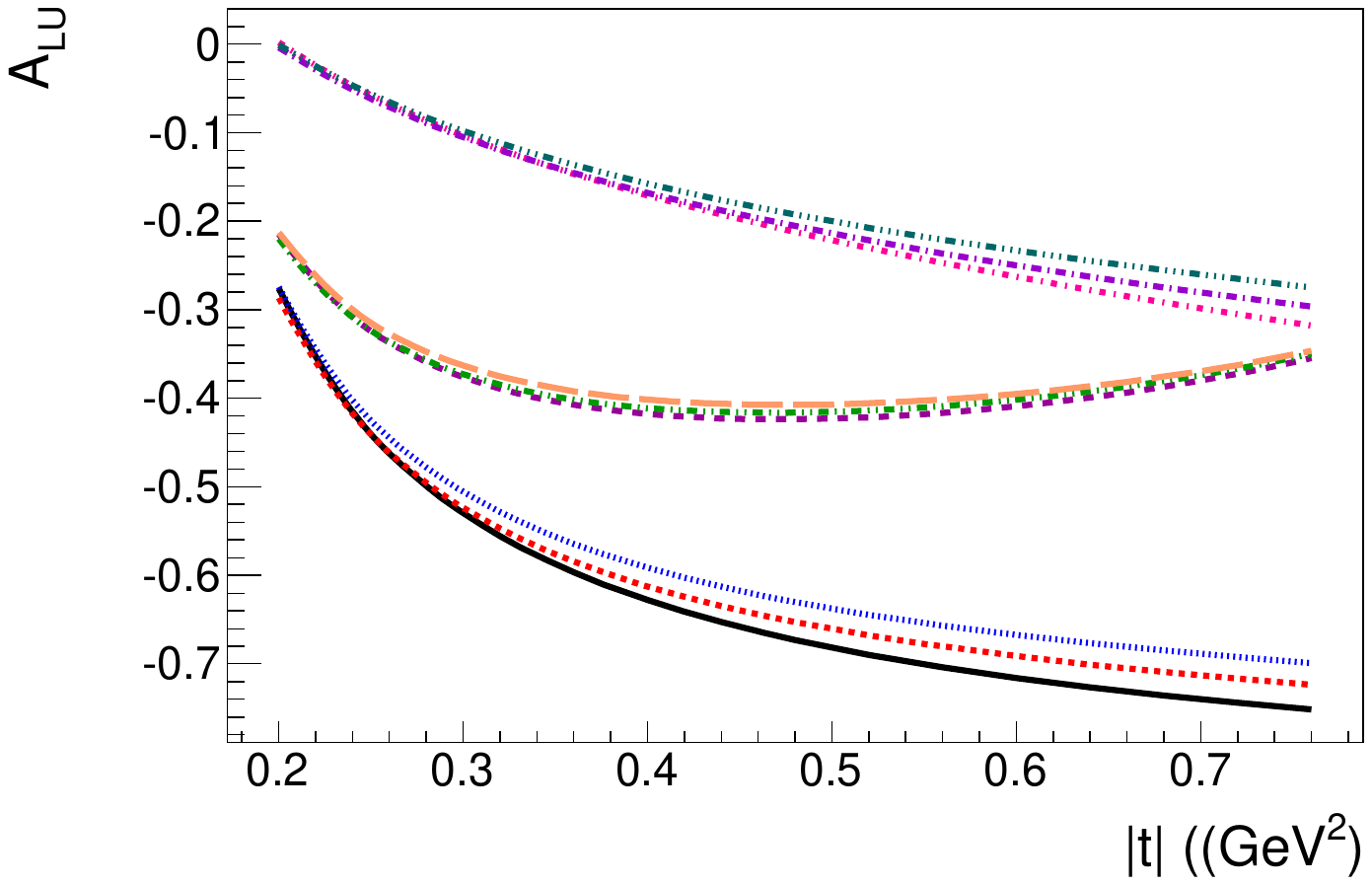}
%\vspace*{-1.5cm}
%\includegraphics[width=8.5cm,height=7.5cm]
%{Figures/compar_t_linpol.pdf}
\caption{$A_{\ell U}$ asymmetry as a function of $t$ for $\xi$=0.2, 
$Q'^2=7$ GeV$^2$, $\Psi=0^\circ$, $\phi=0^\circ$ and
$\theta=45^\circ$, $\theta=90^\circ$ and $\theta\in[\pi/4,3\pi/4]$.
}
\label{fig:BSA_psi2}
\end{center}
\end{figure}

In Fig.~\ref{fig:BSA_psi}, we have used only the contribution of
the $H$ GPD for TCS. In Fig.~\ref{fig:BSA_psi2}, we show the 
$t$-dependence of the $A_{\ell U}$ asymmetry for BH alone,
BH+TCS (with only $H$) and BH+TCS (with $H$+$\tilde{H}$). 
Calculations have been done for $\xi$=0.2, $Q'^2=7$ GeV$^2$, 
$\phi=0^\circ$, $\Psi=180^\circ$ and for $\theta=45^\circ$, 
$\theta=90^\circ$ and $\theta$ integrated over $[\pi/4,3\pi/4]$.
Depending on $t$, we notice some sensitivity of the $A_{\ell U}$ 
observable to the GPDs $H$ and $\tilde{H}$. 

\begin{figure}[htbp]
\begin{center}
\includegraphics[width=8.5cm,height=7cm]
{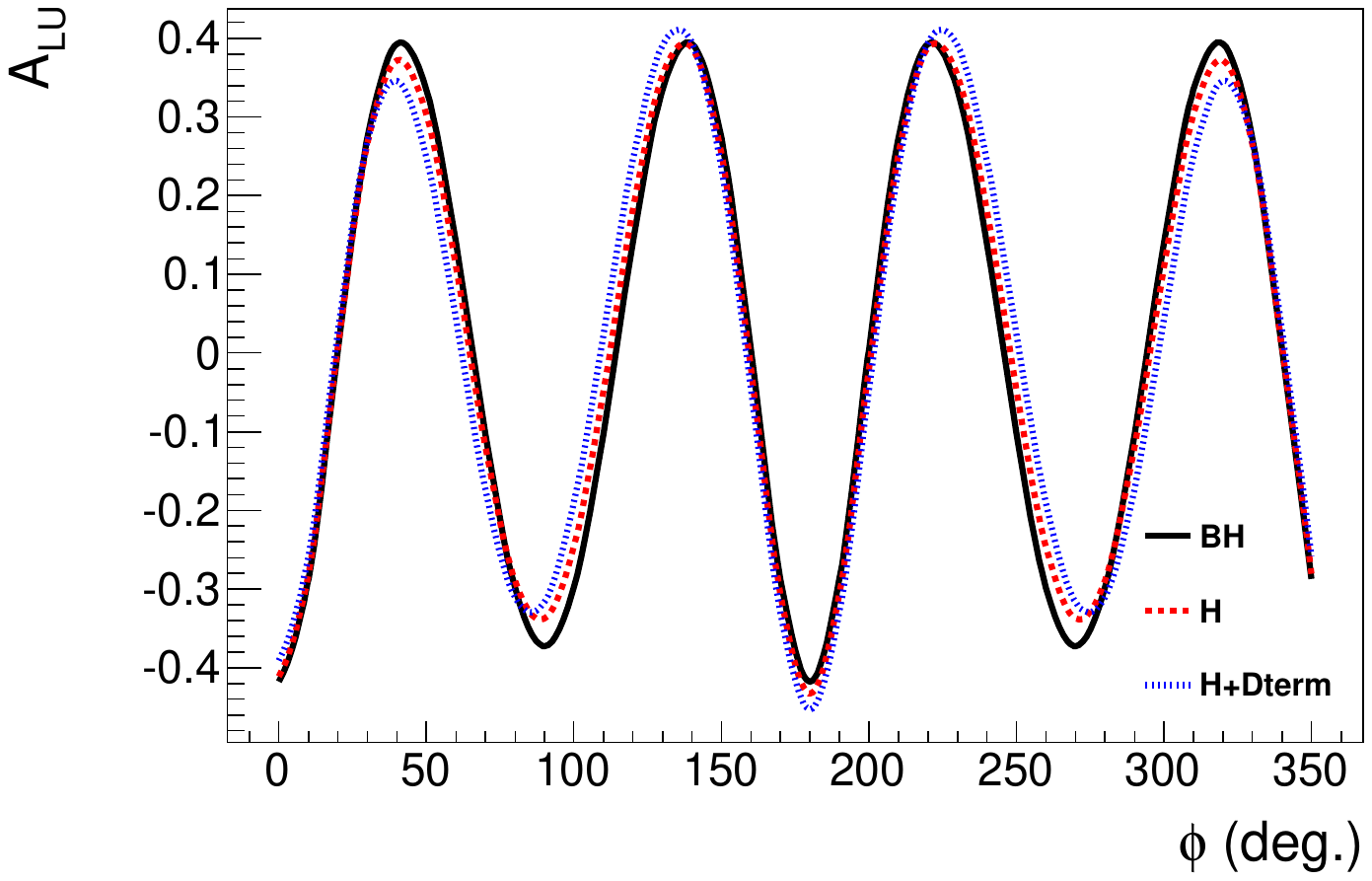}
\caption{
$A_{\ell U}$ asymmetry as a function of $\phi$ for $\xi=0.2$, $Q'^2=7$ GeV$^2$,
$-t=0.4 $ GeV$^2$ and $\theta\in[\pi/4,3\pi/4]$. The black solid curve corresponds to BH only, 
the red dashed 
curve to BH+TCS with only the GPD $H$ contribution and the blue dotted curve to BH+TCS 
with the GPD $H$ and the D-term contributions.
}
\label{fig:BSA_phi_lin}
\end{center}
\end{figure}

In Fig.~\ref{fig:BSA_phi_lin}, we show the peculiar $\phi$-dependence of $A_{\ell U}$
at $\Psi=0^\circ$ at our standard kinematics. We also show the (small) influence
of the D-term.

%%%%%
\subsection{Circularly polarized photons}

We define:
\begin{equation}
A_{\odot U} = \frac{\sigma^{+} - \sigma^{-}}{\sigma^{+} + \sigma^{-}},
\end{equation}
where $\sigma^\pm$ stand  for the 4-fold differential cross sections 
$\frac{d\sigma}{dQ'^2\:dt\:d\phi\:d(cos\theta)}$ 
for the two photon spin states, right and left polarized.

\begin{figure}[htbp]
\begin{center}
\hspace*{-0.5cm}
\includegraphics[width=8.5cm,height=7cm]
{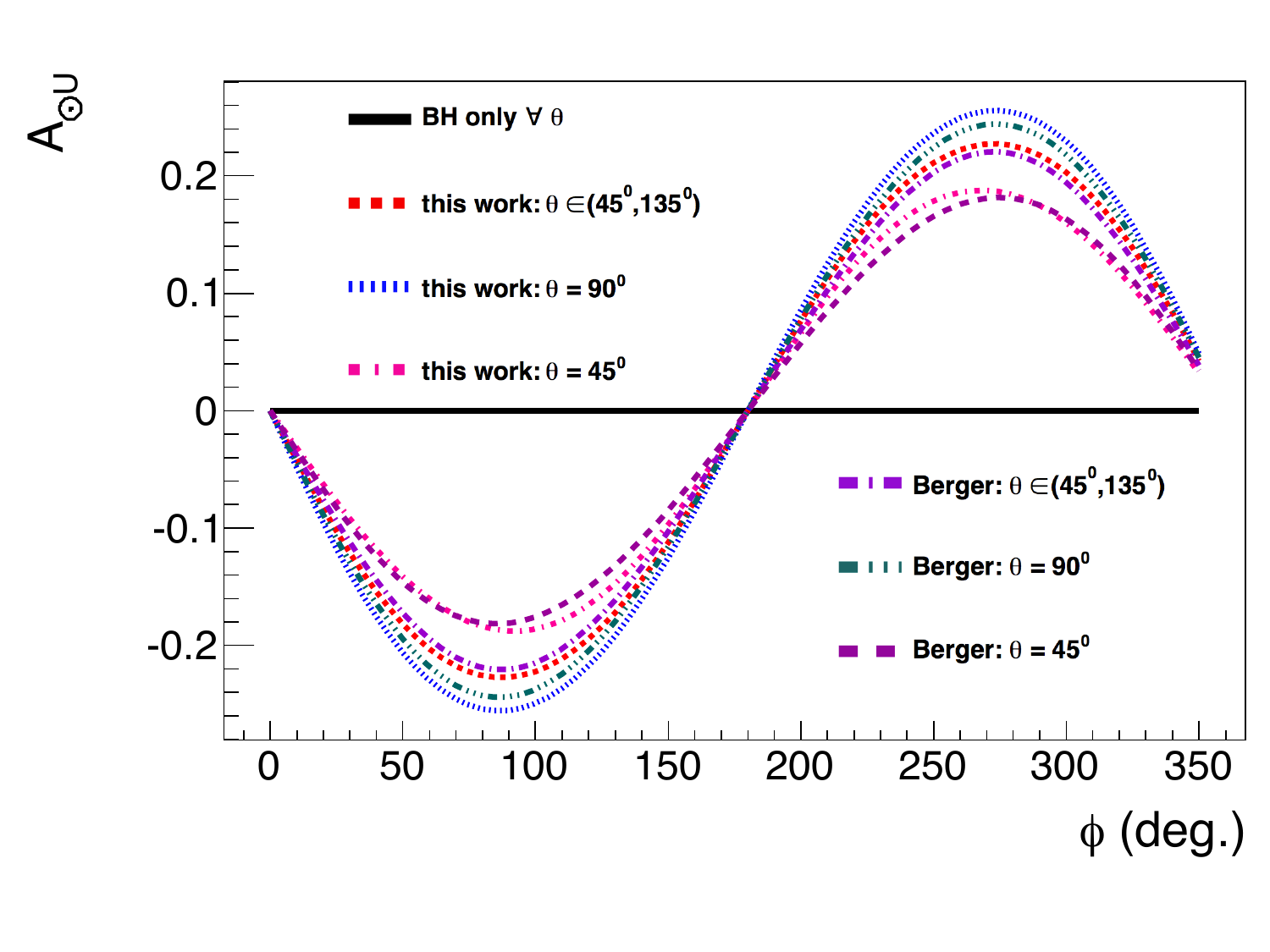}
\includegraphics[width=8.5cm,height=7cm]
{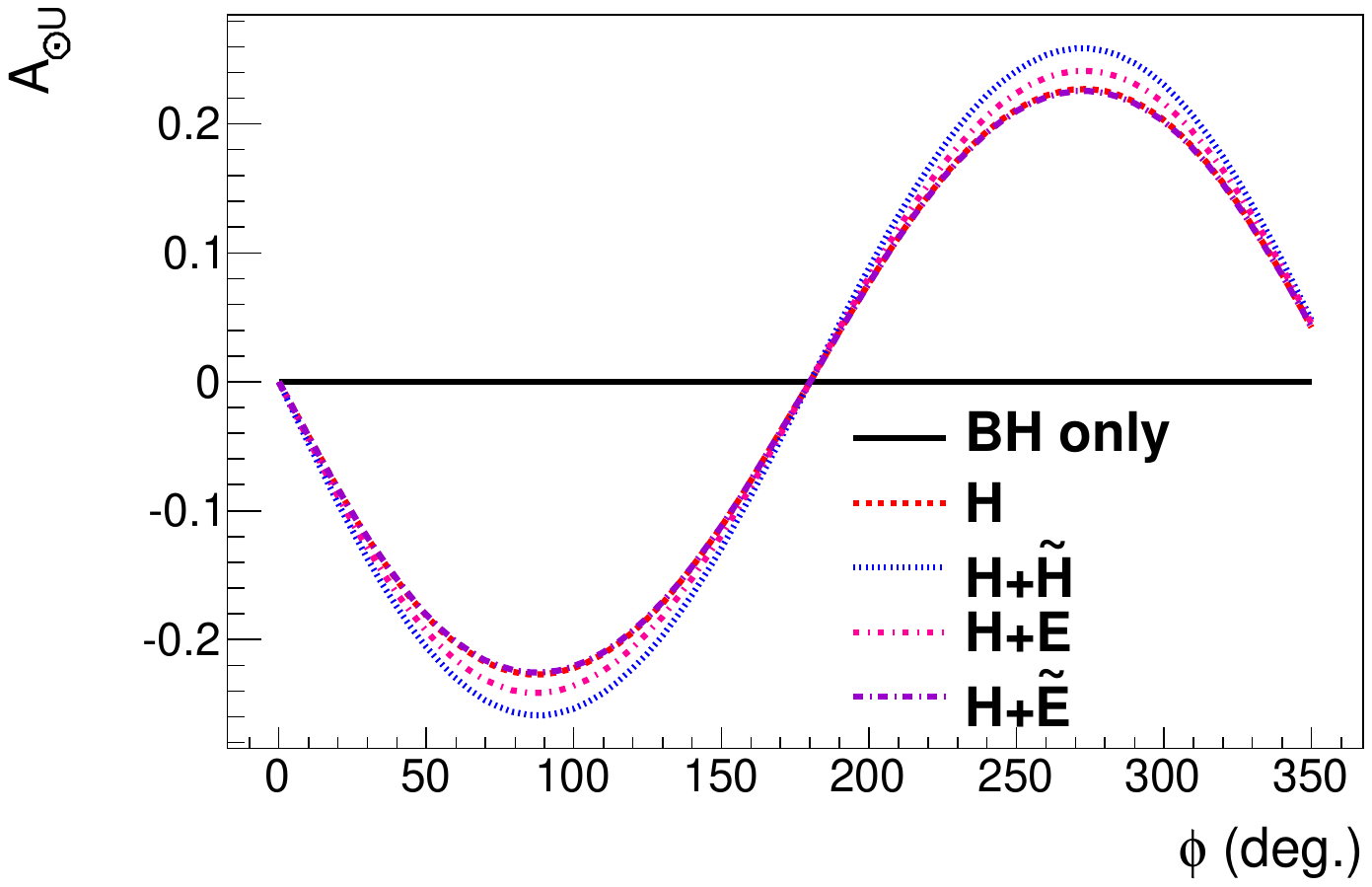}  %BSA_phi_3.pdf}
\caption{Top panel: The $A_{\odot U}$ asymmetry as a function 
of $\phi$ for $\theta=45^\circ$, $\theta=90^\circ$ and $\theta\in[45^\circ,135^\circ]$. 
The TCS is calculated here with only the $H$ GPD.
We compare our calculations to those of Ref.~\cite{Berger:2001xd}. 
Bottom panel: The $A_{\odot U}$ for 
$\theta\in[45^\circ,135^\circ]$ using differents GPDs parametrizations for TCS. 
The calculations are done for $Q'^2=7$ GeV$^2$, $\xi=0.2$, $-t=0.4$ GeV$^2$.
}
\label{fig:BSA}
\end{center}
\end{figure}

We display in Fig.~\ref{fig:BSA} (top panel) our results for $A_{\odot U}$ as a 
function of $\phi$ at $Q'^2=7$ GeV$^2$, $\xi=0.2$, 
$-t=0.4$ GeV$^2$ for $\theta$=45$^\circ$, 90$^\circ$
and $\theta$ integrated over $[45^\circ,135^\circ]$. The TCS is 
calculated here with only the $H$ GPD. 
In all kinematics, we obtain a $\sin(\phi)$ 
shape with a significant amplitude, up to $\approx$ 25\%.
We observe that the BH doesn't produce any asymmetry. Any signal therefore 
reflects a contribution from TCS. This is due to the fact that, as was shown 
in Ref.~\cite{Berger:2001xd}, this observable is sensitive to the 
imaginary part of the amplitude and that the BH amplitude is purely real.
The amplitude of the asymmetry depends on $\theta$. It is maximal for 
$\theta$=90$^\circ$ where BH is minimal and it decreases as $\theta$ 
tends to $\theta$=0$^\circ$ (or 180$^\circ$).
Since the BH does not produce on its own an asymmetry, one sees
that the integration over $\theta$ does not strongly reduce the signal.
Such integration allows to maximize count rates.

In Fig.~\ref{fig:BSA} (top panel), we also compare our results to those
of Ref.~\cite{Berger:2001xd}. In all cases, our calculations
produce amplitudes a few percents larger. This might be attributed
to some mass correction terms to the TCS amplitude which are present 
in our calculation and not in Ref.~\cite{Berger:2001xd}.

Fig.~\ref{fig:BSA} (bottom  panel) shows the $A_{\odot U}$  asymmetry
for  $\theta$ integrated over $[45^\circ,135^\circ]$ using different GPDs  parametrizations for TCS.

\begin{figure}[htbp]
\begin{center}
\includegraphics[width=8.5cm,height=7cm]
{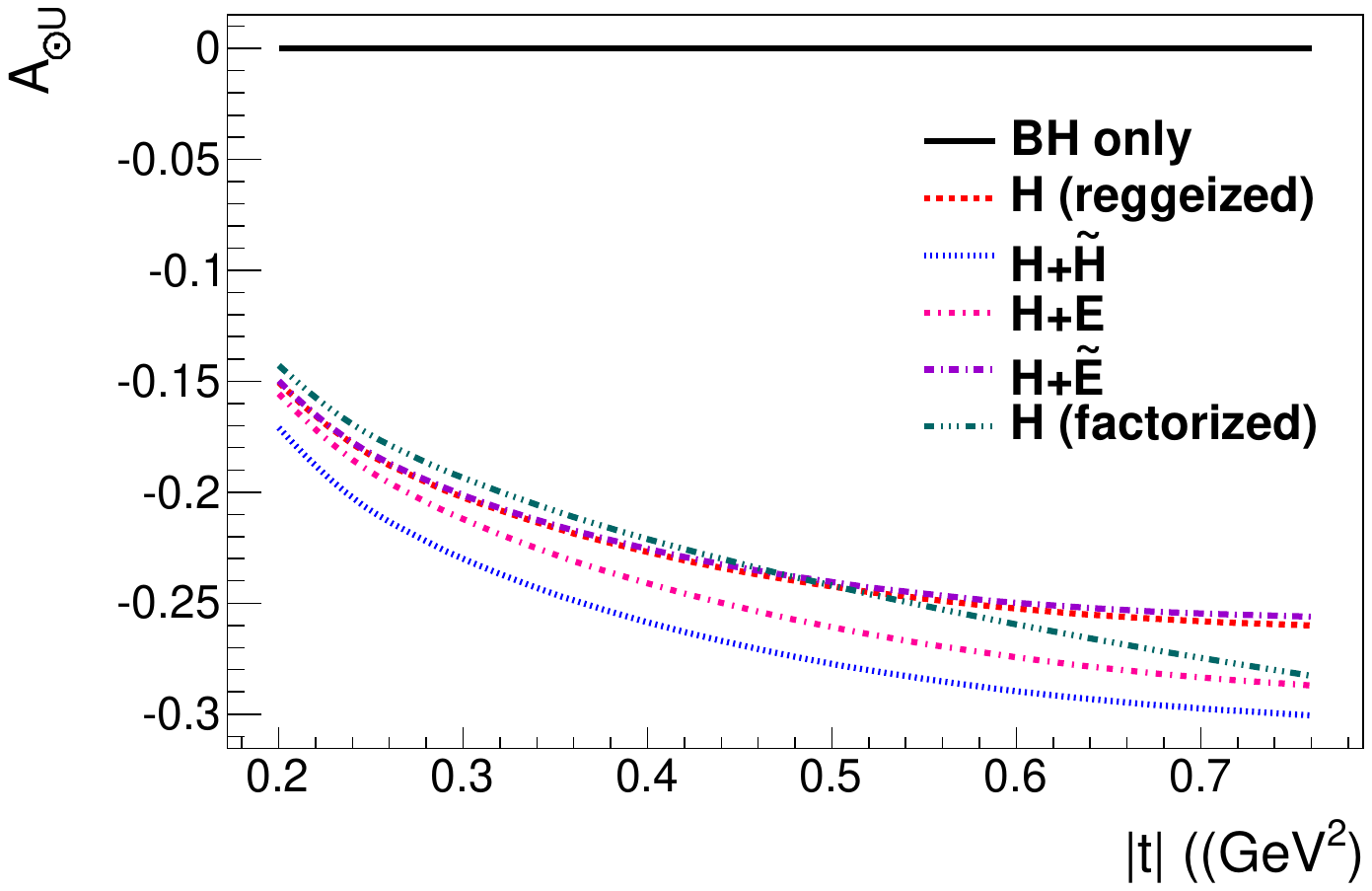} 
%{Figures/BSA_t_9.pdf} 
\caption{
The $A_{\odot U}$ asymmetry as a function of $t$ for BH+TCS at $\xi=0.2$, 
$Q'^2=7$ GeV$^2$, $\phi=90^\circ$ and $\theta$ integrated over
$[45^\circ,135^\circ]$. TCS is calculated with different GPD contributions.
}
\label{fig:BSAvst}
\end{center}
\end{figure}

In Fig.~\ref{fig:BSAvst}, we show for $\xi=0.2$, $Q'^2=7$ GeV$^2$,
$\phi=90^\circ$ and $\theta$ integrated over $[45^\circ,135^\circ]$,
the $t$-dependence of $A_{\odot U}$ and its sensitivity to different GPDs.
We notice that the magnitude
of $A_{\odot U}$ increases with $|t|$ and that there is a sensitivity
of this observable to all four GPDs, especially at large $\mid t\mid$.
We also display in this figure our calculation with the factorized
ansatz for the $t$-dependence of the $H$ GPD in order to illustrate the 
model-dependence of our results.
 
%%%%%
\subsection{Polarized targets}
\label{subsec:TSA}

We define:
\begin{equation}
A_{Ui} = \frac{\sigma^{+} - \sigma^{-}}{\sigma^{+} + \sigma^{-}},
\end{equation}
where $\sigma^\pm$ stands for the four-fold differential cross sections 
$\frac{d\sigma}{dQ'^2\:dt\:d\phi\:d(cos\theta)}$ 
for the two target spin orientations $+$ and $-$ along the axis $i=x,y$ or $z$.

\begin{center}
\begin{figure}[htbp]
\hspace*{-0.8cm}
\includegraphics[width=6.5cm,height=6cm]
{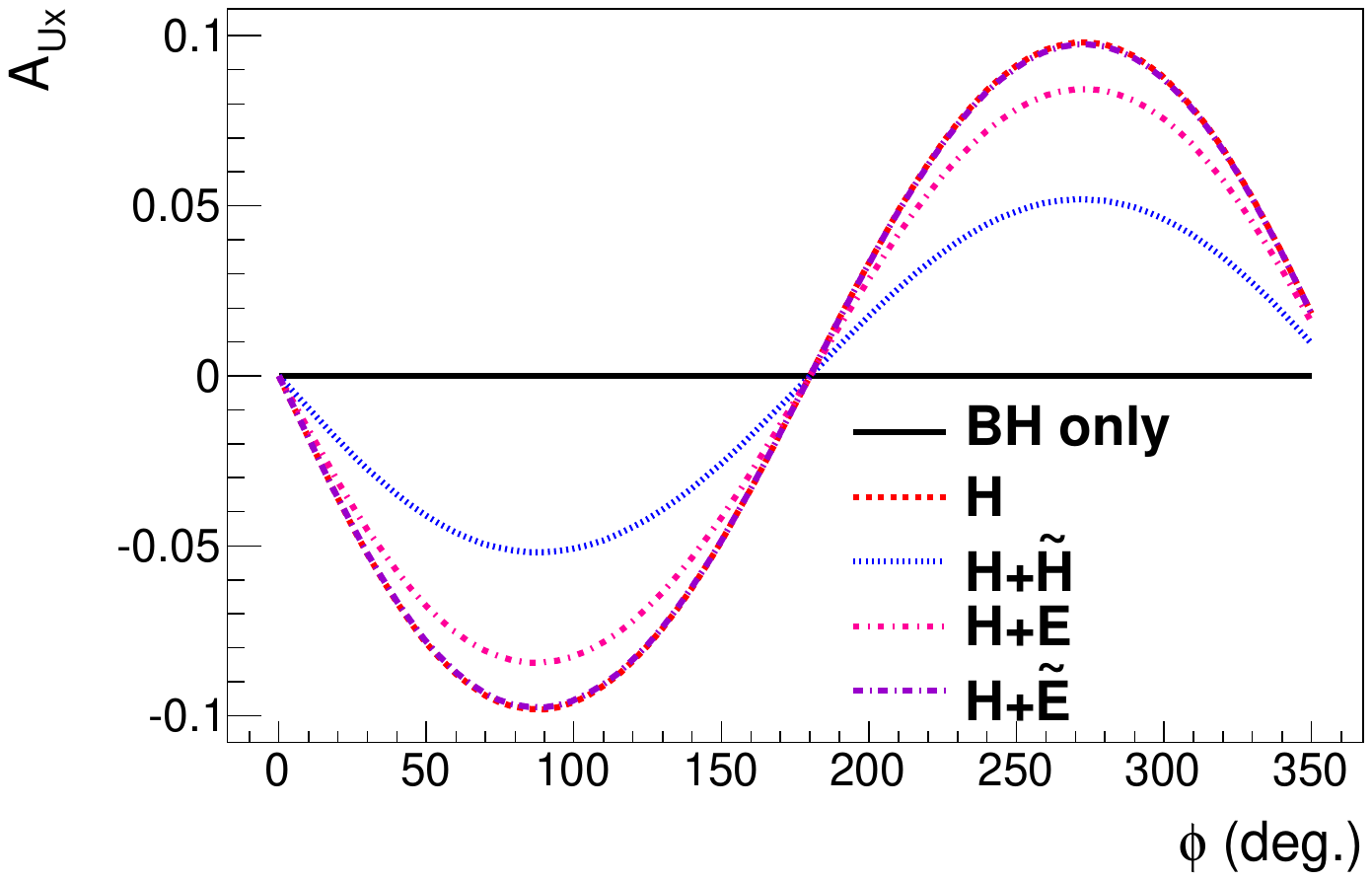}
%TSAx_phi_3.pdf}
\hspace*{-0.8cm}
\includegraphics[width=6.5cm,height=6cm]
{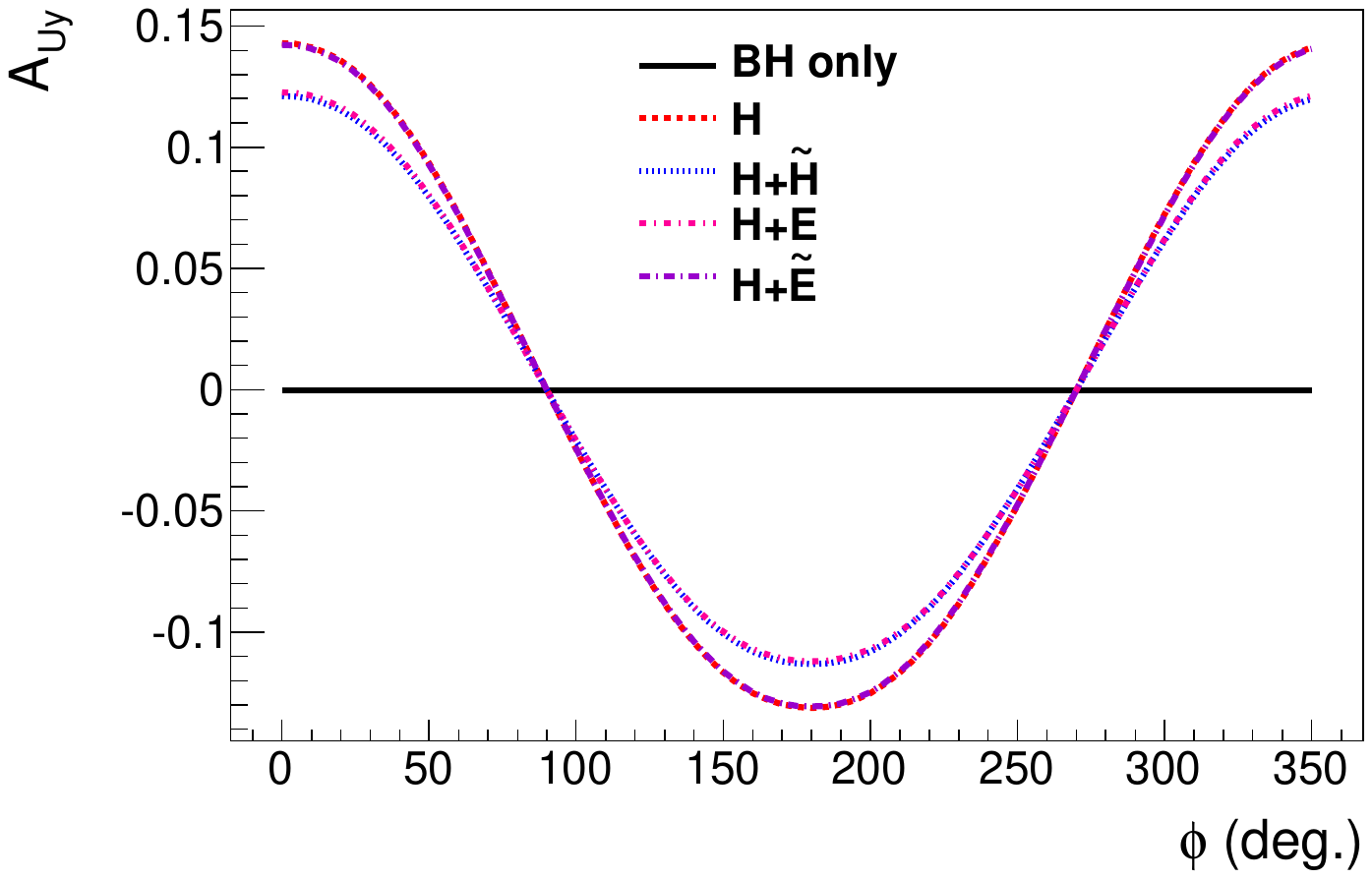}
%TSAy_phi_1.pdf}
\hspace*{-0.8cm}
\includegraphics[width=6.5cm,height=6cm]
{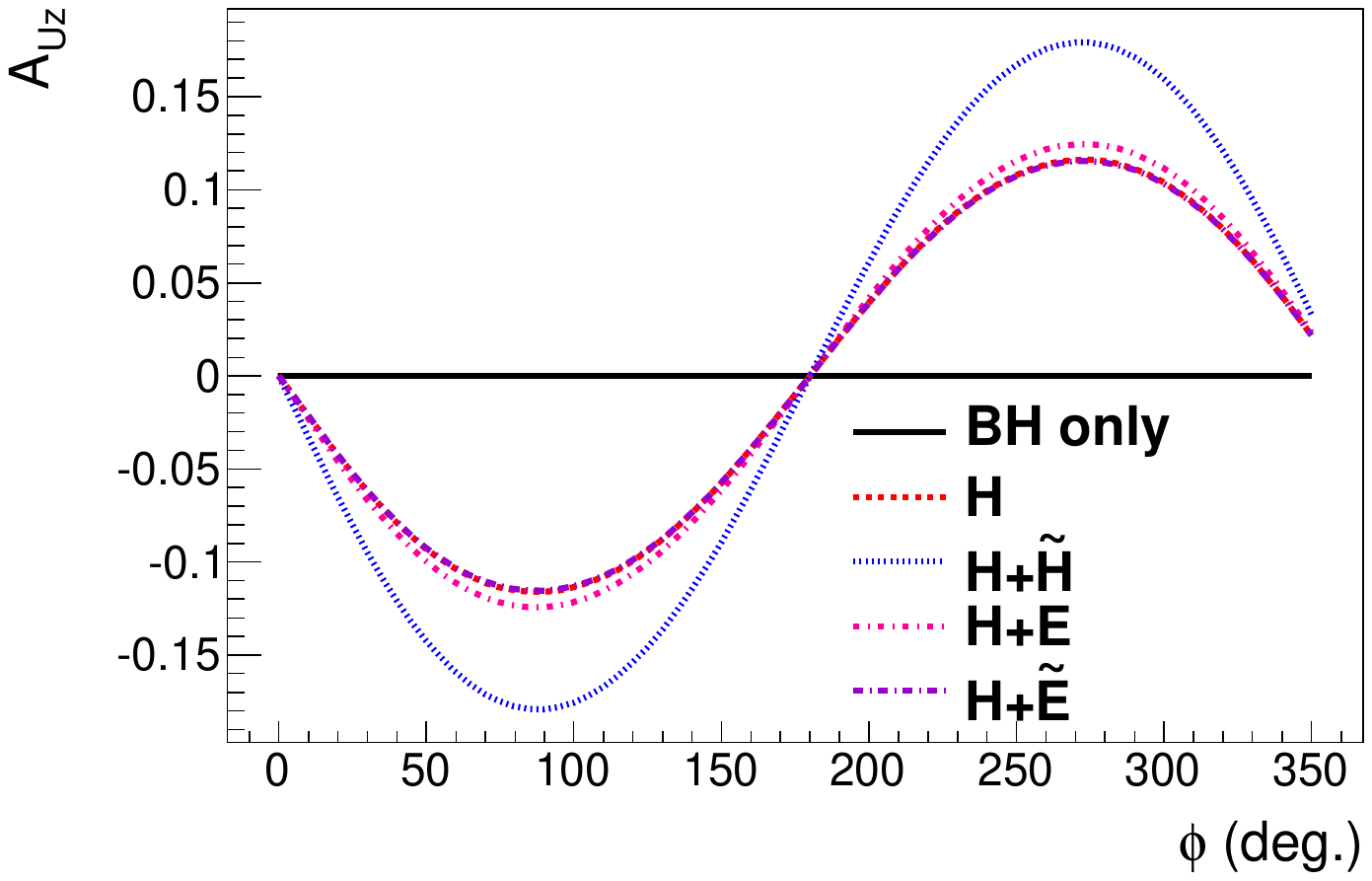}
%TSAz_phi_3.pdf}
\hspace*{-0.8cm}
\caption{The $A_{Ux}$ (left panel),  $A_{Uy}$ (central panel) 
and $A_{Uz}$ (right panel)
asymmetries as a function of $\phi$ for $\xi=0.2$, $Q'^2=7$ GeV$^2$,
$-t=0.4$ GeV$^2$ and for 
%$\theta$=45$^\circ$, 90$^\circ$ and
 $\theta$ 
integrated over $[\frac{\pi}{4},\:\frac{3\pi}{4}]$. The TCS is calculated 
%here with only the $H$ GPD.
with different GPDs contributions.}
\label{fig:TSA}
%\end{figure}
%\end{center}
%
%\begin{center}
%\begin{figure}[htbp]
\hspace*{-0.8cm}
\includegraphics[width=6.5cm,height=6cm]
{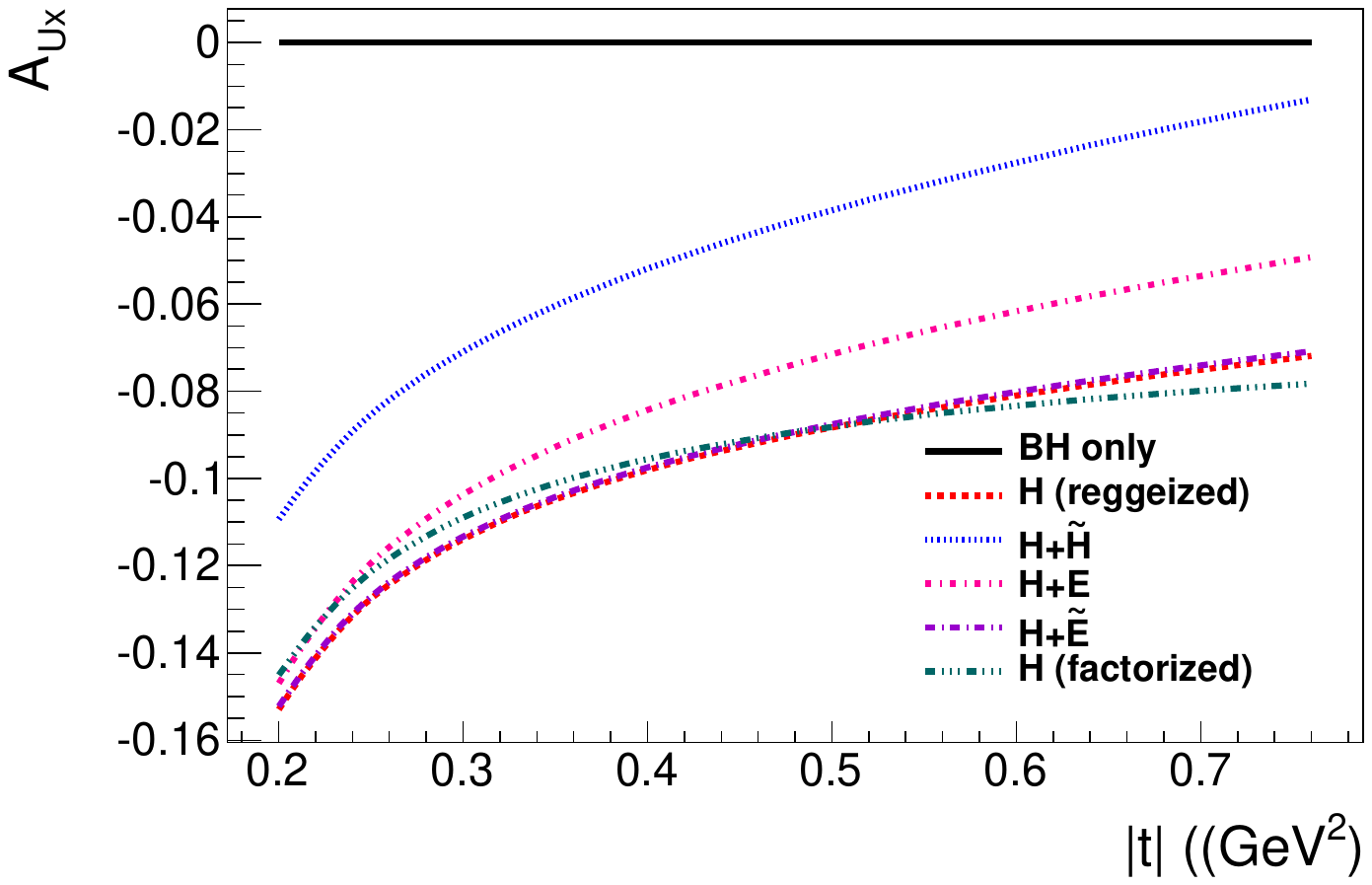}
%TSApx_t_7.pdf}
\hspace*{-0.8cm}
\includegraphics[width=6.5cm,height=6cm]
{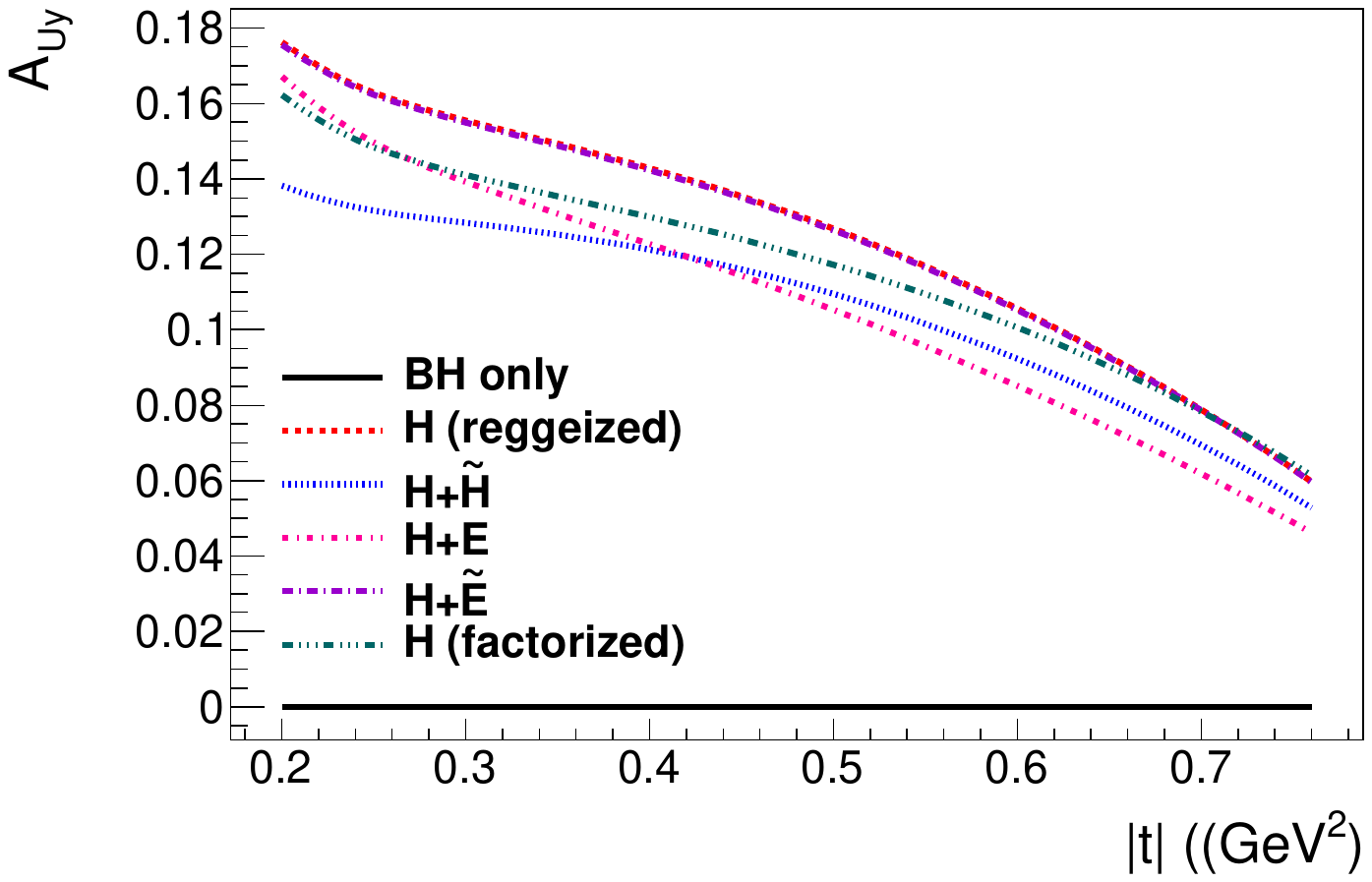}%TSApy_t_10.pdf}
\hspace*{-0.8cm}
\includegraphics[width=6.5cm,height=6cm]
{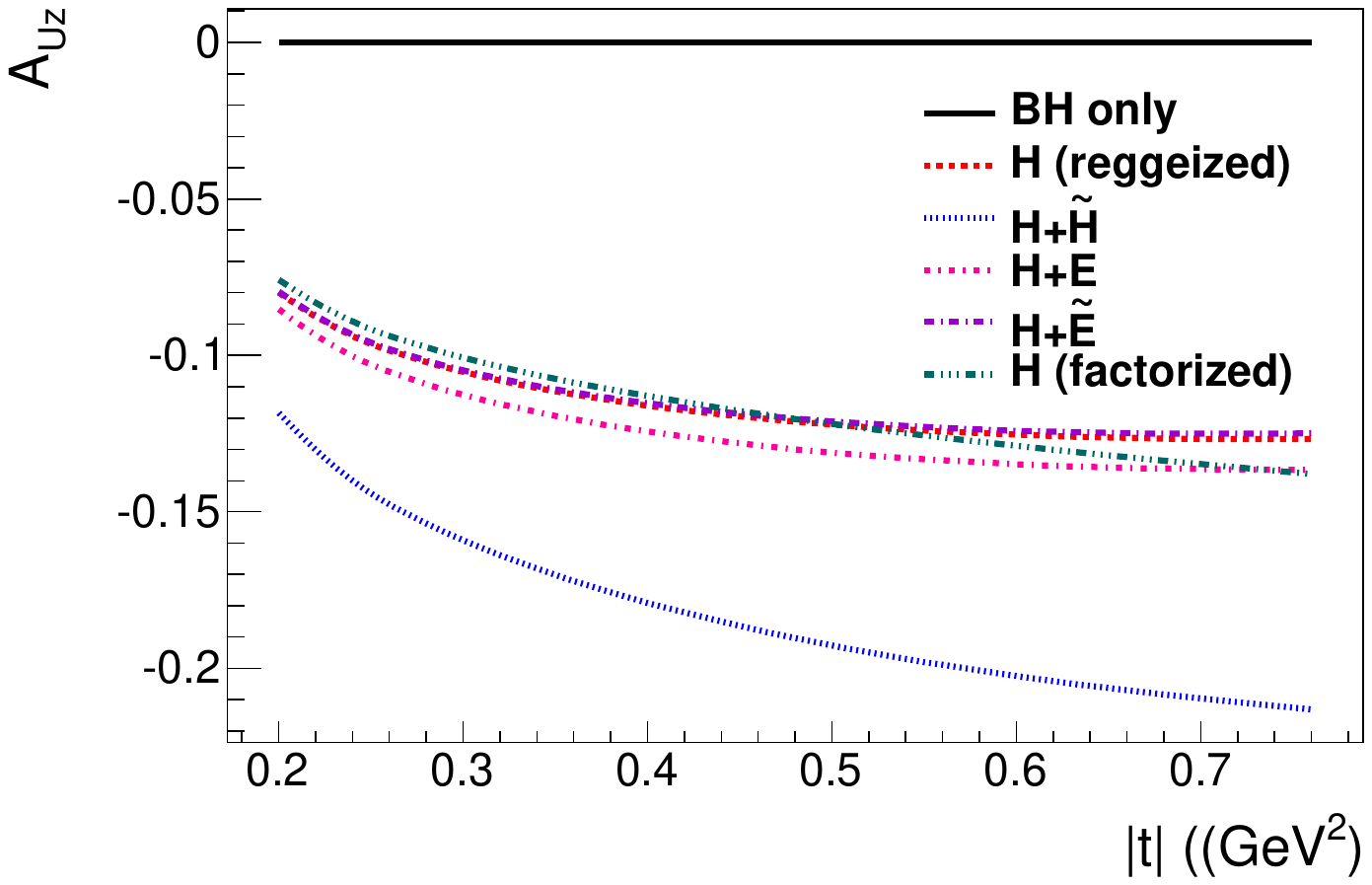}
%TSApz_t_7.pdf}
\hspace*{-0.8cm}
\caption{The $A_{Ux}$ (left panel),  $A_{Uy}$ (central panel) 
and $A_{Uz}$ (right panel) asymmetries as a function of $t$,
at $\phi$=90$^\circ$, 0$^\circ$ and 90$^\circ$ respectively, and for 
$\xi=0.2$, $Q'^2=7$ GeV$^2$, $-t=0.4$ GeV$^2$ and $\theta$ integrated over 
$[\frac{\pi}{4},\:\frac{3\pi}{4}]$. TCS is calculated with different GPD contributions.}
\label{fig:TSAvst}
\end{figure}
\end{center}

We show in Fig.~\ref{fig:TSA} our results for the $\phi$-dependence of 
$A_{Ux}$, $A_{Uy}$ and $A_{Uz}$ 
for $Q'^2=7$ GeV$^2$, $\xi=0.2$, $-t=0.4$ GeV$^2$ 
for %$\theta$=45$^\circ$, 90$^\circ$ and
 $\theta$ integrated over 
$[\frac{\pi}{4},\:\frac{3\pi}{4}]$. 
Like for $A_{\odot U}$, it is advantageous to integrate over $\theta$,
in order to maximize count rates, since the signal is barely reduced.
The TCS is calculated with different GPD contributions.
 We observe
$\sin\phi$ or $\cos\phi$ shapes with amplitudes between 10 and 15\%.
Like for $A_{\odot U}$, the BH doesn't produce any asymmetry 
and any non-zero asymmetry directly reflects the strength of GPDs.

We show in Fig.~\ref{fig:TSAvst} the $t$-dependence of $A_{Ux}$, $A_{Uy}$ 
and $A_{Uz}$ at $\phi$=90$^\circ$, 0$^\circ$ and 90$^\circ$ respectively,
for the kinematics $\xi=0.2$, $Q'^2=7$ GeV$^2$ and $\theta$ integrated 
over $[\frac{\pi}{4},\:\frac{3\pi}{4}]$. In this figure,
TCS is calculated with different GPDs. Depending on the value of $t$,
the three asymmetries are sensitive to the GPDs $H$, $\tilde{H}$ and $E$
in various proportions. We also display in this figure our calculation 
with the factorized
ansatz for the $t$-dependence of the $H$ GPD in order to illustrate the 
model-dependence of our results.

\newpage
\onecolumngrid

\twocolumngrid
\newpage

%%%%%
\section{Double spin asymmetries}
\label{sec:BTSA}

We define the double-spin asymmetries:
\begin{equation}
A_{ij}= \frac{ (\sigma^{++} + \sigma^{--}) - 
( \sigma^{+-} +  \sigma^{-+}) }{\sigma^{++} + \sigma^{--}+
\sigma^{+-} +  \sigma^{-+}},
\end{equation}
where $\sigma^{\pm\pm}$ stand for the four-fold differential cross sections 
$\frac{d\sigma}{dQ'^2\:dt\:d\phi\:d(cos\theta)}$ 
for the two photon beam spin states $+$ and $-$ (first index) 
and the two target spin orientations $+$ 
and $-$ (second index) along the target polarization axis. The first index of $A$ refers
to the polarization nature of the photon beam ($i=\ell$ for a linear polarization
and $i=\odot$ for a circular polarization) and the second index
refers to the axis polarization of the target $j=x,y$ or $z$.
We present and discuss in the two following subsections our results
for  $A_{\odot j}$ and $A_{\ell j}$.

\subsection{Circularly polarized photons and polarized target}

\begin{figure}[htbp] 
\begin{center}
\hspace*{-0.8cm}
\includegraphics[width=6.5cm,height=6cm]
{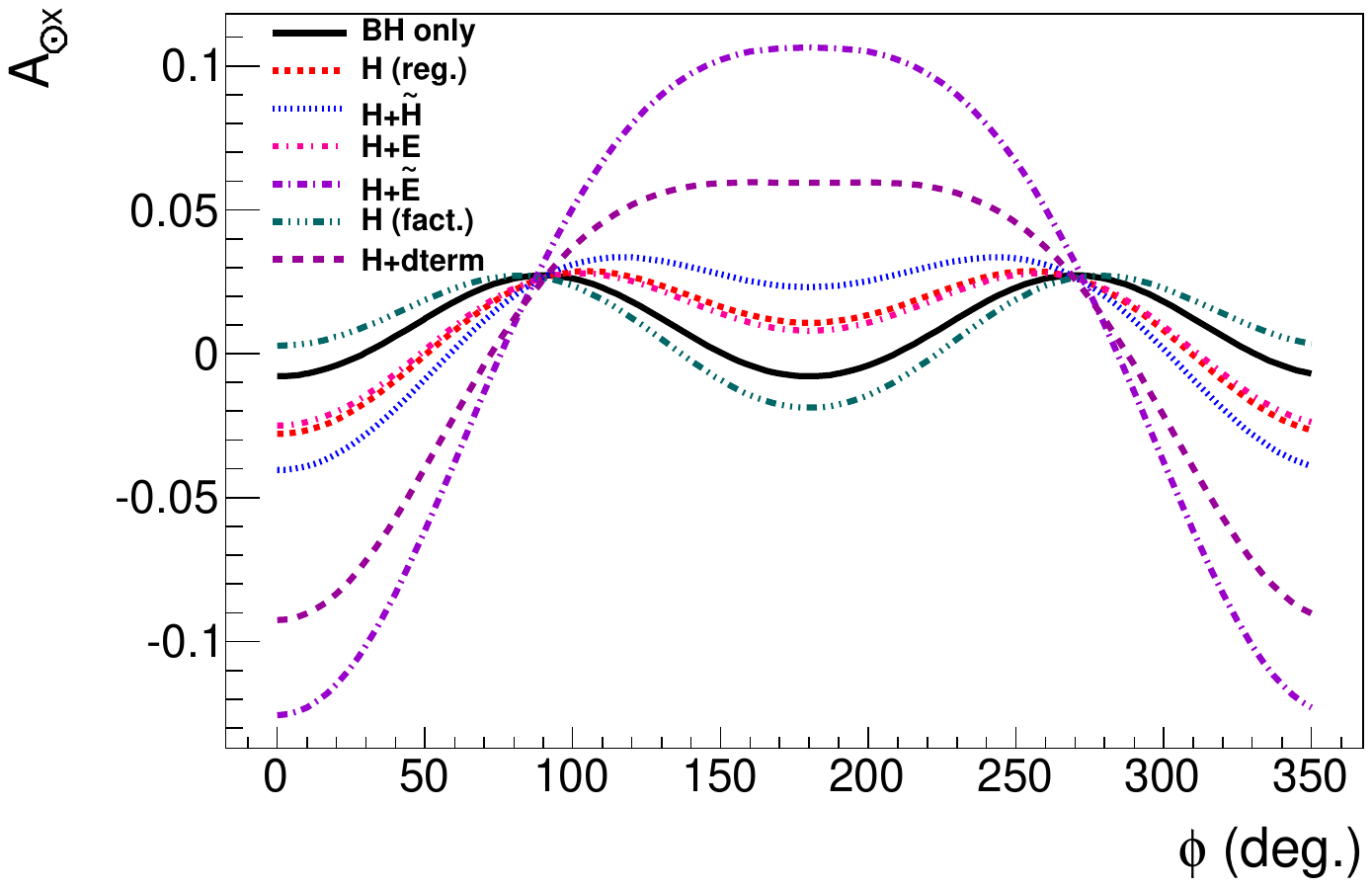}
\hspace*{-0.8cm}
\includegraphics[width=6.5cm,height=6cm]
{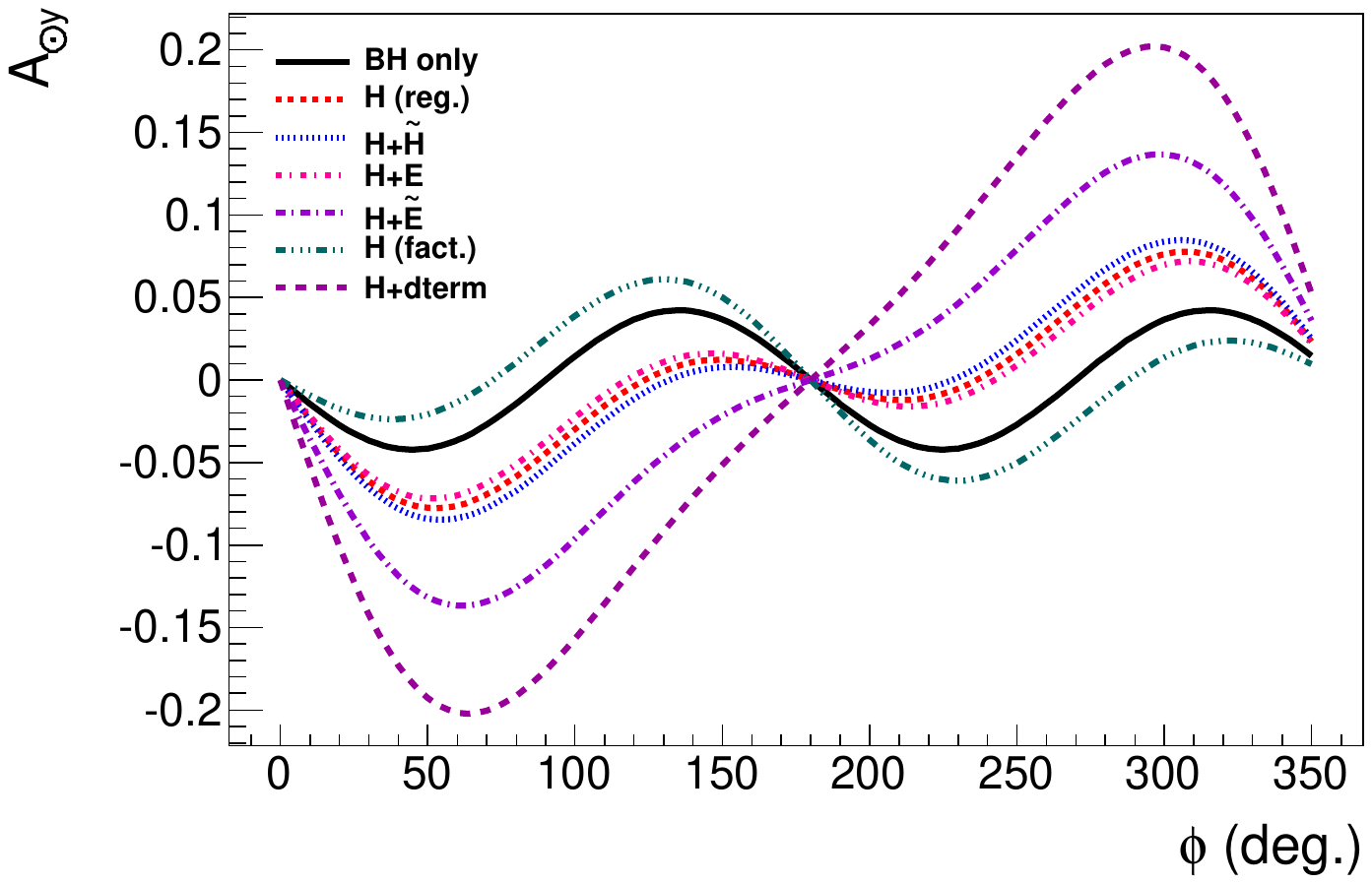}
\hspace*{-0.8cm}
\includegraphics[width=6.5cm,height=6cm]
{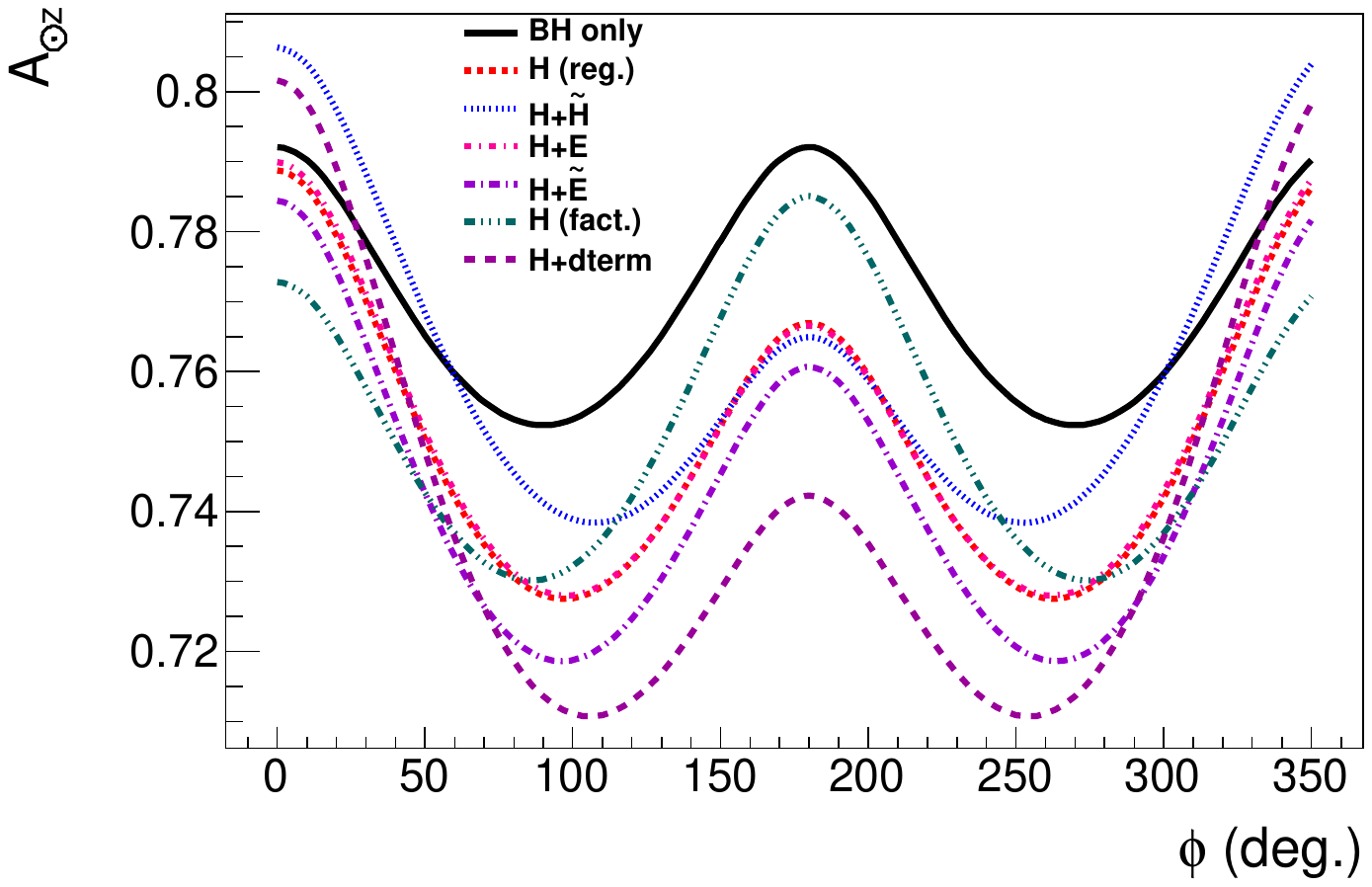} 
\hspace*{-0.8cm}
\\
\hspace*{-0.8cm}
\includegraphics[width=6.5cm,height=6cm]
{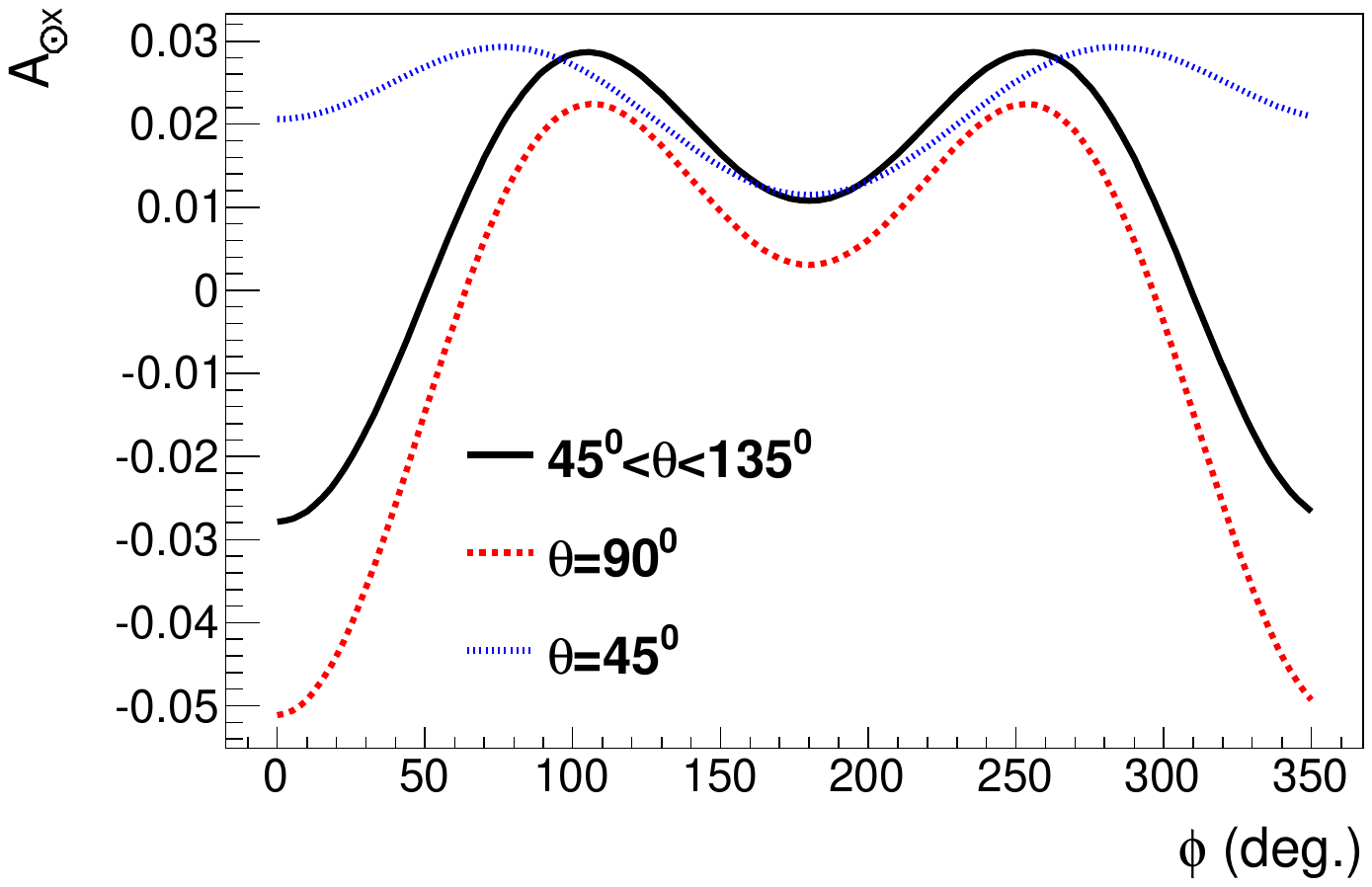} %BTSApx_phi_compar_2.pdf}
\hspace*{-0.8cm}
\includegraphics[width=6.5cm,height=6cm]
{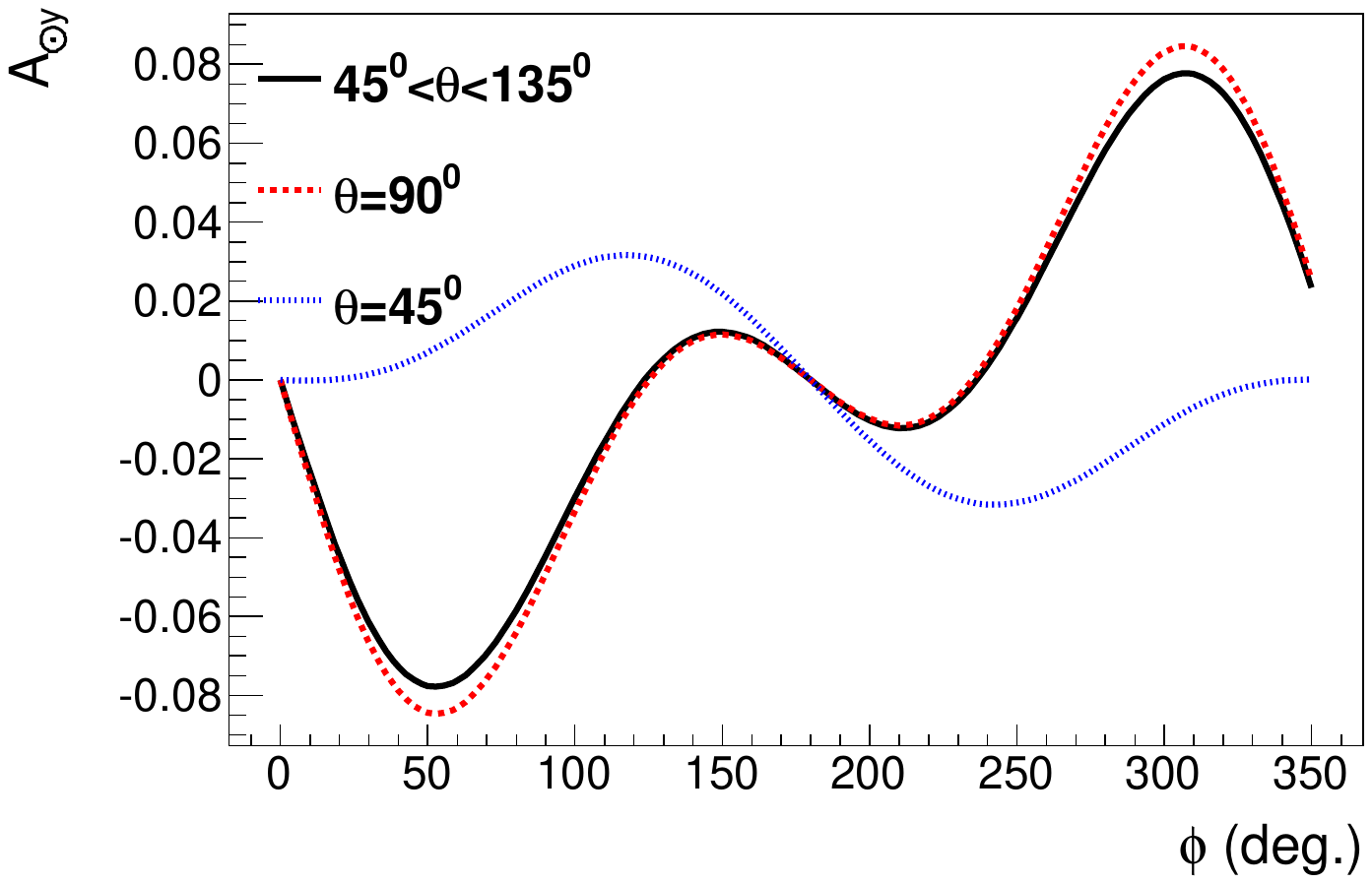} %BTSApy_phi_compar_5.pdf}
\hspace*{-0.8cm}
\includegraphics[width=6.5cm,height=6cm]
{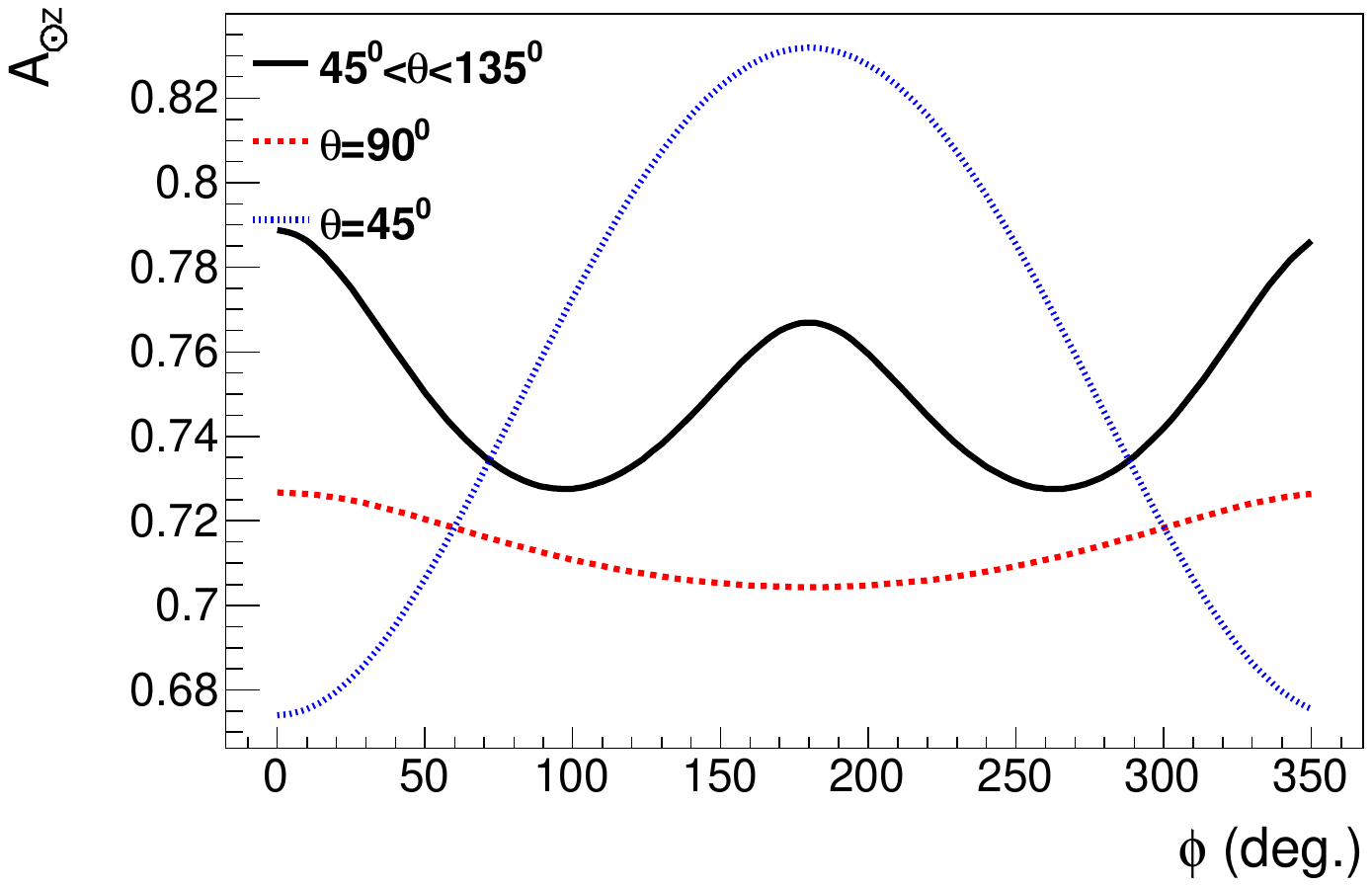} %BTSApz_phi_compar_1.pdf}
\hspace*{-0.8cm}
\caption{Circularly polarized beam-target double spin asymmetries
as a function of $\phi$ at $\xi=0.2$, $-t=0.4$ GeV$^2$ 
and $Q'^2=7$ GeV$^2$.
Left column: $A_{\odot x}$, central column:  $A_{\odot y}$, 
right column:  $A_{\odot z}$.
In the top row panels, 
$\theta$  integrated over $[\frac{\pi}{4},\:\frac{3\pi}{4}]$ 
and the calculations are done for different GPD contributions to the TCS process.
In the bottom row panels, the 
calculations are done for different $\theta$ angles:
$\theta=90^\circ$ (red dashed curve), $\theta=45^\circ$ (blue dotted curve)
 and $\theta$ integrated over $[\frac{\pi}{4},\:\frac{3\pi}{4}]$ (black solid curve)
 and with the GPD $H$ (Reggeized) contribution only. 
}
\label{fig:BTSA}
%\end{center}
%\end{figure}
%% evolution avec t
%\begin{figure}[htbp]
%\begin{center}
\hspace*{-0.8cm}
\includegraphics[width=6.5cm,height=6cm]
{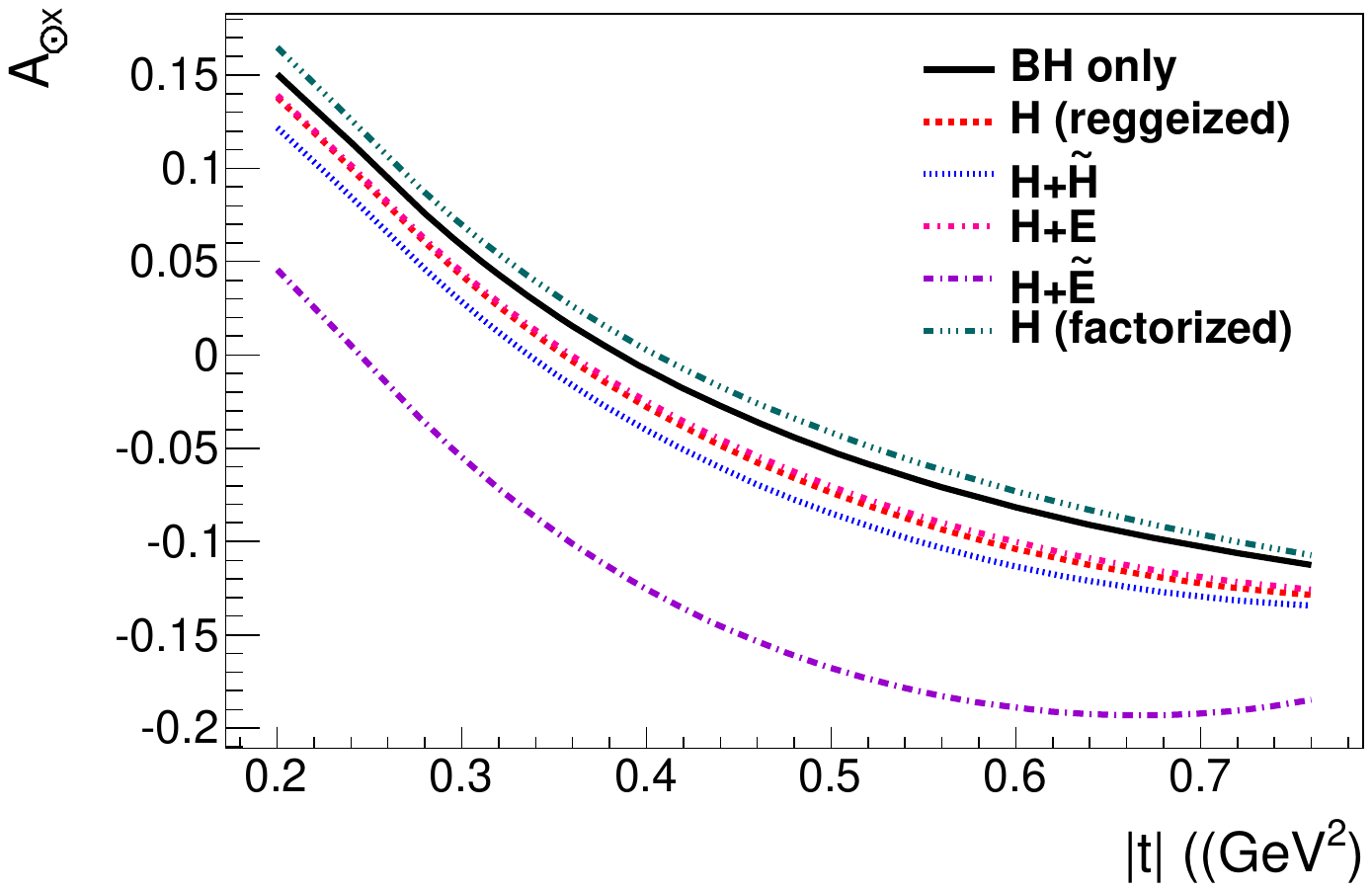} %BTSApx_t_9.pdf}%
\hspace*{-0.8cm}
\includegraphics[width=6.5cm,height=6cm]
{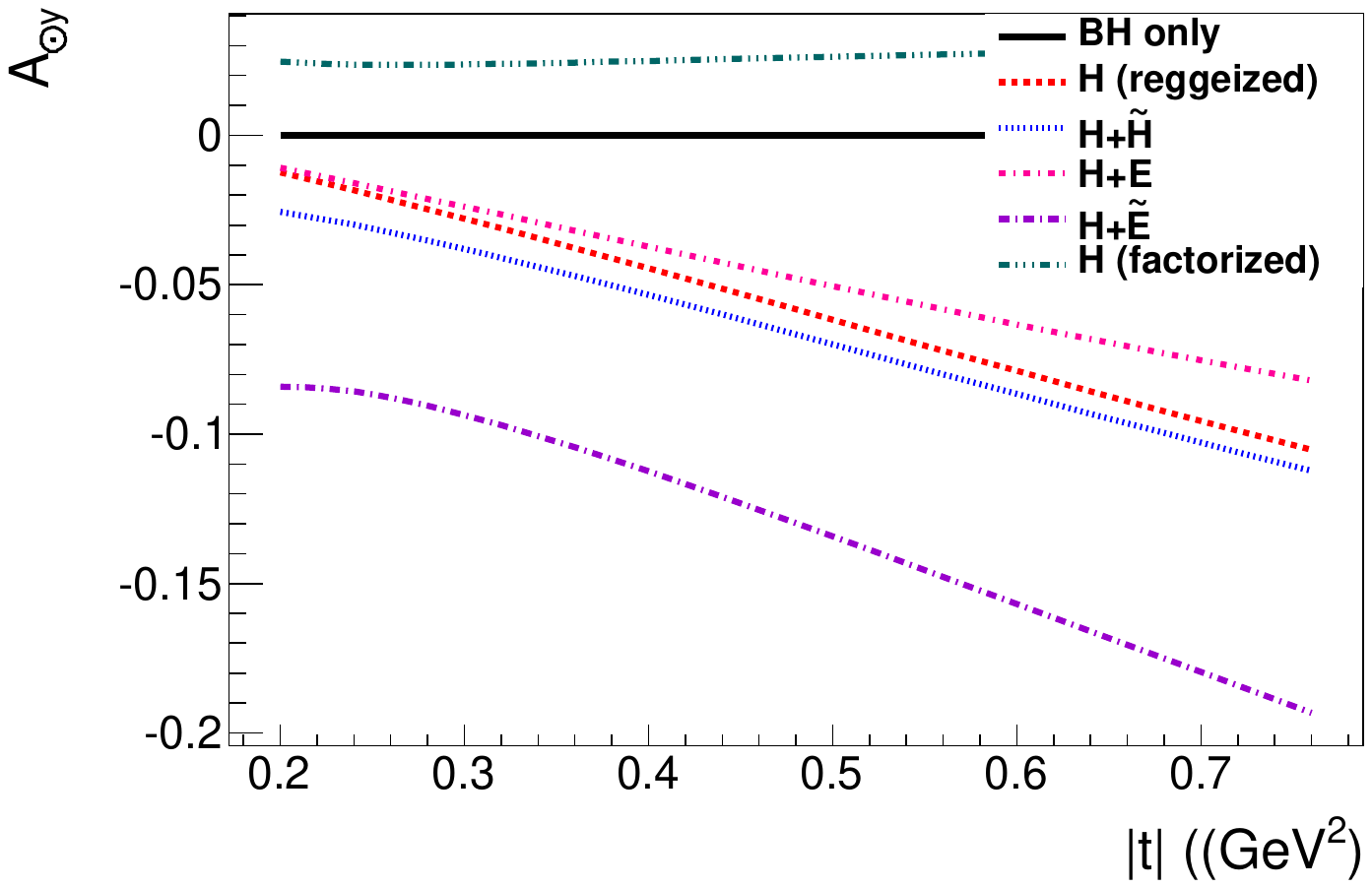} %BTSApy_t_13.pdf}
\hspace*{-0.8cm}
\includegraphics[width=6.5cm,height=6cm]
{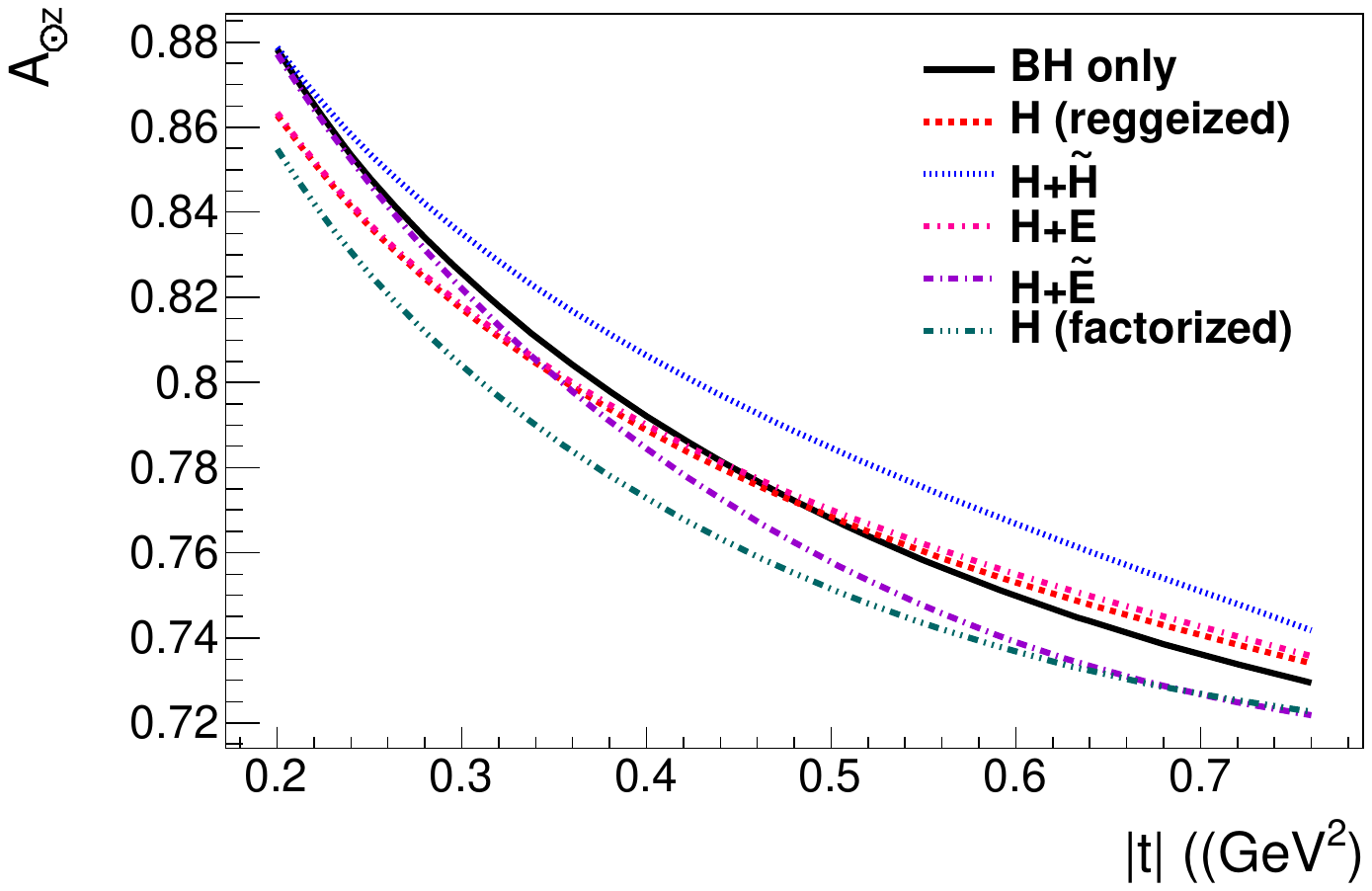}  %BTSApz_t_10.pdf}
\hspace*{-0.8cm}
\caption{Circularly polarized beam-target double spin asymmetries
as a function of $t$. Left: $A_{\odot x}$, center:  $A_{\odot y}$, 
right: $A_{\odot z}$. Calculations are done at the kinematics: 
$\xi=0.2$, and $Q'^2=7$ GeV$^2$ and $\phi$=0$^\circ$ ($A_{\odot x}$), 
$\phi=90^\circ$ ($A_{\odot y}$) and $\phi=0^\circ$ ($A_{\odot z}$).
$\theta$ is integrated over $[\frac{\pi}{4},\:\frac{3\pi}{4}]$
in these calculations.}
\label{fig:BTSAvst}
\end{center}
\end{figure}

Fig.~\ref{fig:BTSA} shows our results for $A_{\odot x}$,
$A_{\odot y}$ and $A_{\odot z}$, from left to right, 
as a function of $\phi$ at the kinematics
$\xi=0.2$, $-t=0.4$ GeV$^2$, $Q'^2=7$ GeV$^2$. The top row shows the result of
our calculations for $\theta$ integrated over $[\frac{\pi}{4},\:\frac{3\pi}{4}]$
with different GPD contributions to the TCS amplitude. The bottom row shows
the same observables with only the GPD $H$ contribution for different
$\theta$ angle sets.

One notes that the BH process alone produces asymmetries in all cases. 
The $\phi$-shapes of the asymmetries are complex and very dependent on $\theta$.
Also, in contrast to the single spin asymmetries that we
studied in the previous section, the $\phi$-shapes are also very dependent 
on the specific GPDs entering the TCS process. One notes in particular
important sensitivities to the $H$, $\tilde{H}$, $\tilde{E}$ GPDs 
as well as to the D-term. Furthermore, one also notes the sensitivity 
of these double-polarization observables to the ansatz used for the
$t$-dependence.

%When $\theta$ is integrated over, this sensitivity remains for
%$A_{\odot x}$ and $A_{\odot y}$ but it tends to vanish for $A_{\odot z}$.

Fig.~\ref{fig:BTSAvst} shows the $t$-dependence of the double spin asymmetries 
$A_{\odot x}$,  $A_{\odot y}$ and $A_{\odot z}$ at $\phi=0^\circ$, 
$\phi=90^\circ$ and $\phi=0^\circ$ respectively, at the kinematics
$\xi=0.2$, $Q'^2=7$ GeV$^2$ and $\theta$ integrated over $[\pi/4,3\pi/4]$.
We also display in this figure our calculation with the factorized
ansatz for the $t$-dependence of the $H$ GPD in order to illustrate the 
model-dependence of our results. The change of sign for $A_{\odot y}$ for the
factorized ansatz compared to the Reggeized ansatz is in particular remarkable.
As can be seen in the top panel of Fig.~\ref{fig:BTSA}, this is due to the
fact that the factorized model crosses the ``zero" line before $\phi=90^\circ$
while the Reggeized model crosses it after, thus producing respectively
positive and negative $A_{\odot y}$'s at $\phi=90^\circ$.

\subsection{Linearly polarized photons and polarized target}

\begin{figure}
\begin{center}
\hspace*{-0.8cm}
\includegraphics[width=6.5cm,height=6cm]
{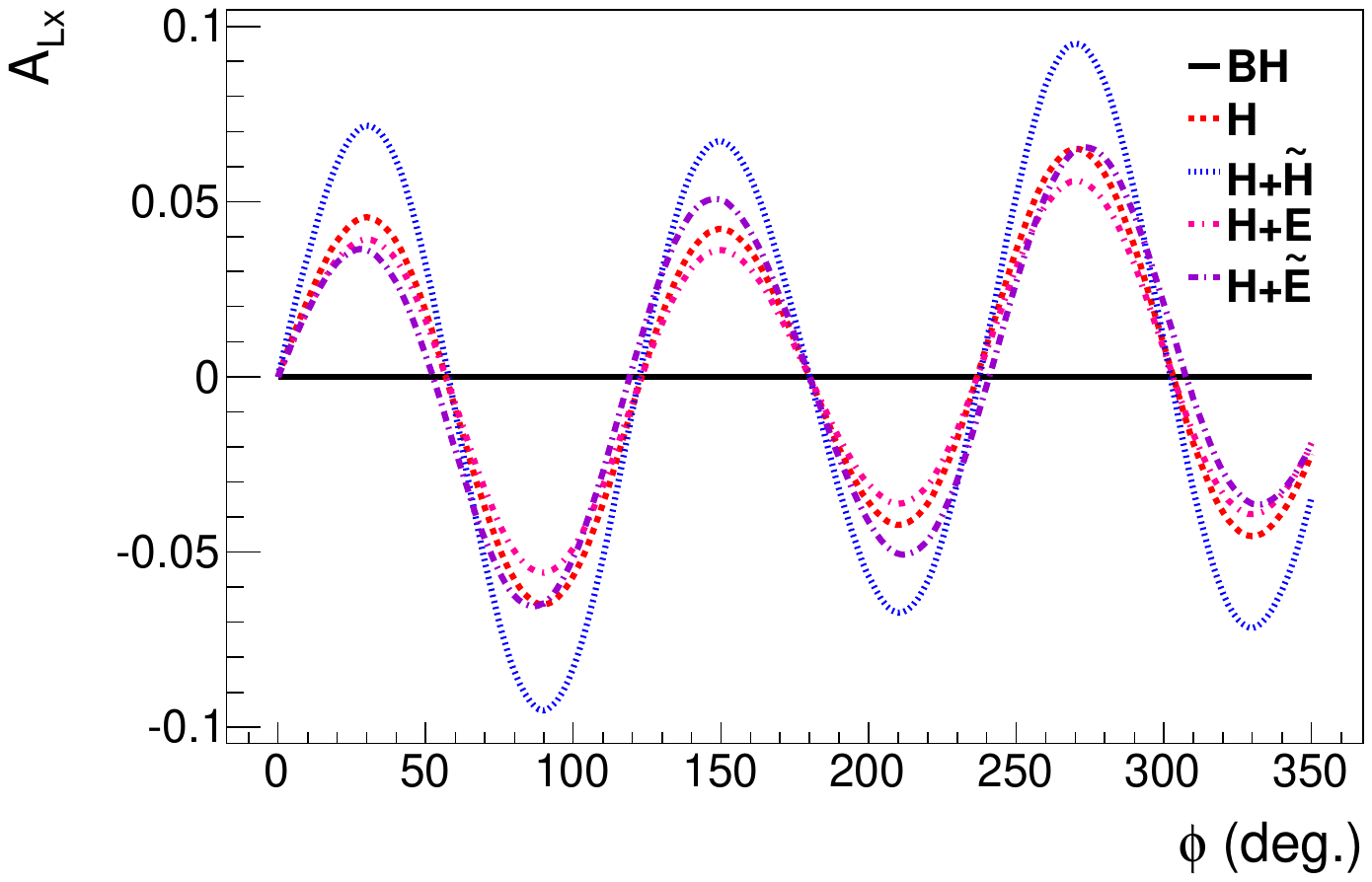}
\hspace*{-0.8cm}
\includegraphics[width=6.5cm,height=6cm]
{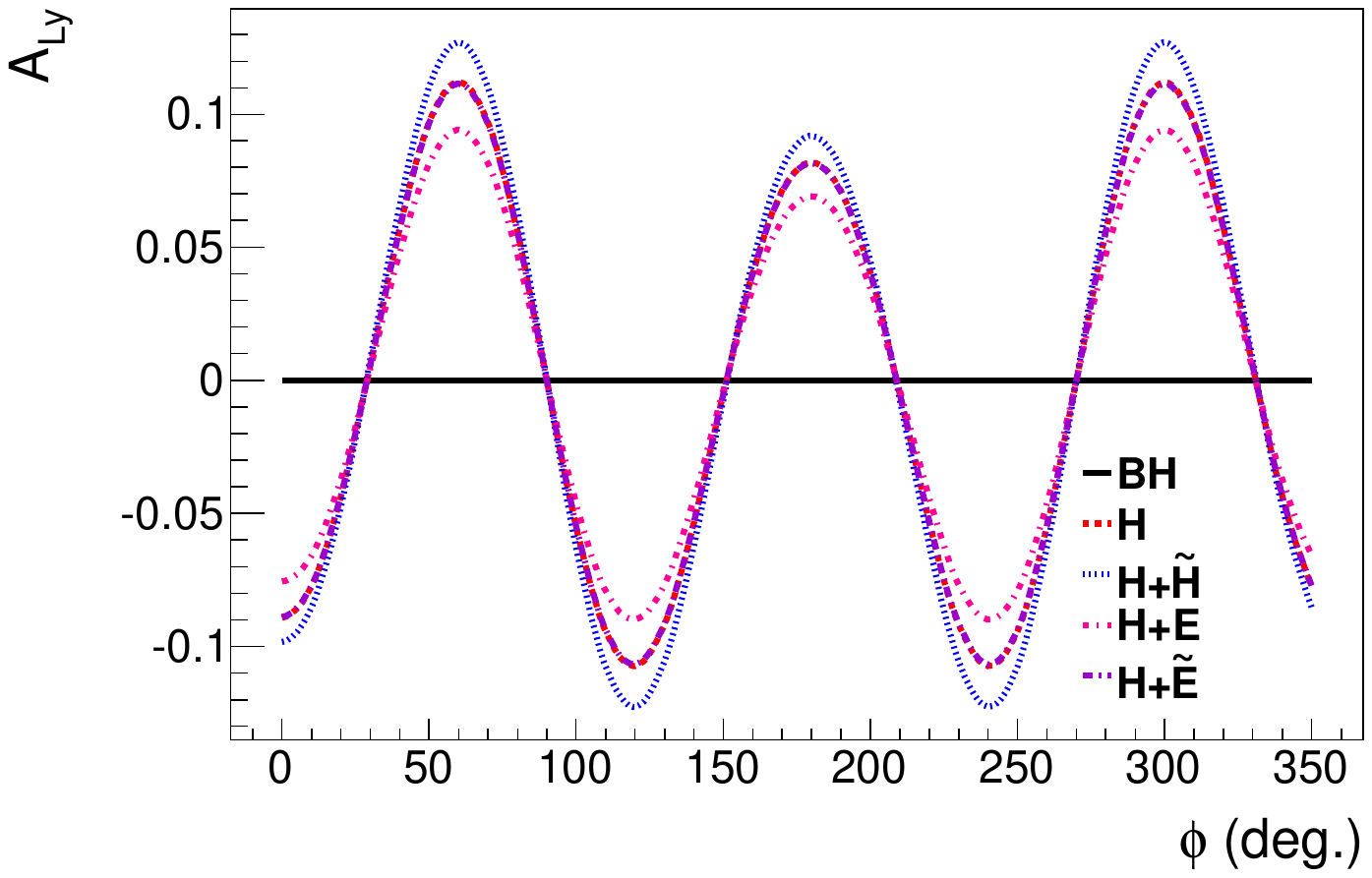}
\hspace*{-0.8cm}
\includegraphics[width=6.5cm,height=6cm]
{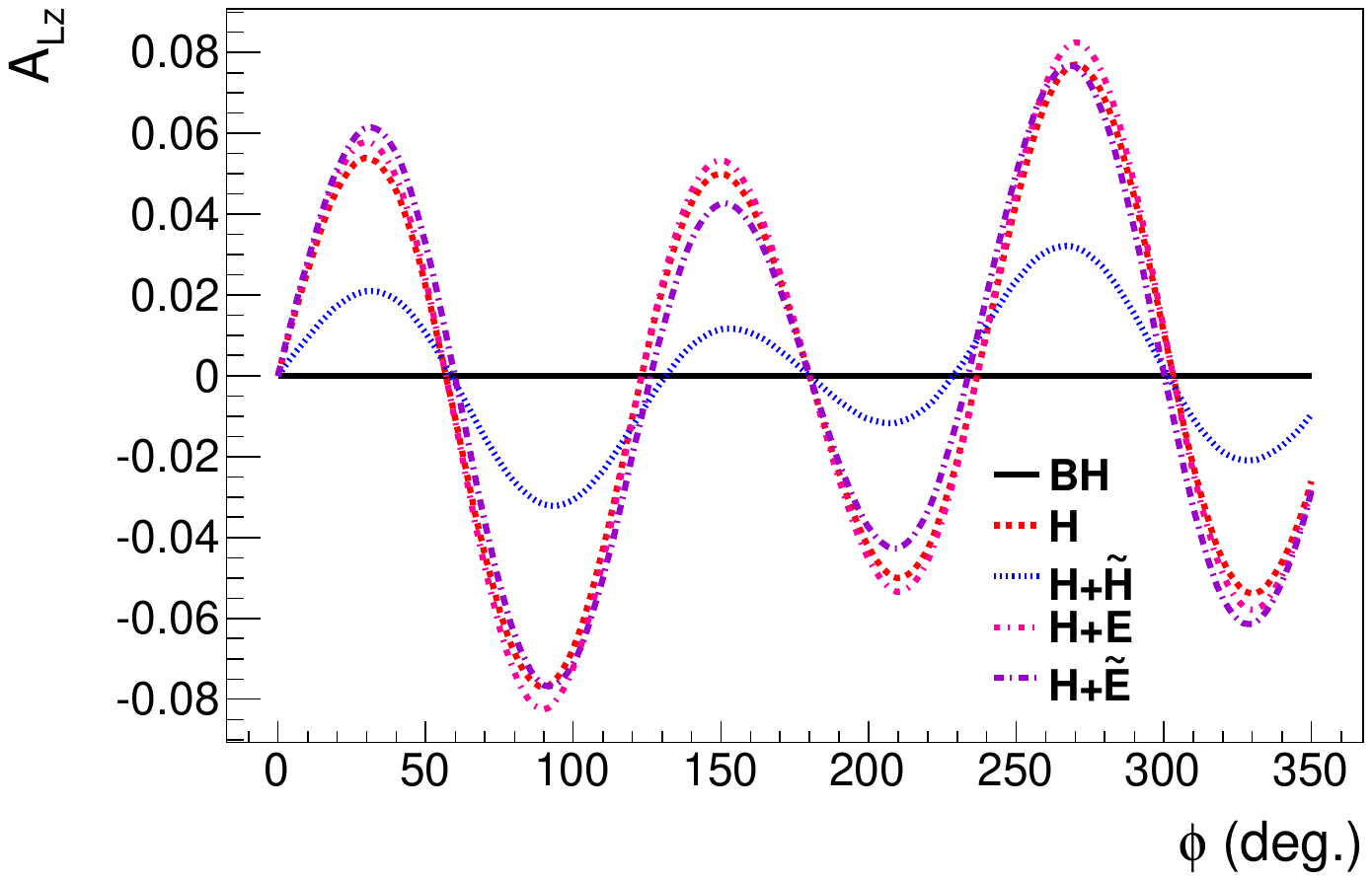}
\caption{Linearly polarized beam-target double spin asymmetries
as a function of $\phi$ at $\xi=0.2$, $-t=0.4$ GeV$^2$ 
and $Q'^2=7$ GeV$^2$ and $\theta$ integrated over $[\frac{\pi}{4},\:\frac{3\pi}{4}]$ 
for different GPD contributions.
Left column: $A_{\ell x}$, central column:  $A_{\ell y}$, 
right column:  $A_{\ell z}$.}
\label{fig:BTSAvst_lin}
\end{center}
\end{figure}

Fig.~\ref{fig:BTSAvst_lin} shows our results for the double-spin asymmetries $A_{\ell x}$,
$A_{\ell y}$ and $A_{\ell z}$, from left to right, 
as a function of $\phi$ at the kinematics
$\xi=0.2$, $-t=0.4$ GeV$^2$, $Q'^2=7$ GeV$^2$ with
$\theta$ integrated over $[\frac{\pi}{4},\:\frac{3\pi}{4}]$.

One notes that the BH process alone produces a null asymmetry in all cases,
making this double-spin asymmetry particularly favorable to study TCS and GPDs. 
$A_{\ell y}$ is mostly sensitive to the GPD $H$ contribution while 
$A_{\ell x}$ and $A_{\ell z}$ show a sensitivity to the GPD $\tilde H$ as well.
 
\newpage
\onecolumngrid

\twocolumngrid
\newpage

\begin{figure}[h!]
\begin{center}
\includegraphics[width=8.5cm,height=7.cm]
{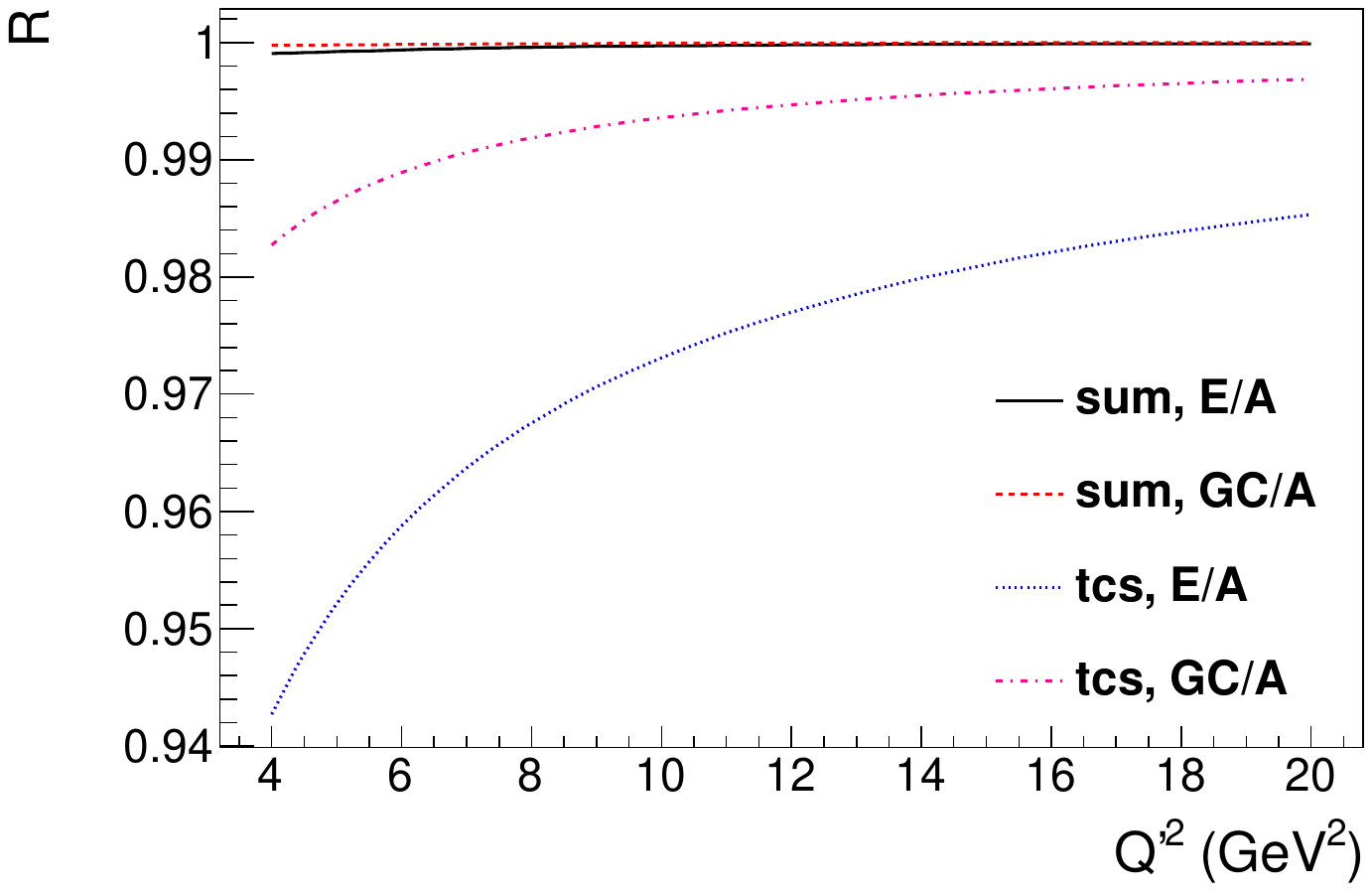}
\hspace*{-0.5cm}
\includegraphics[width=8.5cm,height=7cm]
{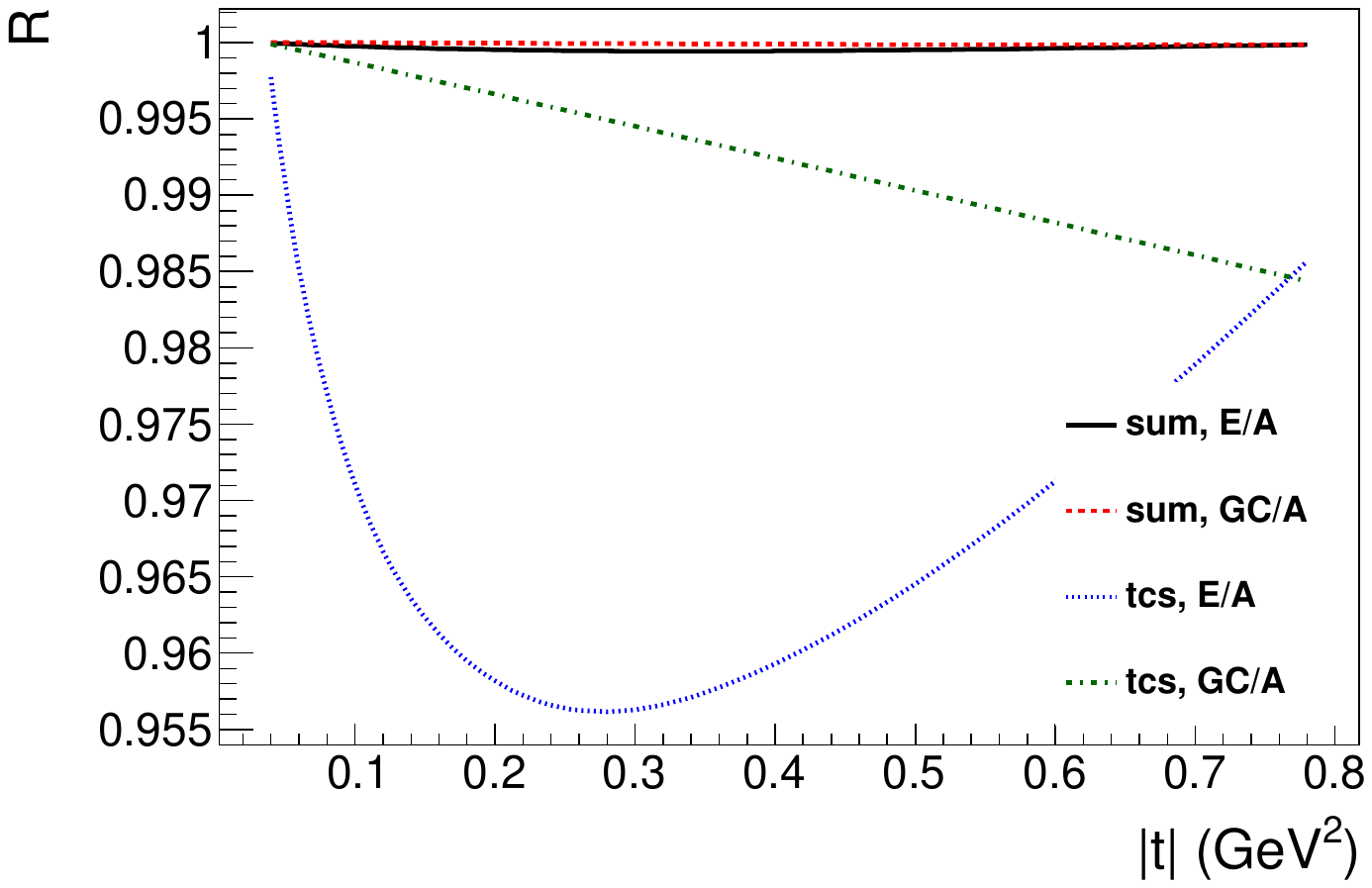}\\
\caption{  
$Q'^2$-dependence (left panel)  and 
$|t|$-dependance  (right panel)  
 of the ratio of the 2-fold cross differential cross sections
$d\sigma/dtdQ'^2$ of Bethe-Heitler and of TCS 
(integrated over the decay angles: $\phi\in[0,2\pi]$ 
and $\theta\in[\frac{\pi}{4}, \frac{3\pi}{4}]$) calculated 
with the gauge invariance restoration term (dot-dashed curve for TCS: gauge invariant/asymptotic) and with 
the exact kinematics (dotted curve for TCS: "exact"/asympototic) to the asymptotic limit
(i.e. Bjorken limit) calculation. The calculation has been 
carried out for $\xi=0.1$,  $\theta$ is integrated
      over $[\pi/4,3\pi/4]$, $\phi$ is integrated over $[0,2\pi]$ and at $-t=0.4$ GeV$^2$ (left panel) or at $Q'^2=7$ GeV$^2$ (right panel). 
}
\label{fig:corr_cross}
\end{center}
\end{figure}

\begin{figure}[h!]
\begin{center}
\includegraphics[width=8.5cm,height=7.cm]
{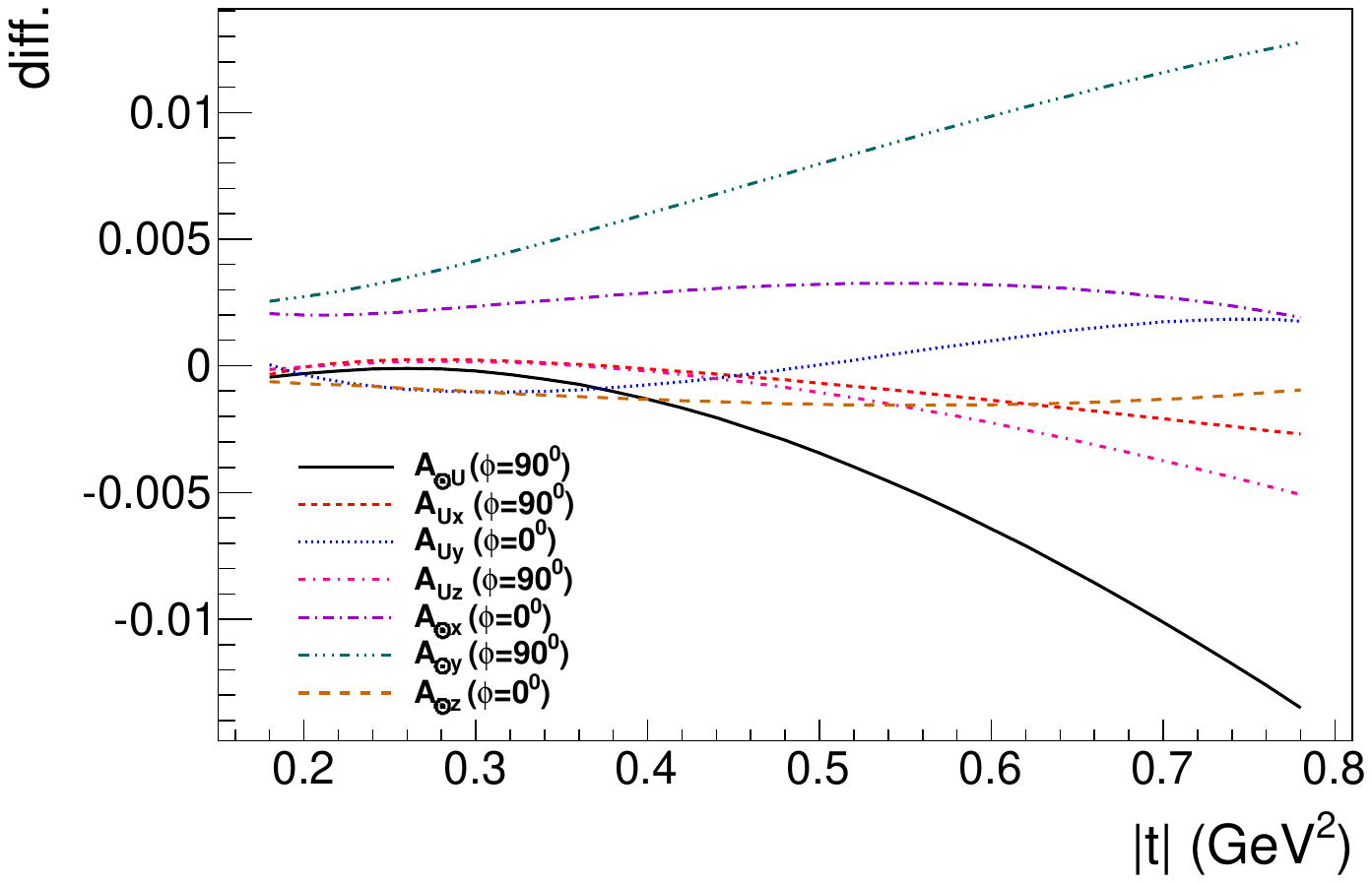}
\hspace*{-0.5cm}
\includegraphics[width=8.5cm,height=7cm]
{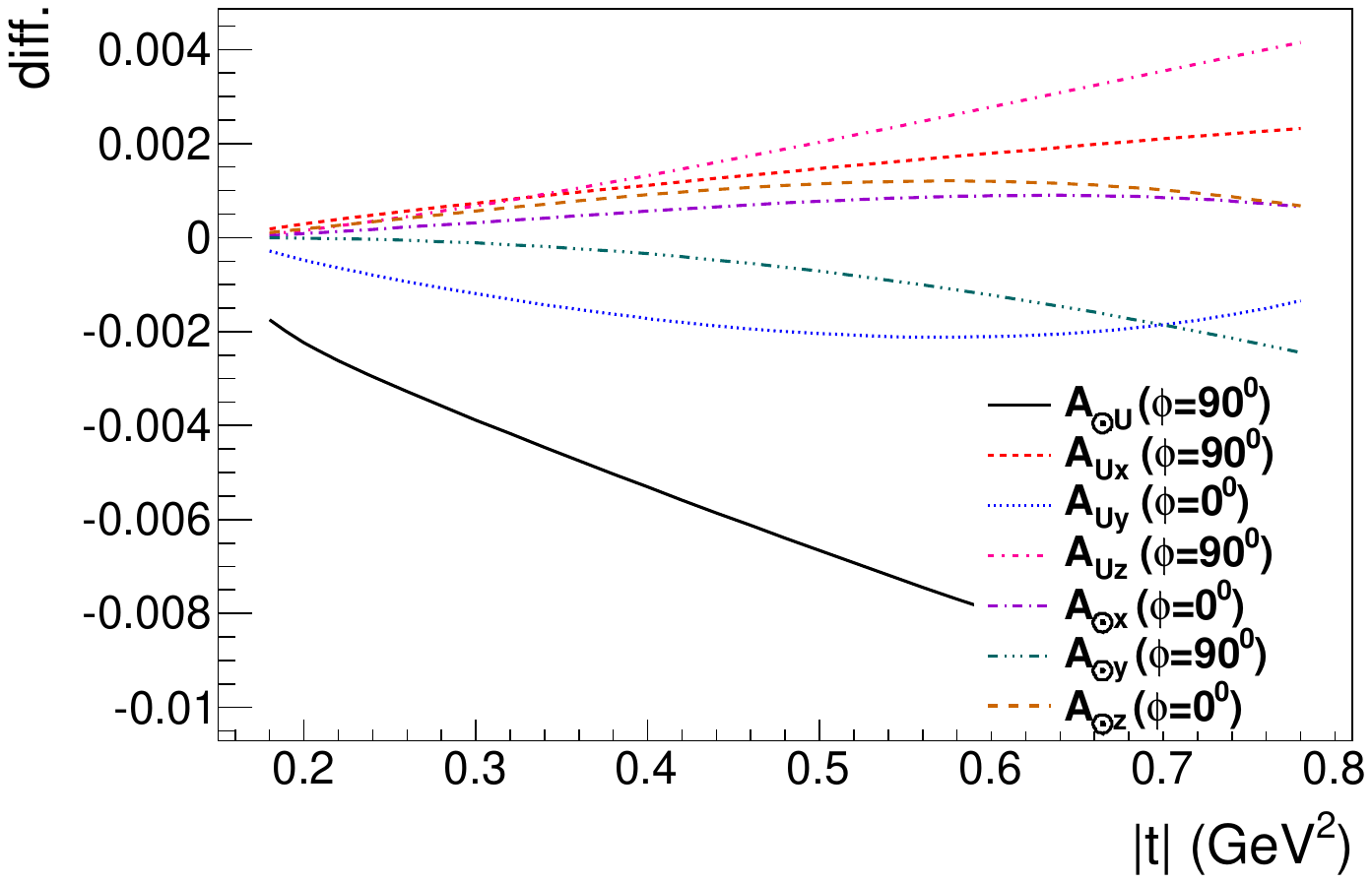} \\
\caption{ Difference of the single and 
double spin asymmetries (circularly polarized photon)  as a function of $|t|$ before and after the corrections. Left panel:
with the exact kinematic compared  to the asymptotic limit. 
Right panel:
 with the gauge invariance restoration term compared to the asymptotic limit. 
%1/ exact-asymp, 2/ gauge correction - asymp
The calculations have been carried out for $\xi=0.2$, $-t=0.4$ GeV$^2$, $Q'^2=7$ GeV$^2$
and $\theta$ has been integrated over $[\pi/4,3\pi/4]$. The different $\phi$ angles are indicated on the figure.
}
\label{fig:corr_asym}
\end{center}
\end{figure}

\twocolumngrid
\newpage

%%%%%%%%%%%%%%%%%%%%%%%%%ùù
\section{Corrections to the leading-twist calculations}
\label{sec:Corrections} 

We evaluate in this final section two types of higher-twist corrections: keeping
the exact kinematics presented in section~\ref{subsec:kin}, i.e. 
$\mid\tilde{\xi\mid} \neq \mid\xi\mid \neq \mid\tilde{\xi}'\mid \neq \mid\xi'\mid$,
and adding the gauge correction term of section~\ref{subsec:gi} to the   
TCS tensor.

Fig.~\ref{fig:corr_cross} shows the $Q'^2$-dependence and the $|t|$-dependence 
of the ratio of the 2-fold cross differential cross sections
$d\sigma/dtdQ'^2$ for Bethe-Heitler and for TCS 
(integrated over the decay angles: $\phi\in[0,2\pi]$ 
and $\theta\in[\frac{\pi}{4}, \frac{3\pi}{4}]$) calculated 
with the gauge invariance restoration term (dot-dashed curve for TCS) and with 
the exact kinematics (dotted curve for TCS) to the asymptotic limit
(i.e. Bjorken limit) calculation that we have presented so far.
%The upper panel of the figure shows the $Q'2$-dependence at $-t=0.4$ GeV$^2$ 
%and the bottom panel the $t$-dependence at $Q'^2=7$ GeV$^2$ of this ratio. Both
The calculation has been carried out for $\xi=0.1$ and at $-t=0.4$ GeV$^2$ (left panel) or 
at $Q'^2=7$ GeV$^2$ (right panel). 
One sees that these ratios tend to 1 as $Q'^2$ increases and as $|t|$ decreases, as expected. 
The effects of these corrections in the
domains of current interest, covered in this figure, are of the order of a few percents for TCS. 
The exact kinematic corrections are more important (always less than 6\% though)
than the gauge invariance restoring corrections.

We show in Fig.~\ref{fig:corr_cross}
that the impact of these corrections on the asymmetries that we discussed earlier
is of the order of 0.1\% to 1\% and don't affect the conclusions that we drew earlier.\\

%%%%
\section{Conclusion}
\label{sec:conclusion}

In this work, we have studied the $\gamma P\to P' e^+e^-$ process 
in terms of GPDs in a regime where one can expect to access them. 
We have presented our derivations of the TCS
and BH amplitudes both contributing to the $\gamma P\to P' e^+e^-$ process
and calculated all unpolarized, single beam, single target and 
double beam-target spin observables. 

We show that, since the TCS process is lower by several factors in 
the unpolarized cross section compared to the BH process, it is judicious 
to measure spin asymmetries which reveal a more direct sensitivity to GPDs. 
In particular, the BH process alone doesn't produce any signal for
the target single spin asymmetries, for the circularly polarized
beam single spin asymmetries and for the linearly polarized photons and polarized target
double spin asymmetries. These observables are therefore
particularly favorable to directly measure
GPD strength. We have shown in general that the various single-
and double-polarization observables that we have calculated show
different sensitivities to the four GPDs, which should ultimately allow
to disentangle them with some adequate GPD fitting algorithms. 
We also provided first estimations of higher-twist
contributions such as keeping the exact kinematics and including a
gauge invariance restoring term. The effects are at the level of a few percent 
on cross sections and spin asymmetries.
 
A rich new experimental TCS program can be envisioned with the forthcoming 
JLab energy upgrade, which would complement the DVCS program already 
approved in order to access GPDs. All of the TCS observables that we calculated
in this work should be measurable and can serve as a basis for developping experimental proposals.
This work might also find some applications at higher energies, like at hadron colliders,
such as LHC and RHIC, in ultraperipheral collisions~\cite{Pire:2008ea} or at
the projected electron-ion collider EIC~\cite{Accardi:2012qut}.
 
\section*{Acknowledgments}

We are very thankful to A.~T.~Goritschnig, B. Pire, L. Szymanowski, S. Wallon and J. Wagner 
for useful discussions and comments on this work.
The work of M.V. is supported by the Deutsche Forschungsgemeinschaft DFG 
through the Collaborative Research Center ``The Low-Energy Frontier of the
Standard Model" (SFB 1044) and the Cluster of Excellence ``Precision Physics, Fundamental
Interactions and Structure of Matter" (PRISMA). 
M. G. and M. V. are also supported by the Joint Research Activity ``GPDex" of the
European program Hadron Physics 3 under the Seventh Framework Programme of the European Community.
M.B. and M.G. also benefitted from the GDR 3034 ``PH-QCD" and the ANR-12-MONU-0
008-01 ``PARTONS" support.

%########################################################################
% Bibliographie
%########################################################################
% \newpage
 
%\onecolumn
%\thispagestyle{headings} 

%%%%%%%%%%%%%%%%%%%%%%%%%%%%%%%%%%%%%%%%%%%%%%%%%%%%%%%%%%%%%%%%%%%%%%%%%%%%%%%%%%%
%%%%%%%%%%%%%%%%%%%%%%%%%%%%%%%%%%%%%%%%%%%%%%%%%%%%%%%%%%%%%%%%%%%%%%%%%%%%%%%%%%%

%%%%%%%%%%%
\end{document}